\documentclass[
    balance,
    colorlinks,
    upint,
    subscriptcorrection,
    varvw,
    german,
    spanish,
    vietnamese,
    russian,
    greek,
    article
]{asmeconf}

\usepackage[cal=boondox]{mathalfa}

\usepackage{graphicx}
\usepackage{subcaption}
\usepackage{tcolorbox}
\usepackage{booktabs}  
\usepackage{multirow}  
\tcbuselibrary{skins}
\usepackage{overpic}
\usepackage{xcolor}
\usepackage{tabularx}
\usepackage{booktabs}
\usepackage{array}
\usepackage{multirow}
\usepackage{siunitx}
\usepackage{tikz}
\usepackage{float}
\usepackage{makecell}
\usetikzlibrary{shapes.geometric, arrows,fit,backgrounds,positioning,calc}
\usetikzlibrary{arrows.meta, positioning}
\usepackage{amsmath}
\hypersetup{%
	pdfauthor={author1},									  
	pdftitle={ASME Conference Paper LaTeX Template},                  
	pdfkeywords={ASME conference paper, LaTeX template, BibTeX style},
	pdfsubject = {Describes the asmeconf LaTeX template},			  
}

\begin{document}

\ConfName{Proceedings of the \linebreak International Design Engineering Technical Conferences and Computers and Information in Engineering Conference}
\ConfAcronym{IDETC-CIE 2025}
\ConfDate{Agust 17–20, 2025}
\ConfCity{California, USA}
\PaperNo{IDETC2025-168885}


\title{A Comparative Analysis of Robust and Reliable Designs Using the Compromise Decision Support Problem: A Case Study in Hot Rod Rolling Processes} 

\SetAuthors{%
	Maryam Ghasemzadeh\affil{1}, H M Dilshad Alam Digonta\affil{2}, Anand Balu Nellippallil\affil{2}, Anton van Beek\affil{1}\CorrespondingAuthor{anton.vanbeek@ucd.ie}
}

\SetAffiliation{1}{School of Mechanical and Materials Engineering, University College Dublin, Dublin, Ireland}

\SetAffiliation{2}{Department of Mechanical and Civil Engineering,Florida Institute of Technology, Melbourne, Florida, USA}
\maketitle




\begin{abstract}
Design under uncertainty is a challenging problem, as a system’s performance can be highly sensitive to variations in input parameters and model uncertainty. A conventional approach to addressing such problems is robust optimization, which seeks to enhance design performance by reducing sensitivity to uncertainty. Alternatively, reliability-based design focuses on optimizing performance while ensuring that failure constraints are satisfied with a specified probability. While both methods are well established, their integration into multi-objective and multi-stakeholder decision-making frameworks remains a challenging problem. In this study, we extend the Compromise Decision Support Problem (cDSP) framework to incorporate reliability-based design considerations and evaluate its performance in comparison to the conventional robust-based cDSP formulation. The developed framework has been validated on a multidisciplinary hot rod rolling process including parametric and model uncertainties. The results compare the predicted performance under robust and reliable scenarios, validating the efficiency of the approach in managing uncertainties for complex, multidisciplinary systems. Specifically, we found that the two methods exhibit markedly different performance when the predicted performance follows a non-normal distribution, a situation that arises in non-linear systems with parametric uncertainty. Based on this insight, we offer guidance to designers on the conditions under which each method is most appropriate.

\end{abstract}

\section{Introduction}

Multidisciplinary systems present a set of interconnected subsystems that form a network of interdependent relationships where the outputs of some subsystems serve as the input to others \cite{lin2004sequential}. Variations in subsystem parameters, caused by manufacturing tolerances and environmental influences, result in uncertainties propagating through the system \cite{madsen2006methods, du2002collaborative}, influence the overall performance \cite{lee2009comparative}. Additionally, these parametric uncertainties are often compounded by model uncertainties (e.g., uncertainties that arise from using incomplete and/or data-driven models), which originate from the inherent non-linearity of each subsystem and the limitations of available data \cite{sullivan2015introduction}. To establish high-performing systems, designers need to account for these sources of uncertainty and the interconnectedness of the subsystems. 

System-level uncertainty quantification often requires a large number of time-consuming or monetary exhaustive model evaluations\cite{lin2004sequential, lewis1998other}. To this end, designers often data-driven surrogate models, which effectively capture the interactions between subsystems and integrate them into a multidisciplinary system. This approach is facilitated through Meta-Design, a framework that allows designers to interact with the system and refine it as necessary \cite{mistree1991designing}. However, modeling these interdisciplinary systems can be computationally expensive, often requiring the decomposition of the Meta-Design into subsystem-level components. Using small training data sets of subsystem evaluations enables designers to establish surrogate models \cite{forrester2008engineering}, such as radial basis functions, Neural networks, and Gaussian processes (GP) \cite{van2023digital,bostanabad2018leveraging,williams2006gaussian}, to approximate the behavior of individual subsystems. The accuracy of these surrogate models depends on both the quality and quantity of the available data, as well as the complexity of the subsystem. As a result, uncertainties can arise due to the performance of the surrogate model, affecting the overall uncertainty at the system level \cite{simpson2001metamodels}. 

The combination of the simulation uncertainty at the subsystem level and parametric uncertainty, as they propagate through the network, can significantly impact the accuracy of the decision-making process. This is because minor changes in design parameters can result in substantial variations in system outputs, making it difficult to achieve robust decisions. In such scenarios, it is often necessary to compromise on achieving an optimal decision in favor of a design that performs well under uncertainty. To address this challenge, Mistree et al. \cite{mistree1981optimization} introduced the cDSP framework, which provides a structured approach to model and solve engineering decision problems involving trade-offs between optimality and robustness. The cDSP framework integrates principles from Suh’s Independence Axiom and Information Content Axiom \cite{suh2001axiomatic}, as well as Taguchi’s signal-to-noise ratio \cite{unal1990taguchi}, to assess and improve the quality of decisions. These principles enable designers to systematically evaluate trade-offs, manage uncertainties, and arrive at decisions that balance performance and robustness.

Within the framework of cDSP for robust design optimization, assessing the robustness of a decision involves striking a balance between achieving the system's expected performance and effectively managing the overall uncertainty. To address this challenge, Choi et al. \cite{choi2005approach} introduced the Error Margin Index (\textit{EMI}) as a formulation for robust design. The EMI is defined as the ratio of the difference between the mean system output and the target value to the response variation (i.e., the numerator and denominator in Taguchi’s signal-to-noise ratio, respectively). This index provides a mathematical framework for assessing how closely the mean system output aligns with the design target while simultaneously capturing the effects of variability. The EMI-based multiobjective design operates under the assumption that system output variation follows a normal distribution, formulating the EMI equation accordingly. This framework offers a structured method for evaluating and improving design robustness by simultaneously accounting for performance targets and the inherent variability within the system \cite{choi2005approach}.

In this paper, we consider the challenge of non-normal output distributions in defining the admissible design space by integrating reliability-based design principles with the cDSP framework. To achieve this, we first establish the reliability index based on the target EMI value \cite{ nellippallil2020inverse}, which is determined during the robust design phase in the cDSP. Next, we show has this can be used for inverse design problems by exploring the admissible design spaces using both robust design and reliability constraints. This involves identifying the range of design variables that satisfy the target EMI value while ensuring that the probability of failure remains within acceptable limits. It is worth noting that the admissible space for robust and reliable designs can be the same when the output distribution is normal. The admissible space is therefore a subset of the feasible design space, explicitly defined for both reliable and robust design under these conditions. Finally, we compare the solutions obtained through robust design and reliability-based design to assess their alignment and trade-offs. This comparison provides a comprehensive perspective on design, balancing performance variation with achieving the desired reliability or robustness level. Additionally, this approach enables designers to assess the risks and benefits of selecting a reliability-based design strategy compared to a robust design approach. By providing a clear comparison between the two methods, we offer valuable insights into the trade-offs involved, supporting informed and improved decision-making in engineering design. This analysis helps designers determine which approach, or combination of approaches, best aligns with the specific requirements and constraints of their design problem.

The remainder of this paper is structured as follows: In Section~\ref{Sec:background}, we provide background information on GPs as surrogate models for simulating multidisciplinary systems and introduce the concept of cDSP. Section~\ref{sec:method} establishes a connection between cDSP, originally formulated for robust design, and reliable design by presenting a method to compare these two scenarios within the cDSP framework. In Section~\ref{sect:Result}, we compare the admissible design spaces for robust and reliable scenarios to explore the goal of the cDSP, which is to identify optimal decisions under compromised conditions for the practical application of a multidisciplinary hot rod rolling process. Finally, We summarize the key findings and provide concluding remarks in Section~\ref{sect:summary}.

\section{Background}\label{Sec:background}
In this section, we provide a background on GPs as a method to approximate each subsystem's response within a multidisciplinary system. In addition, we also introduce the cDSP, a decision-making framework for multi-objective and multi-stakeholder design under uncertainty. 

\subsection{Subsystem Response Approximation via GP}\label{subsection:GP}

The standard formulation of GPs to approximate subsystem response is defined as
\begin{equation}
y(\boldsymbol{x})=f(\boldsymbol{x})+Z(\boldsymbol{x}),
\label{eq:Gp}
\end{equation}
where $y(\boldsymbol{x})$ represents output observed for a $D$-dimensional design variable  $\boldsymbol{x} = \{x_1, x_2, \ldots, x_D\}^T$, $f(\boldsymbol{x})$ is a mean function that capture the general trend of a subsystem, and $Z(\boldsymbol{x})$ is a stationary stochastic process with zero mean. With the assumption that the prior mean of the process is zero ($f(\boldsymbol{x})=0$), the ordinary GP formulation is simplified to its stochastic process, $Z(\boldsymbol{x})$. The covariance function $cov(\cdot)$, also referred to as the kernel function, is a fundamental component of GPs and is defined as a measure of similarity between random variables $\boldsymbol{x}$ and $\boldsymbol{x}^\prime$, given by
\begin{equation}
cov(\boldsymbol{x},\boldsymbol{x^\prime})=\sigma^2R(\boldsymbol{x},\boldsymbol{x}^\prime),
\label{eq:kernel}
\end{equation}
where $R(\cdot)$ is the correlation function often defined using the square exponential function given in Equation~\eqref{eq:correlation}, and $\sigma^2$ represents the prior variance.
\begin{equation}
R(\boldsymbol{x}, \boldsymbol{x}') = \exp\left\{-\sum_{i=1}^D 10^{\boldsymbol{\omega}_i} \left(x_i - x_i'\right)^2\right\},
\label{eq:correlation}
\end{equation}

The vector of parameters, \(\boldsymbol{\omega}\), governs the smoothness of the correlation function and plays a critical role in defining the behavior of the GP. Moreover, \(\boldsymbol{\omega}\) are often referred to as the roughness parameters, that determine how effectively the GP model captures variations caused by specific input variables. Finally, experimental uncertainty is often assumed to be constant over the design space )i.e., homescedastic) \cite{vanbeek2021scalable}. Under this assumption, the observed response surface \(\boldsymbol{y}\) is modeled as

\begin{equation}
y_i = f(\textbf{x}_i) + \epsilon_i,
\label{eq:noisy_response}
\end{equation}

where $\epsilon_i \sim \mathcal{N}(0, \delta^2)$ for $i = 1, \ldots, n$ represents normally distributed experimental uncertainty with constant variance $\delta^2$, observed across $n$ training samples. This noise is directly incorporated into the correlation function, modifying it as follows
\begin{equation}
\boldsymbol{R}_\delta = \boldsymbol{R} + \delta^2 \boldsymbol{I},
\label{eq:noisy_correlation}
\end{equation}

where $\boldsymbol{R}$ is the original correlation matrix between $\boldsymbol{x}$ and $\boldsymbol{x}^\prime$ as it is defined in Equation \ref{eq:correlation}, and $\boldsymbol{I}$ denotes an identity matrix. The hyperparameters of the covariance matrix—$\boldsymbol{\omega}$, $\sigma^2$, and $\delta$—must be carefully estimated during GP training to ensure accurate predictions. A common approach for determining these parameters is maximum likelihood estimation, which optimizes the likelihood function to fit the observed data, as detailed in \cite{tao2021multi}. Once the optimal hyperparameters are determined, the mean value of the posterior prediction for the response under study, denoted by \(\hat{f}(\boldsymbol{x})\), is obtained. 

\subsection{cDSP: A Strategy for Compromise Solutions}\label{subsect:cDSP}
The foundational decision support construct used in this work is the cDSP \cite{mistree1993compromise}. The cDSP is anchored in the robust design paradigm first proposed by Genichi Taguchi \cite{taguchi1986introduction} and states that the models we work with are typically incomplete, inaccurate, and not of equal fidelity \cite{taguchi1986introduction,bras1993robust}. The cDSP is a hybrid of mathematical programming and goal programming. Specifically, by setting target goals a designer is able to restrict the space of admissible designs and prioritize objectives consistent with their preference. This is achieved by seeking multiple solutions through trade-offs among multiple conflicting goals often referred to as satisficing solutions. The problem formulation in the cDSP is carried out using four keywords – \textit{given}, \textit{find}, \textit{satisfy} and \textit{minimize}. Using the cDSP, the designer minimizes a deviation function – a function formulated using the deviation variables that exist for each of the goal targets. The details regarding formulating and solving the cDSP are available in \cite{mistree1993compromise,bras1993robust}. 

The focus in robust design is to improve the quality of systems by reducing their sensitivity to uncertainty without eliminating the sources \cite{taguchi1986introduction,taguchi1990robust,nair1992taguchi,tsui1992overview}. Three categories of information interact with the system model in robust design \cite{mcdowell2009integrated}: i) control factors, also known as design variables, are parameters that the designer adjusts to move toward a desired product, ii) noise factors are exogenous parameters that affect the performance of product/process but cannot be controlled by the designer, iii) responses are performance measures for the product or process. Managing these sources of uncertainty using the cDSP is carried out using the following types of robust design formulations \cite{mcdowell2009integrated}: (i) Type I Robust Design \cite{taguchi1993robust} - to identify control factor (design variable) values that satisfy a set of performance requirement despite variations in noise factors; (ii) Type II Robust Design \cite{chen1996procedure,chen1999satisfying} -  to identify control factor values that satisfy a set of performance requirements target despite variation in control factors themselves; and (iii) Type III Robust Design \cite{choi2005approach} - to obtain design solutions that are insensitive to uncertainty embedded within the model(s) used. Using the EMI, designers can determine balance the mean response and the spread of the response, considering the variability associated with design variables and the system models.

\section{Reliable and Robust Decision-Making Under Uncertainty}\label{sec:method}

In this section, we present the reliability-based cDSP. First, we will introduce the overarching framework, an explanation for how to establish system-level models from a network of subsystem surrogates. Finally, we will introduce the updated design formulation to enable reliability-based cDSP. 
\subsection{System Level Uncertainty}
\begin{figure}[t]
\centering
\scalebox{0.7}{ 
\begin{tikzpicture}[node distance=2cm, every node/.style={scale=0.7}]
\tikzstyle{data} = [rectangle, rounded corners, minimum width=3cm, minimum height=1cm,text centered, draw=black, fill=none]
\tikzstyle{MDS} = [rectangle,rounded corners, minimum width=3cm, minimum height=1cm, text centered, draw=black, fill=none]
\tikzstyle{Output} = [rectangle, rounded corners, minimum width=3cm, minimum height=1cm,text centered, draw=black, fill=none]
\tikzstyle{DSP} = [rectangle, rounded corners, minimum width=3cm, minimum height=1cm,text centered, draw=black, fill=none]
\tikzstyle{rcDSP} = [rectangle, rounded corners, minimum width=3cm, minimum height=1cm,text centered, draw=black, fill=none]
\tikzstyle{robcDSP} = [rectangle, rounded corners, minimum width=5cm, minimum height=0.5cm,text centered, draw=black, fill=none]
\tikzstyle{decision} = [diamond, minimum width=3cm, minimum height=0.5cm, text centered, draw=black, fill=none]
\tikzstyle{arrow} = [thick,->,>=stealth, line width=0.5mm]
\newcommand{\circled}[1]{
  \tikz[baseline=(char.base)]{
    \node[shape=circle, draw, line width=2pt,inner sep=2pt, fill=white,minimum size=2em] (char) {\huge #1};
  }
}

\node (data) [data,font=\Large,align=center] 
 {Training Data \\ $\boldsymbol{x}\pm \epsilon \in \mathbf{R}^d $\\
 $f(\boldsymbol{x}) \in \mathbf{R}^O$};
\node(model)[MDS, below of=data,font=\Large] {$\mathcal{GP}(\mu,\, k_{\ell}(x, x'))$};

\begin{pgfonlayer}{background}
    \node[fit=(data)(model), draw=none, fill=gray!20, inner sep=11mm] (backgroundbox) {};
    \node[anchor=west, xshift=0cm, yshift=1.7cm, font=\Large] at (backgroundbox.west) {\circled{1}};
\end{pgfonlayer}

\node[fit=(data)(model), draw, rounded corners, line width=0.6mm, dashed, inner sep=11mm, 
      label={[font=\Large]\hspace{0cm}\textbf{DSS}}, fill=none] {};

\node(output)[Output,below of= model,font=\Large,align=center]{\circled{2} System Output\\ $\boldsymbol{y} \sim F \quad (F \neq \mathcal{N})$};
\node(dsp)[DSP,below of= output,font=\Large,align=center] {\circled{3} cDSP };


\node(rcdsp) [rcDSP, left=2cm of dsp,font=\Large,align=center] {\circled{4} RcDSP\\$P(\hat{f(\boldsymbol{x}})\geq y_{target})\geq \alpha_{T}$};
\node(robcdsp) [robcDSP, right=2cm of dsp,font=\Large,align=center] {\hspace{0cm}\circled{5} rcDSP\hspace{1.5cm}\\\hspace{1cm}$EMI\geq1$};

\draw[arrow] (dsp) -- (rcdsp);
\draw[arrow] (dsp) -- (robcdsp);

\node(decision) [decision, below of=dsp, yshift=-0.7cm, font=\Large,align=center] {\circled{6}\\ Decision};

\draw[arrow] (rcdsp) -- (decision);
\draw[arrow] (robcdsp) -- (decision);

\node[fit=(dsp)(rcdsp)(robcdsp)(decision), draw, rounded corners, line width=0.6mm, dashed,
      minimum width=19cm, minimum height=5.5cm,yshift=0cm, 
      label={[font=\Large]\hspace{11cm}\textbf{Design}}, fill=none] {};

\node[fit=(data)(model)(output), draw, rounded corners, line width=0.6mm, dashed, 
      minimum width=6cm, minimum height=7.4cm, 
      label={[font=\Large]\hspace{0cm}\textbf{Meta-Design}}, fill=none, 
      yshift=0.5cm] {};

\draw[arrow](data)--(model);
\draw[arrow](model)--(output);
\draw[arrow](output)--(dsp);

\end{tikzpicture}
} 
\caption{Decision-making under reliability and robustness constraints}
\label{FlowChart}
\end{figure}
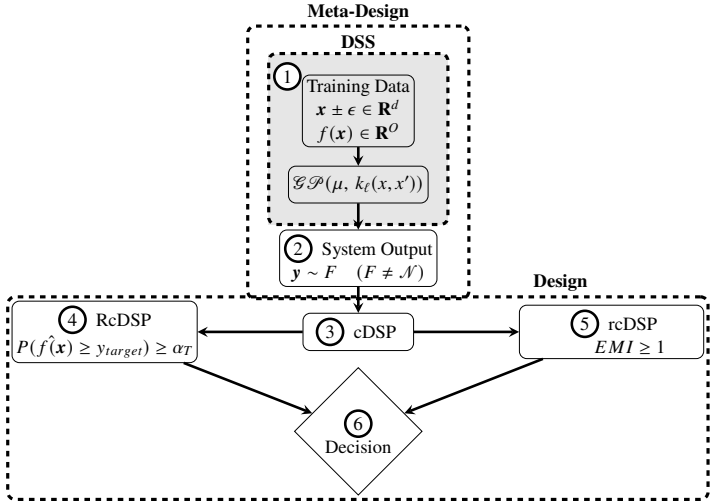

 We establish a link between cDSP and Meta-Design \cite{fischer2006meta}, emphasizing the structured development of both the design process and the final design solution. The relationship between Meta-design, decision-making, and their integration is illustrated in the flowchart shown in Figure \ref{FlowChart}. Meta-design empowers users to actively participate in the design process by enabling them to modify the decision support system (DDS) \cite{bonczek2014foundations} and interpret its outputs \cite{fischer2000meta,fischer2006meta}.
 
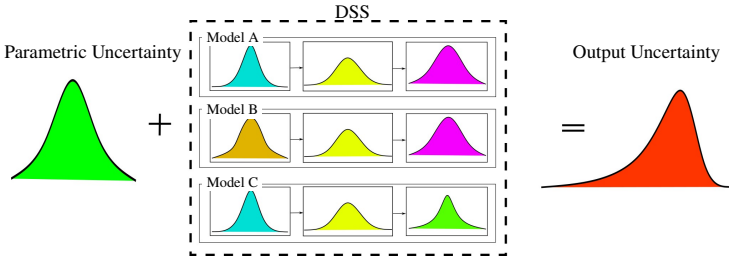
\begin{figure}[t]
    \centering
    \scalebox{0.5}{ 
        \includegraphics[width=\textwidth]{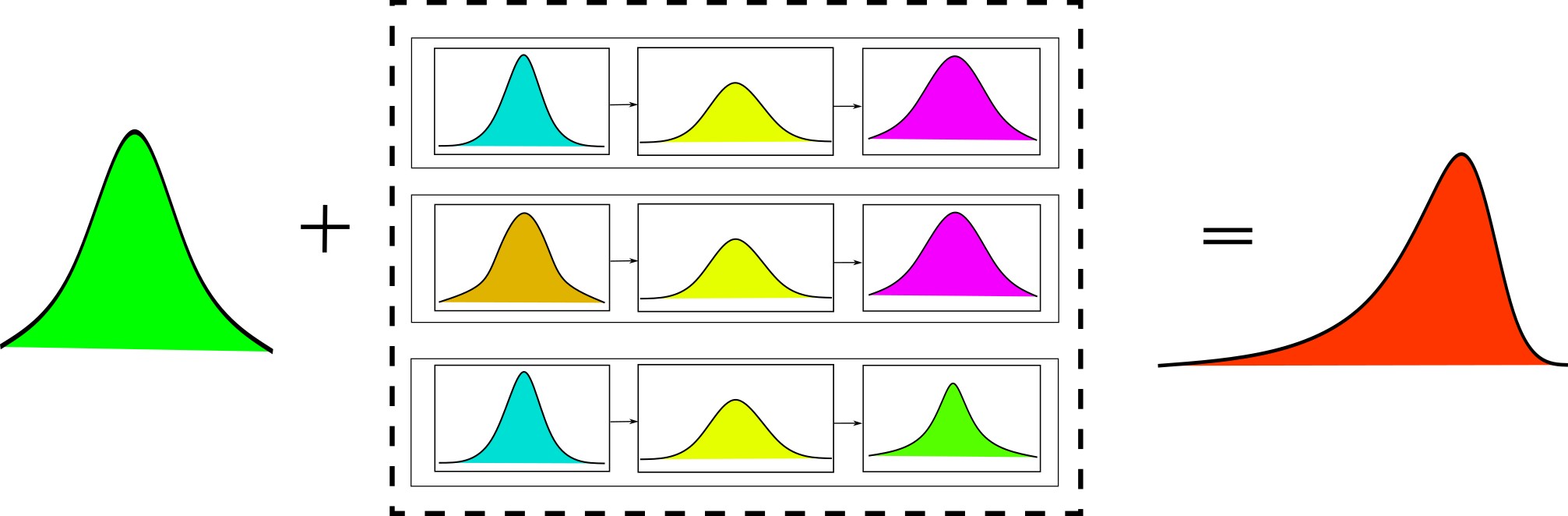}
        \begin{picture}(0,0)
            \put(-550,150){\Large{Parametric Uncertainty}}
            \put(-300,180){\Large{DSS}}
            \put(-400,162){\large\colorbox{white}{Model A}}
            \put(-400,107){\large\colorbox{white}{Model B}}
            \put(-400,52){\large\colorbox{white}{Model C}}
            \put(-120,150){\Large{Output Uncertainty}}
        \end{picture}
    }
    \caption{Uncertainty Propagation}
    \label{Uncertainty}
\end{figure}

As shown in Step \scalebox{1.2}{\small\textcircled{\raisebox{-0.1em}{1}}}, DSS \cite{bonczek2014foundations} highlights the role of computers and data-driven models in the decision-making process, for which is this study we rely on GPs. Additionally, computers can contribute by generating data for decision-making through simulations based on existing physical equations. We model each of the subsystems' integrated functions to establish the DSS using GPs, as shown in Figure \ref{Uncertainty}, to effectively address parametric uncertainty. Without loss of generality, we assume that this uncertainty follows a normal distribution with a standard deviation of $\sigma_{pa}$ (i.e., we assume normal distributed parametric uncertainty but the method works for any distribution). The subsystem models interact and collaborate sequentially, with the output of one system serving as the input for the next. The performance of these subsystems introduces an additional layer of uncertainty, particularly when the available data is insufficient. This uncertainty follows a normal distribution at the subsystem level due to the implementation of GP for simulation and it is referred to as model uncertainty denoted as $\sigma_{pr}$.  

Although uncertainties at the subsystem level individually follow normal distributions, the combined system-level uncertainty does not necessarily retain this property. To illustrate this, consider a system formed by a chain of two subsystems ,$\mathbf{y}_1$ and $\mathbf{y}_2$ as 
\begin{align}
\mathbf{y}_1 &= \mathbf{f}_1(\mathbf{x}_1), \quad \hat{\mathbf{f}}_1 \sim \mathcal{GP}(\mathbf{\mu}_1(\textbf{x}_1), \mathbf{\sigma}^2_{\mathrm{pr}_1}(\textbf{x}_1)) \notag \\
\textbf{x}_2 &= \mathrm{concat}(\textbf{x}_2', \textbf{y}_1), \quad \textbf{y}_2 = \textbf{f}_2(\textbf{x}_2), \quad \hat{\textbf{f}}_2 \sim \mathcal{GP}(\mu_2(\textbf{x}_2), \sigma^2_{\mathrm{pr}_2}(\textbf{x}_2))
\label{eq:subsystems}
\end{align}

Where, \( \mathbf{x}_2' \) represents the components of \( \mathbf{x}_2 \) that are not \( \mathbf{y}_1 \), and \( \textbf{y}_1 \in \mathbf{x}_2 \). For a fixed input \( \mathbf{x}' = \mathbf{x} + \varepsilon, \varepsilon \sim \mathcal{N}(0,\sigma_{pa}^2 )  \), the output \( \mathbf{y}_1 \) is a normal random variable due to the GP prior. However, \( \mathbf{y}_2 \) is dependent on \( \mathbf{y}_1 \), which is itself a random variable. Consequently, the distribution of \( \mathbf{y}_2 \) given \( \mathbf{x}'  \) is obtained by integrating over the uncertainty in \( \mathbf{y}_1 \)
\begin{equation}
P(\mathbf{y}_2 \mid \mathbf{x}' ) = \int P(\mathbf{y}_2 \mid \mathbf{y}_1) \cdot P(\mathbf{y}_1 \mid \mathbf{x}' ) \, d\mathbf{y}_1
\label{eq:conditional_integration}
\end{equation}
While \( \mathbf{P}(\mathbf{y}_1 \mid \mathbf{x}')\) is normal, \( \mathbf{P}(\mathbf{y}_2 \mid \mathbf{y}_1) \) involves a nonlinear mapping via the second GP. As a result, the integral in Equation~\ref{eq:conditional_integration} yields a distribution that is generally non-normal. This compounding effect becomes more pronounced in deeper chains of GPs, leading to increasingly complex and non-normal output distributions at the system level.

As shown in Figure \ref{Uncertainty}, we identify that the system output at Step  \scalebox{1.2}{\small\textcircled{\raisebox{-0.1em}{2}}} carries system-level uncertainty, which arises from the interaction between parametric uncertainty and uncertainties at the subsystem level. Additionally, we introduce model uncertainty, which includes uncertainty in choosing the data source and constructing sub-models, where the user must decide whether to consider individual models or a combination of them.

\subsection{Design}
The Design Section in Figure~\ref{FlowChart} begins with defining the design goal, denoted as $d$, at Step~\scalebox{1.2}{\small\textcircled{\raisebox{-0.1em}{3}}}. This goal represents the deviation from the target value for the system, $y_{\text{target}}$, as described in \cite{nellippallil2020inverse}.

\begin{equation}  
    d=1-\frac{EMI}{EMI_{target}},
\label{equation:target}
\end{equation}
where we define the EMI for the system modeled using GP as
\begin{equation}
    EMI=\frac{\hat{f}(\boldsymbol{x})-y_{target}}{\sigma_{pr}+\sigma_{pa}}.
    \label{EMI}
\end{equation}
When the overall uncertainty at the system level $(\sigma_{pr}+\sigma_{pa})$ follows a normal distribution, many uncertainty propagation techniques can be employed to calculate the mean and variance of the output distribution, as demonstrated in \cite{lee2009comparative}. However, as uncertainties propagate through sequential subsystems and interact with performance uncertainties, the resulting output distribution is more likely to deviate from normality \cite{ghanem2017handbook} as shown in Figure \ref{Uncertainty}. This is because subsystem nonlinearities and inter-dependencies are coupled with performance uncertainty. These factors can result in non-normal distributions, even when the input uncertainties follow a normal distribution which has been extensively studied in the literature, as highlighted in \cite{du2002efficient,ghanem2017handbook}. 

When defining the deviation from $y_{target}$ in Equation \ref{equation:target}, we assume that the system-level output follows a normal distribution by only including the first two statistical moments. However, we account for deviations from the normal distribution when applying design restrictions in combination with the cDSP. In Step~\scalebox{1.2}{\small\textcircled{\raisebox{-0.1em}{4}}}, we define the target reliability $\alpha_T$ based on the $EMI_{\text{target}}$ specified by the designer. We assume the relationship between these two as $
\alpha_T = \Phi(EMI_{\text{target}}),$
where $\Phi$ is the cumulative distribution function of the standard normal distribution. We then impose the reliability constraint as 
\begin{equation}
  P\left(\hat{f}(\boldsymbol{x}) \geq y_{\text{target}}\right) > \alpha_T,  
  \label{Reliability constraint}
\end{equation}
and refer to this scenario as the Reliability-based cDSP, RcDSP, illustrated in Step~\scalebox{1.2}{\small\textcircled{\raisebox{-0.1em}{4}}} of Figure~\ref{FlowChart}. Table \ref{tabupperborder} offers a structured overview of these two contrasting approaches.

\begin{table}[h]
\caption{Comparative Mathematical Formulation of the $\text{rcDSP}$ and $\text{RcDSP}$}
\label{tabupperborder}

\begin{tabularx}{\linewidth}{X}
\toprule
\makecell{cDSP formulations with EMI and \\ reliability-based constraints for reliable and robust design} \\
\midrule

\textbf{\textit{Given}} \\
$m$, number of subsystems \\
$\sigma_{pa}$, variability in design parameters of the system \\
$EMI_{target}$ \\
$\alpha_{T}$, reliability index \\
$y_{target}$, performance requirement \\
$\mathbf{x}_i$, $i=1,\ldots,m$, training data for each subsystem \\

\textbf{\textit{Find}} \\
$\sigma_{pr}$, performance uncertainty \\
$d$, deviation variable \\
$\hat{f}$, mean system output \\

\textbf{\textit{Satisfy}} \\
System goal: \\
$\dfrac{EMI}{EMI_{target}} + d = 1$ \\
System constraint for rcDSP: \\
$EMI > 1$ \\
System constraint for RcDSP: \\
$P(\hat{f} > y_{target}) > \alpha_T$ \\

\textbf{\textit{Minimize}} \\
$d$, or single objective optimization \\
\bottomrule
\end{tabularx}
\end{table}

In Step~\scalebox{1.2}{\small\textcircled{\raisebox{-0.1em}{5}}}, we impose the restriction of the cDSP for robust design by requiring \( EMI > 1 \) \cite{nellippallil2020inverse}. This condition ensures that the design achieves sufficient robustness against uncertainties. We refer to this scenario as the Robust-based cDSP (rcDSP).  These restrictions define the admissible space at Step ~\scalebox{1.2}{\small\textcircled{\raisebox{-0.1em}{6}}}, allowing us to evaluate whether the achieved optimum target is acceptable or not, considering each of these design scenarios.

\section{Application and Results}\label{sect:Result}

In this section, we compare the results of the cDSP approach under robust design constraints and reliability constraints for a hot rod rolling process. First, we describe the hot rod rolling process as a multidisciplinary system, emphasizing its key characteristics. Next, we analyze the process outputs to evaluate robust and reliable design outcomes, comparing both scenarios to guide designers in selecting the more appropriate design method. 

\subsection{Hot Red Rolling Process: A Multidisciplinary Design Process}

\begin{figure*}
    \centering
    \includegraphics[width=\linewidth]{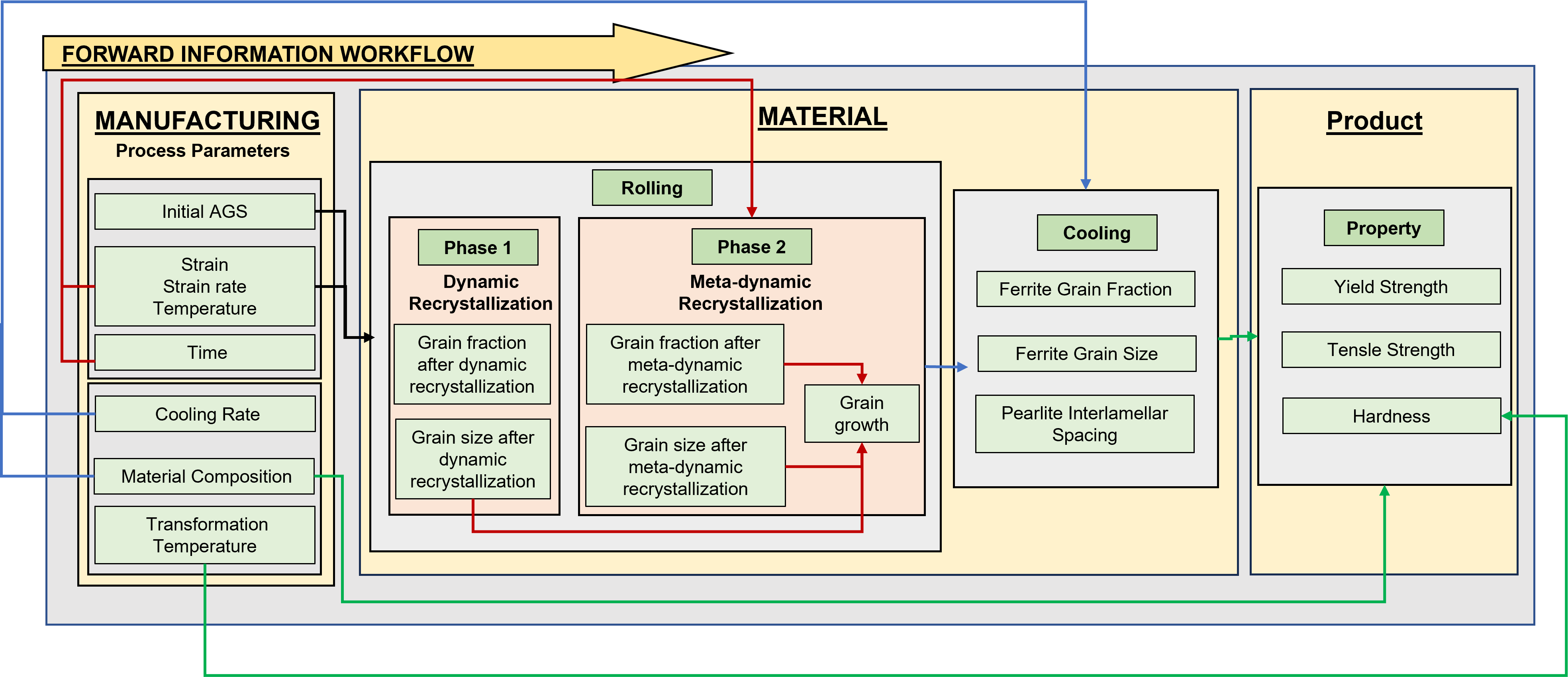}
    \caption{ Information flow for the hot rod rolling process, illustrating the sequence of steps involved in transforming raw material into the final product \cite{digonta2025framework}.}
    \label{information}
\end{figure*}

Hot rod rolling is a series of manufacturing processes that transform cast steel, typically in the form of billets, into steel rods with the desired mechanical properties and performance. This process begins with reheating steel billets in a furnace at high temperatures, ensuring that the steel reaches a state suitable for deformation. This process is known as reheating, which is followed by a series of rolling passes (Phase 1 and Phase 2 in Figure~\ref{information}), where the reheated steel billet passes through a series of rotating rollers, reducing their cross-sections and elongating their lengths. This is referred to as the rolling process. The rolling process plays a crucial role in influencing the material microstructure through process parameters that includes the strain, strain rate, and temperature.

After the final rolling pass, the steel rods are subjected to cooling on the cooling bed and run-out table (Cooling phase in Figure \ref{information}). This process governs the microstructural transformations, including the transition from austenite to ferrite and pearlite, depending on the cooling rate and the steel's composition. The cooling process can significantly affect the mechanical properties, such as tensile strength, yield strength, and hardness, ultimately determining the final product performance (Product level in Figure~\ref{information}). In the hot rod rolling process, thermo-mechanical parameters like temperature, strain, and strain rate directly affect the material's microstructure, which determines the product's performance. Here, the product is the system we aim to design.

According to the Processing-microStructure-Property-Performance (PSPP) relationship \cite{olson1997computational}, the performance of steel rods produced through the hot rod rolling process is determined by their mechanical properties, such as yield strength, tensile strength, and hardness. The system designer defines these properties at the product level based on the intended application of the steel rods. For instance, in applications like gear manufacturing, the system designer focuses on maximizing yield strength, tensile strength, and hardness \cite{nellippallil2020inverse}. However, these mechanical property requirements often conflict with one another in relation to the steel microstructure after cooling. For example, prioritizing the maximization of yield strength by controlling the microstructure could impact the tensile strength and hardness requirements. Therefore, product-level decisions must carefully balance these competing objectives to meet performance requirements. The microstructures are inherently random and embody uncertainty. Uncertainty in microstructures, which originates at the subsystem level, propagates through the process chain and ultimately affects the product's mechanical properties. The PSPP relationship establishes that mechanical properties are directly influenced by the material's microstructural characteristics developed during the cooling process. Specifically, yield strength, tensile strength, and hardness depend on three key microstructural features: the ferrite grain size, ferrite fraction, and pearlite interlamellar spacing \cite{kuziak1997modeling,yada1988prediction}. Material designers define target values for microstructural features at the material (subsystem) level to achieve the desired mechanical properties of the product. These features, in turn, are influenced by factors such as carbon and manganese content, Cooling Rate during the cooling process, and Final austenite grain size after the rolling process \cite{kuziak1997modeling,nellippallil2020inverse,pd1992mathematical}. Thus, subsystem-level decisions must be carefully aligned with product-level objectives to ensure desired microstructural characteristics and the steel rod's performance.

The rolling process defines the final austenite grain size, a critical factor in microstructure formation during cooling. According to the PSPP relationship, processing parameters dictate the material's microstructure. The rolling process encompasses complex thermo-mechanical interactions that affect microstructural development. During rolling, dynamic recrystallization, metadynamic recrystallization, and grain growth govern the evolution of the austenite grain size. As the steel deforms through the rollers, dynamic recrystallization takes place. The dynamic recrystallized grain size and grain fraction are defined in terms of the initial austenite grain size before rolling and the strain, strain rate, and temperature during the rolling process \cite{kuziak1997modeling}. Following dynamic recrystallization, metadynamic recrystallization initiates. The grain size and grain fraction after metadynamic recrystallization are defined in terms of the strain rate, temperature during rolling, and the interpass time \cite{kuziak1997modeling}. This is followed by grain growth, which further refines the grain size and determines the final austenite grain size \cite{lee2002integrated}. The final austenite grain size is a critical factor for the cooling process, where microstructural phase transformations occur. Consequently, decisions regarding manufacturing process parameters—including initial austenite grain size, strain, strain rate, temperature, and time—directly influence the final austenite grain size and, subsequently, the cooling process and the overall performance of the steel rod.
Figure \ref{information} shows the information workflow through this process which leads to the final desired goal which is the yield strength of the product. The information flow for the hot rod rolling process, detailing the sequence of steps is depicted in Figure \ref{information}.

Different researchers have studied the yield strength function over the years, resulting in models that predict yield strength as a function of multiple microstructural parameters associated with rolling and cooling. However, each of these models tends to predict yield strength at different ranges for a given microstructural input. There is uncertainty associated with the predictions from these models for the same input (i.e., another source of model uncertainty). In this paper, we use three yield strength models proposed by researchers from different groups. To demonstrate the approach presented in this paper, we assume the yield strength model by Gladman and co-authors \cite{gladman1976structure,gladman1972some} as the middle-level response model $f_0 (x)$ for our problem. The upper-level model $f_1 (x)$ for yield strength is assumed as the model by Hodgson and Gibbs \cite{pd1992mathematical}. The Hodgson and Gibbs model predicts yield strength higher than the model by Gladman and co-authors for a given input. The lower-level model $f_2 (x)$ for yield strength is assumed as the model by Kuziak and co-authors \cite{kuziak1997modeling}. This model predicts yield strength at a lower level than the other two models for a given input. 

\subsection{Robust and Reliable cDSP Design for Hot Rod Rolling}

We perform an empirical comparison of two design scenarios, robust and reliable, using the cDSP framework, applied to optimize the hot rod rolling process. Our analysis focuses on temperature as the primary design variable with parametric uncertainty in the form of a normal distribution as \( T \sim \mathcal{N}(\mu, \sigma^2) \), where \( \mu \) represents the mean value, and \( \sigma \) denotes the standard deviation. In this study, the Lower Requirement Limit (LRL) serves as the target value in Equation~\ref{EMI}. We assume a parametric uncertainty to follow a Normal distribution with a standard deviation given as \( \sigma^2 = 25 \) for temperature, which propagates through dynamic recrystallization and meta-dynamic recrystallization, indirectly affecting downstream subsystems in the entire system. We conducted our analysis through two fundamental training paradigms. The first scenario employed model-specific training using exclusively the middle yield material model, $f_0 (x)$, assumed to be the baseline performance under ideal model alignment conditions. The second scenario adopted model-agnostic training, incorporating all three candidate material models ($f_0 (x)$,$f_1 (x)$,$f_2 (x)$) to test system performance under non-normal performance distribution. We examine each training paradigm through two distinct robust design frameworks: robust design type 1 focuses on ensuring system robustness under parametric uncertainty, while robust design type 2 extends this by addressing both simulation uncertainty and parametric uncertainty. By incorporating these considerations, we define four test cases

\begin{itemize}
    \item \textbf{Case A}: Employs the middle yield model with abundant training data ($n = 50 \times d$), eliminating model uncertainty while isolating performance uncertainty to inherent process variability.
    
    \item \textbf{Case B}: Utilizes the middle yield model with limited training data (\( n = 5 \times d \)), where \textit{a priori} validation mitigates model uncertainty, while performance uncertainty arises due to inherent process variability.

\item \textbf{Case C}: Utilizes all three material models with abundant training data (\( n = 50 \times d \)), where the model-agnostic approach introduces model uncertainty while isolating performance uncertainty to inherent process variability.
\item \textbf{Case D}: Utilizes all three material models with sparse training data (\( n = 5 \)), where the model-agnostic approach introduces model uncertainty while incorporating performance uncertainty due to inherent process variability.
\end{itemize}
These test cases have been studied using failure probabilities $\alpha$ and $EMI_{target}$ levels that would result in equivalent feasible design spaces if the predicted system output distribution was normal. 

\begin{figure}[t]
  \centering
    \begin{subfigure}[b]{0.48\columnwidth}
      \begin{overpic}[width=\textwidth]{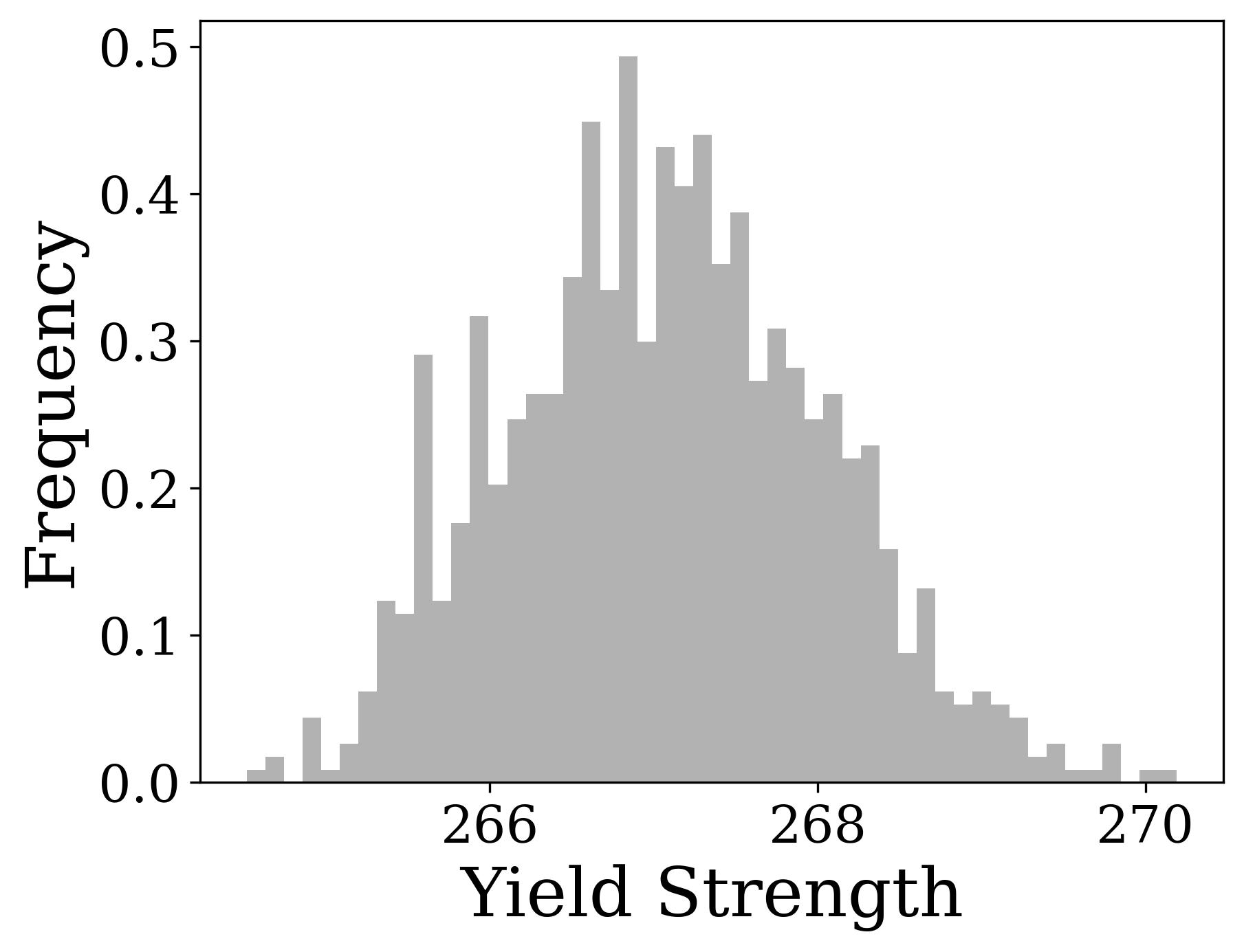}
        \put(18,69){\colorbox{white}{\footnotesize\textbf{a}}} 
      \end{overpic}
      \label{yield-caseA}
    \end{subfigure}
    \hfill
    \begin{subfigure}[b]{0.5\columnwidth}
      \begin{overpic}[width=\textwidth]{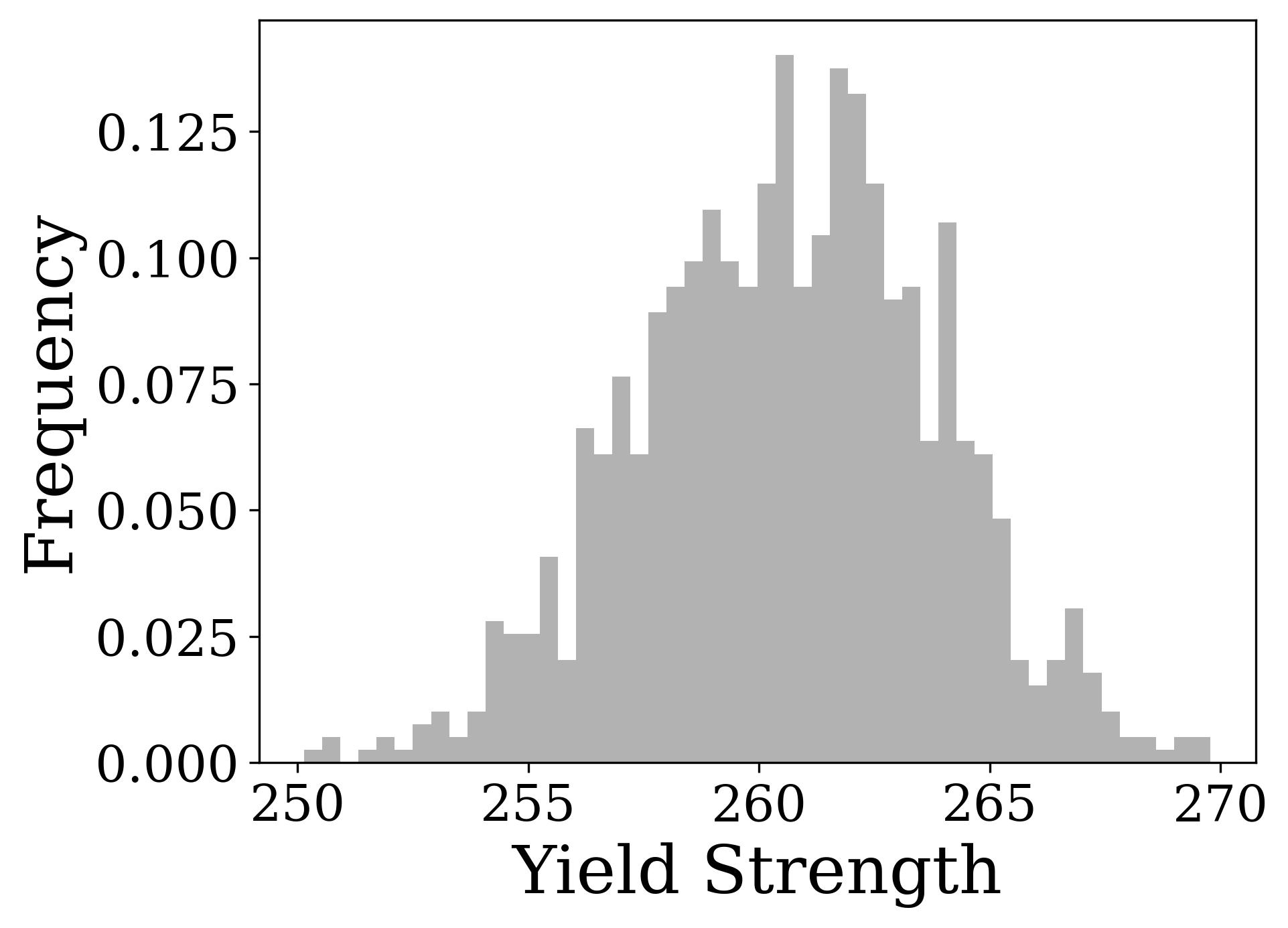}
        \put(21,63){\colorbox{white}{\footnotesize\textbf{b}}}
      \end{overpic}
      \label{yield-caseB}
    \end{subfigure}

    \vspace{0.2cm}
 
    \begin{subfigure}[b]{0.48\columnwidth}
      \begin{overpic}[width=\textwidth]{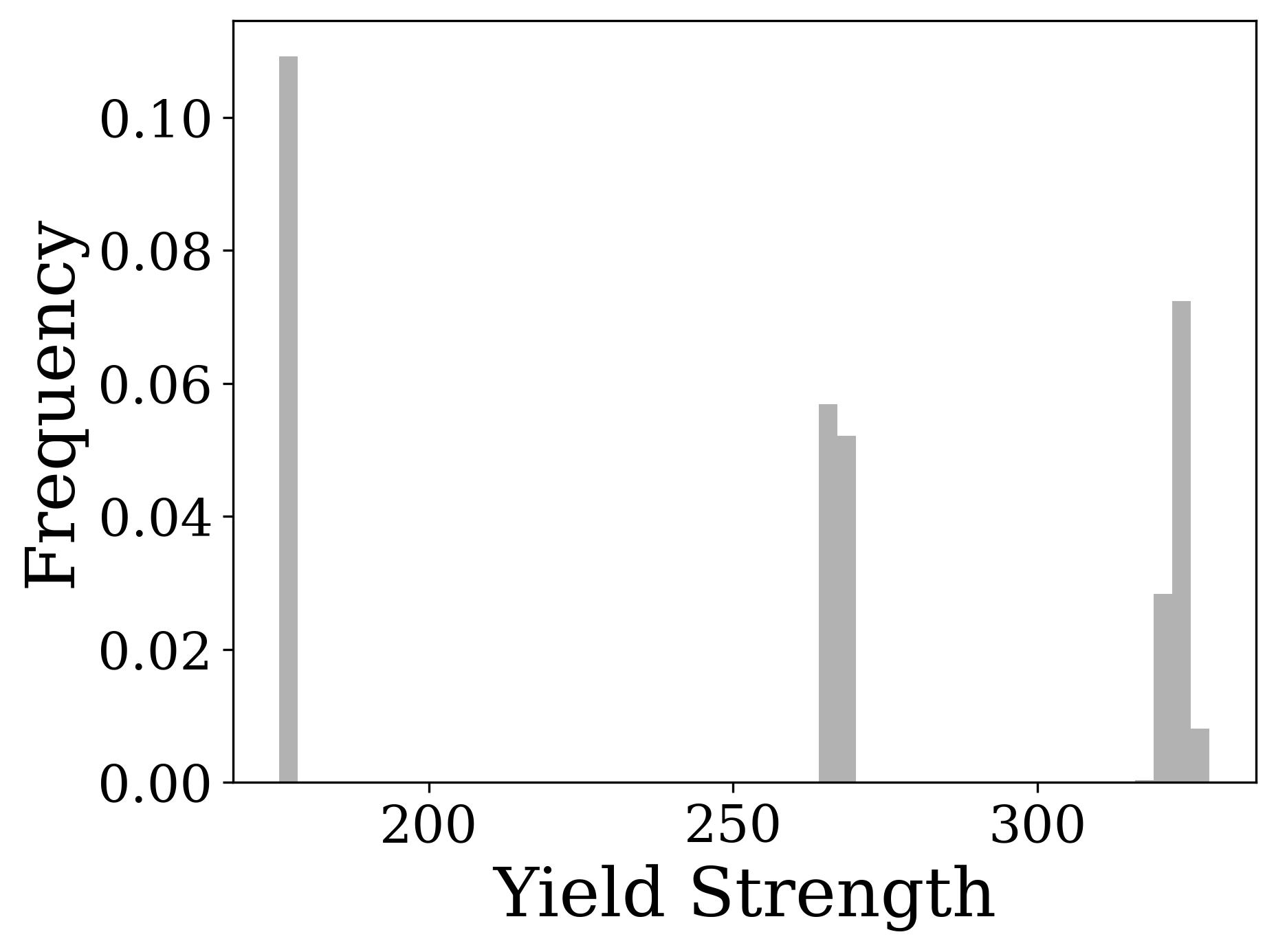}
        \put(20,66){\colorbox{white}{\footnotesize\textbf{c}}}
      \end{overpic}
      \label{yield-caseC}
    \end{subfigure}
    \hfill
    \begin{subfigure}[b]{0.48\columnwidth}
      \begin{overpic}[width=\textwidth]{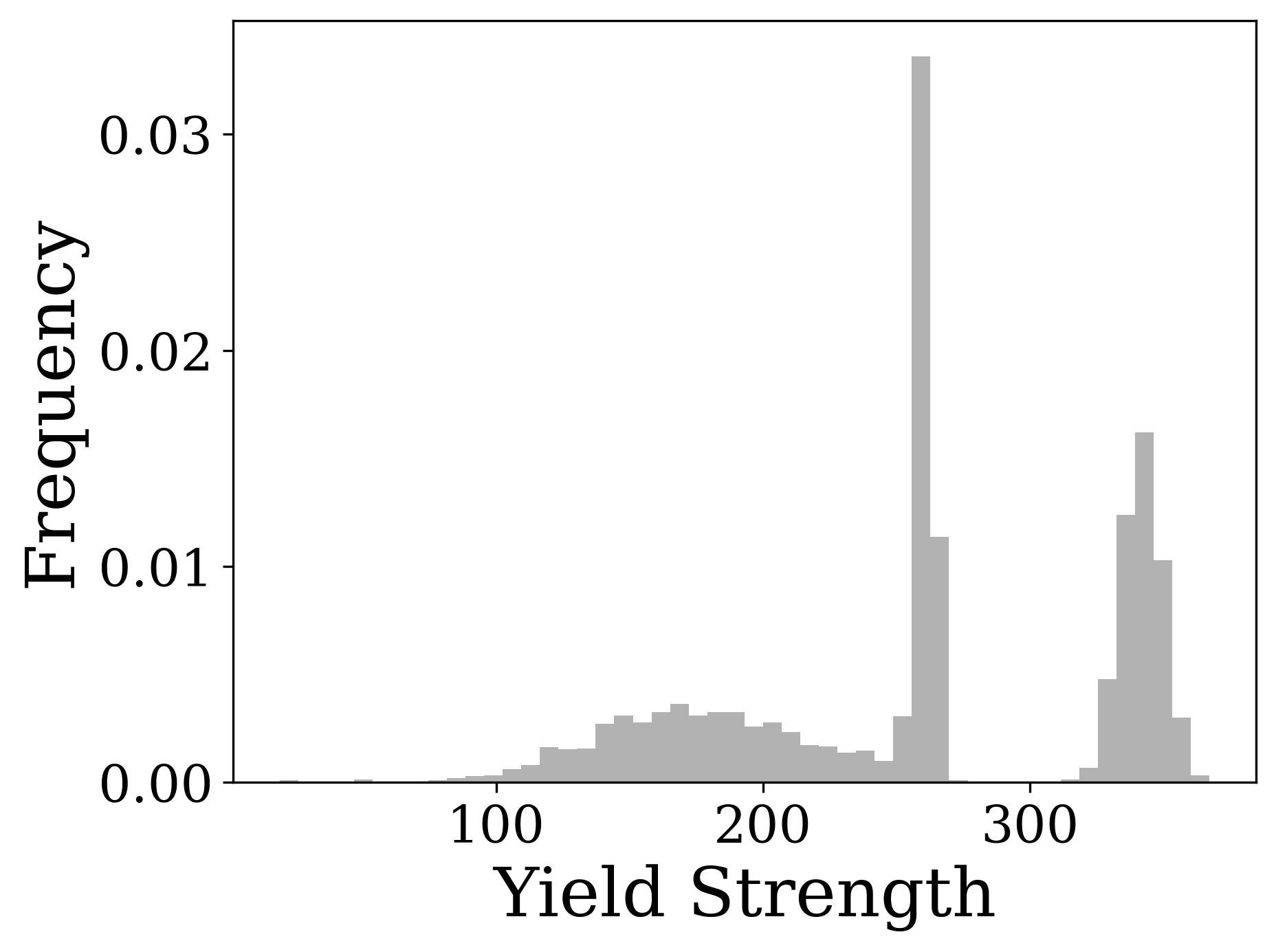}
        \put(20,65.5){\colorbox{white}{\footnotesize\textbf{d}}}
      \end{overpic}
      \label{yieldcased}
    \end{subfigure}
   \caption{Yield strength distributions from hot rod rolling at $1450\,\text{\textdegree}\mathrm{F}$: 
  $\text{(a)}$~\textbf{Case A} -- Middle yield model without uncertainty, 
  $\text{(b)}$~\textbf{Case B} -- Middle yield model with uncertainty, 
  $\text{(c)}$~\textbf{Case C} -- Combined model without uncertainty, 
  $\text{(d)}$~\textbf{Case D} -- Combined model with uncertainty.}
  \label{yield_distributions}
\end{figure}
\begin{figure*}[ht]
  \centering
  \includegraphics[width=0.7\textwidth]{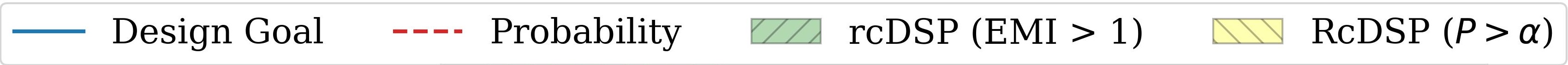}
  \begin{subfigure}{\textwidth}
    \centering
   \begin{subfigure}[b]{0.05\textwidth} 
    \centering
    \rotatebox{90}{\textcolor{black}{\parbox{4cm}{\centering\large \textbf{LRL = 200 MPa}}}}
\end{subfigure}
    \begin{subfigure}[b]{0.3\textwidth}
      \includegraphics[width=\textwidth]{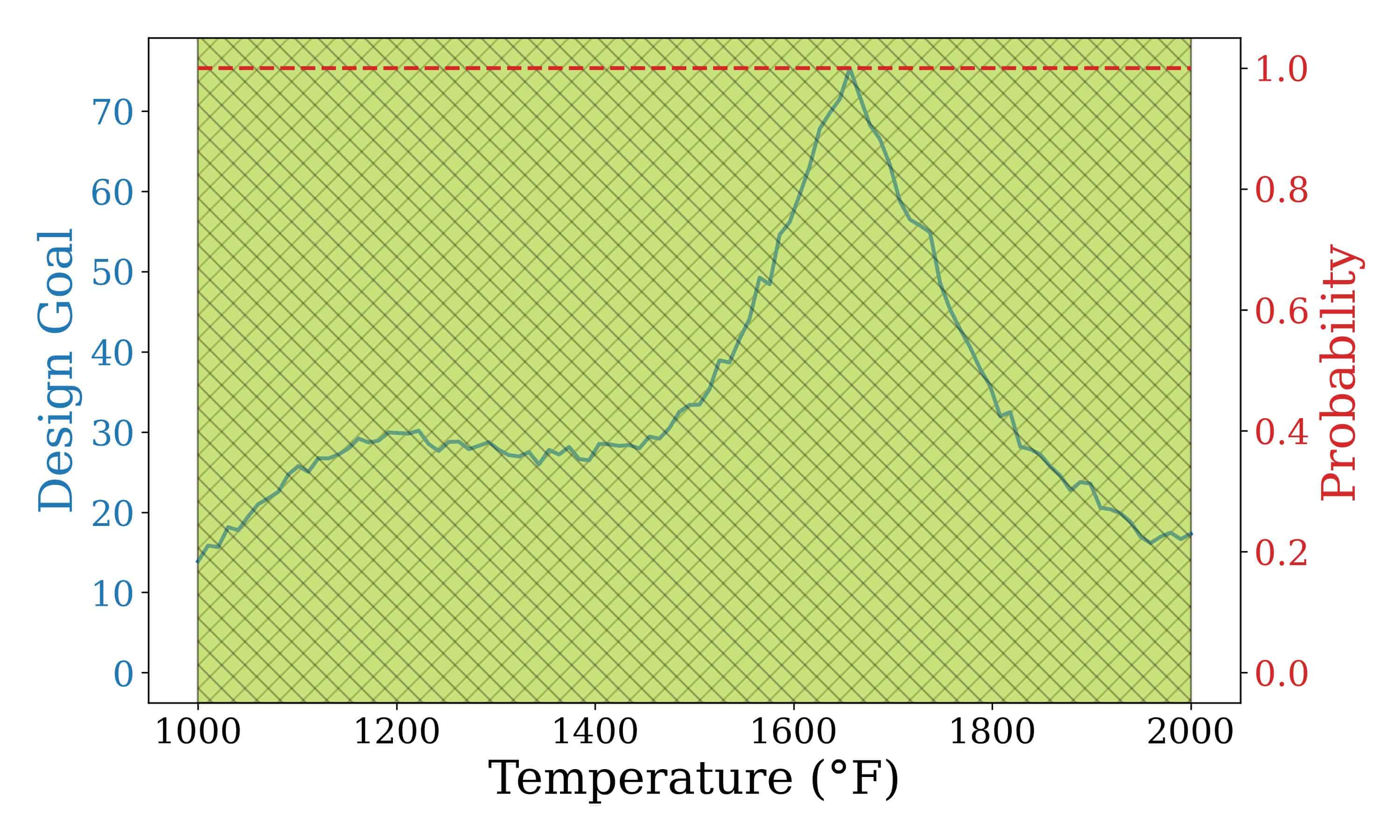}
    \end{subfigure}    
    \hfill
    \begin{subfigure}[b]{0.3\textwidth}
      \includegraphics[width=\textwidth]{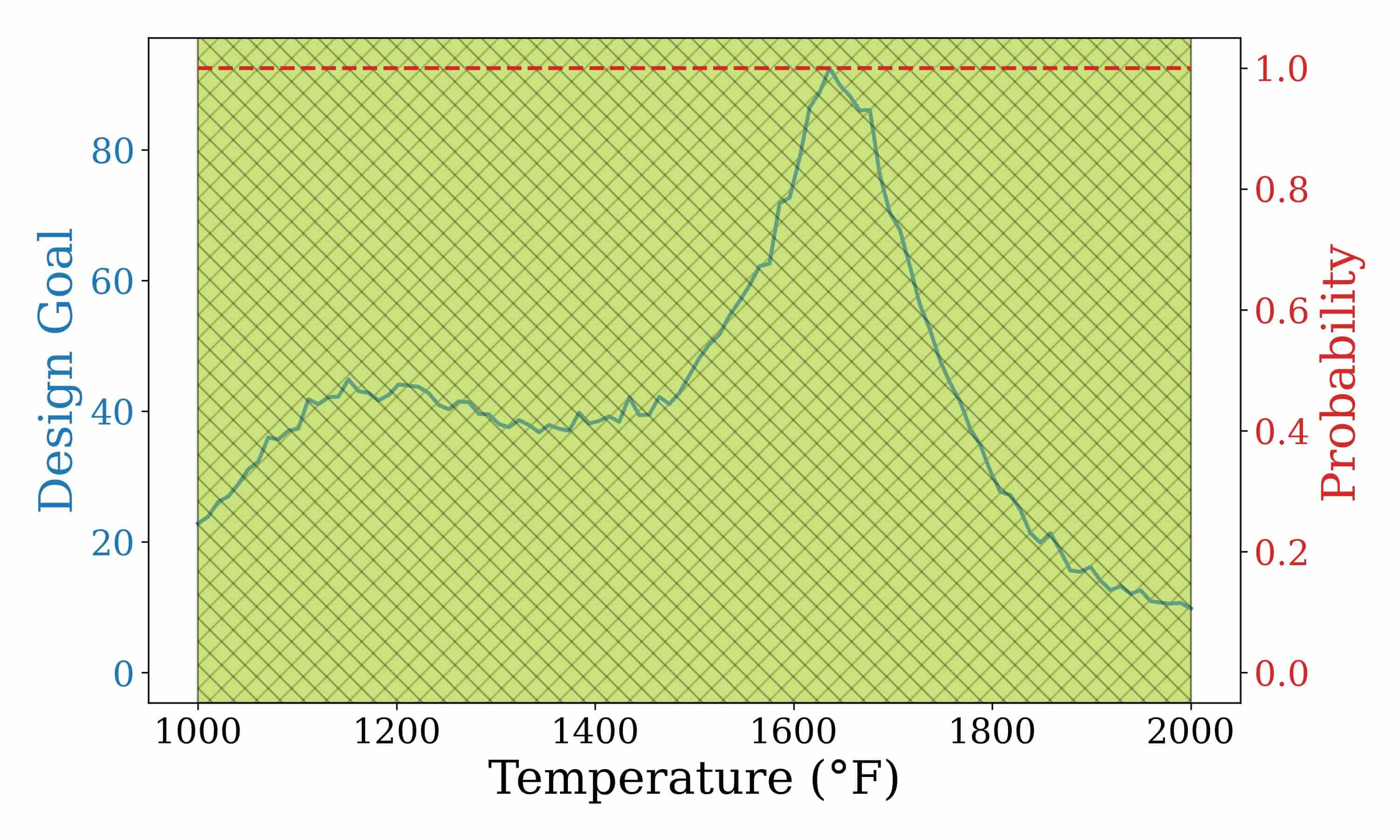}
    \end{subfigure}
    \hfill
    \begin{subfigure}[b]{0.3\textwidth}
       \includegraphics[width=\textwidth]{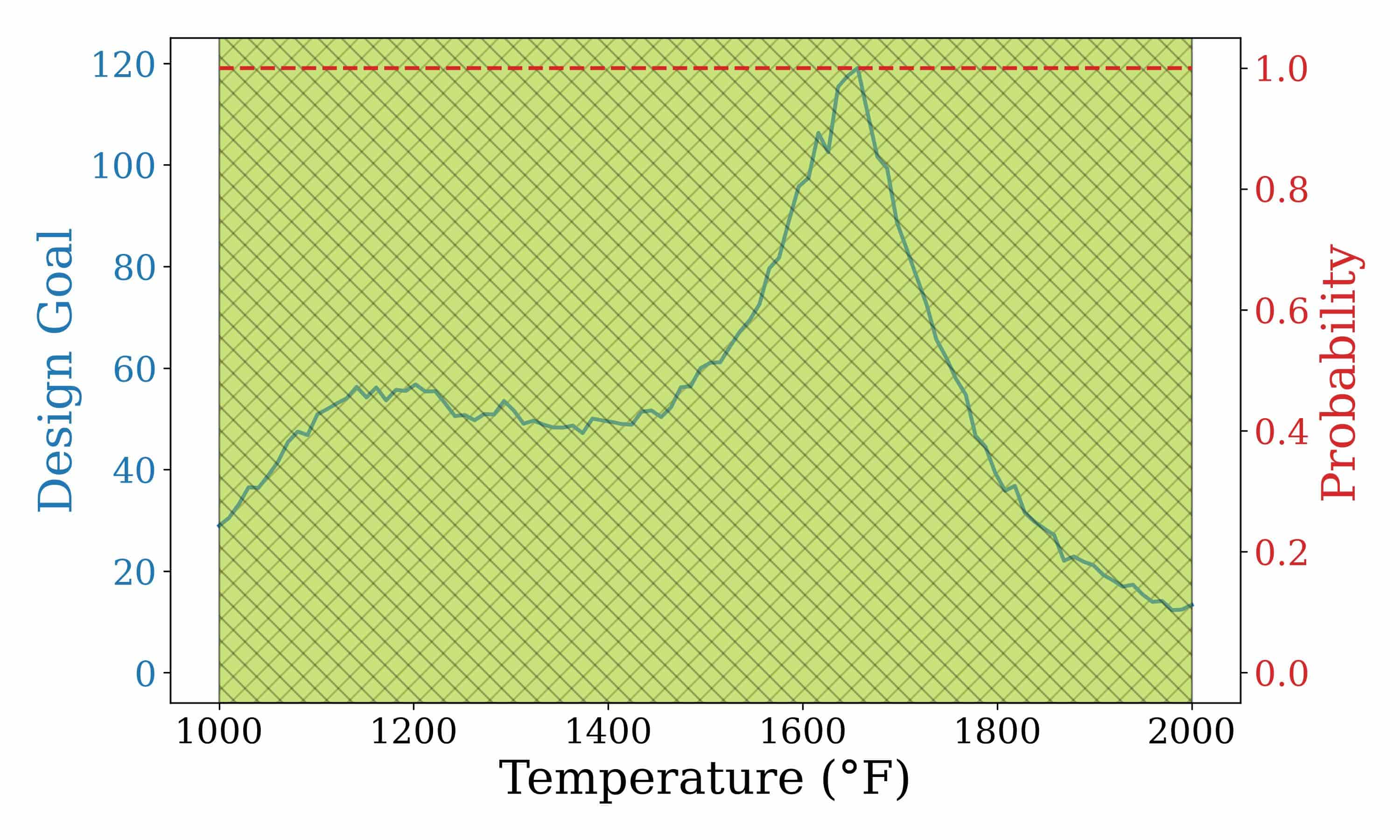}
    \end{subfigure}

    \vspace{0.5cm}
    \begin{subfigure}[b]{0.05\textwidth} 
    \centering
    \rotatebox{90}{\textcolor{black}{\parbox{4cm}{\centering\large \textbf{LRL = 270 MPa}}}}
\end{subfigure}
    \begin{subfigure}[b]{0.3\textwidth}
       \includegraphics[width=\textwidth]{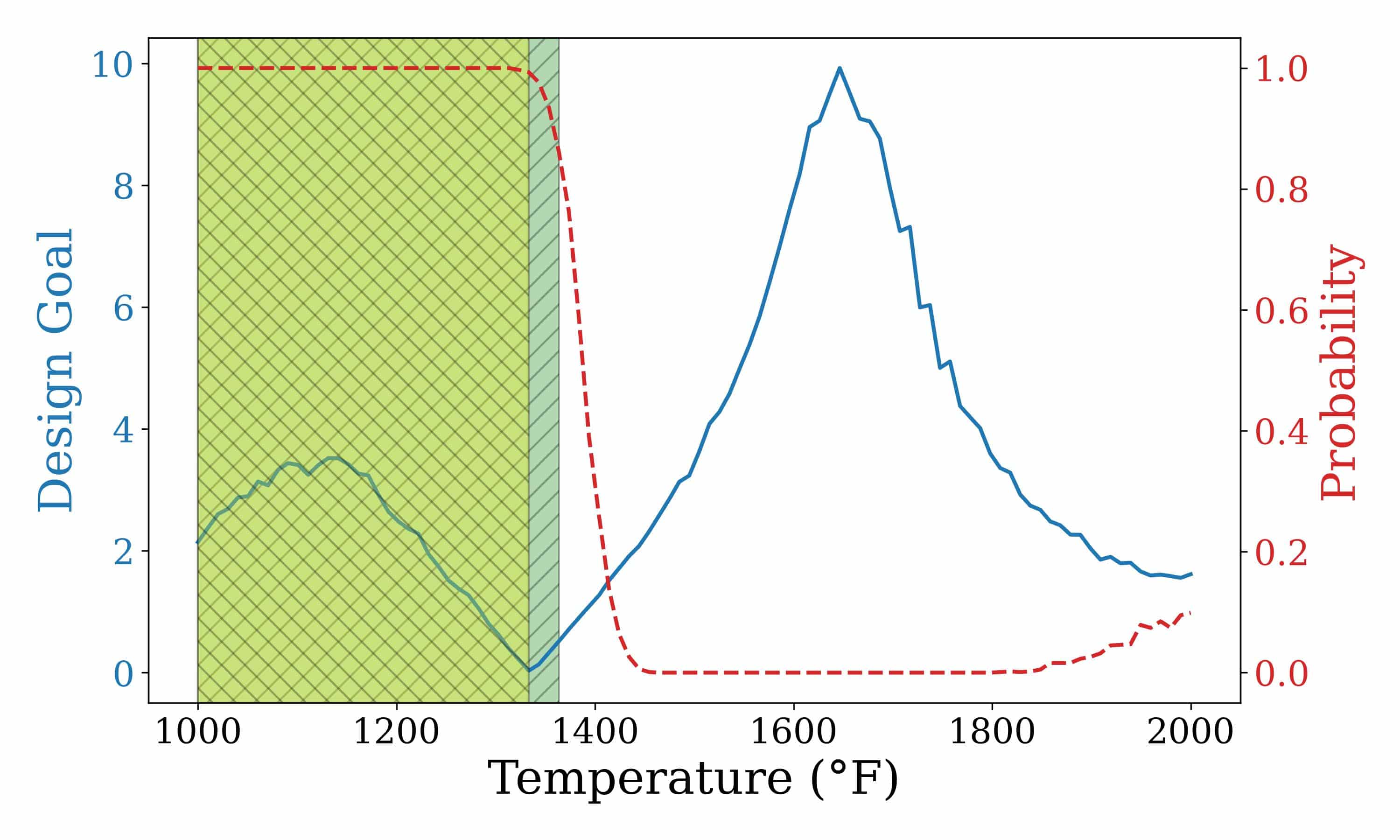}
    \end{subfigure}
    \hfill
    \begin{subfigure}[b]{0.3\textwidth}
      \includegraphics[width=\textwidth]{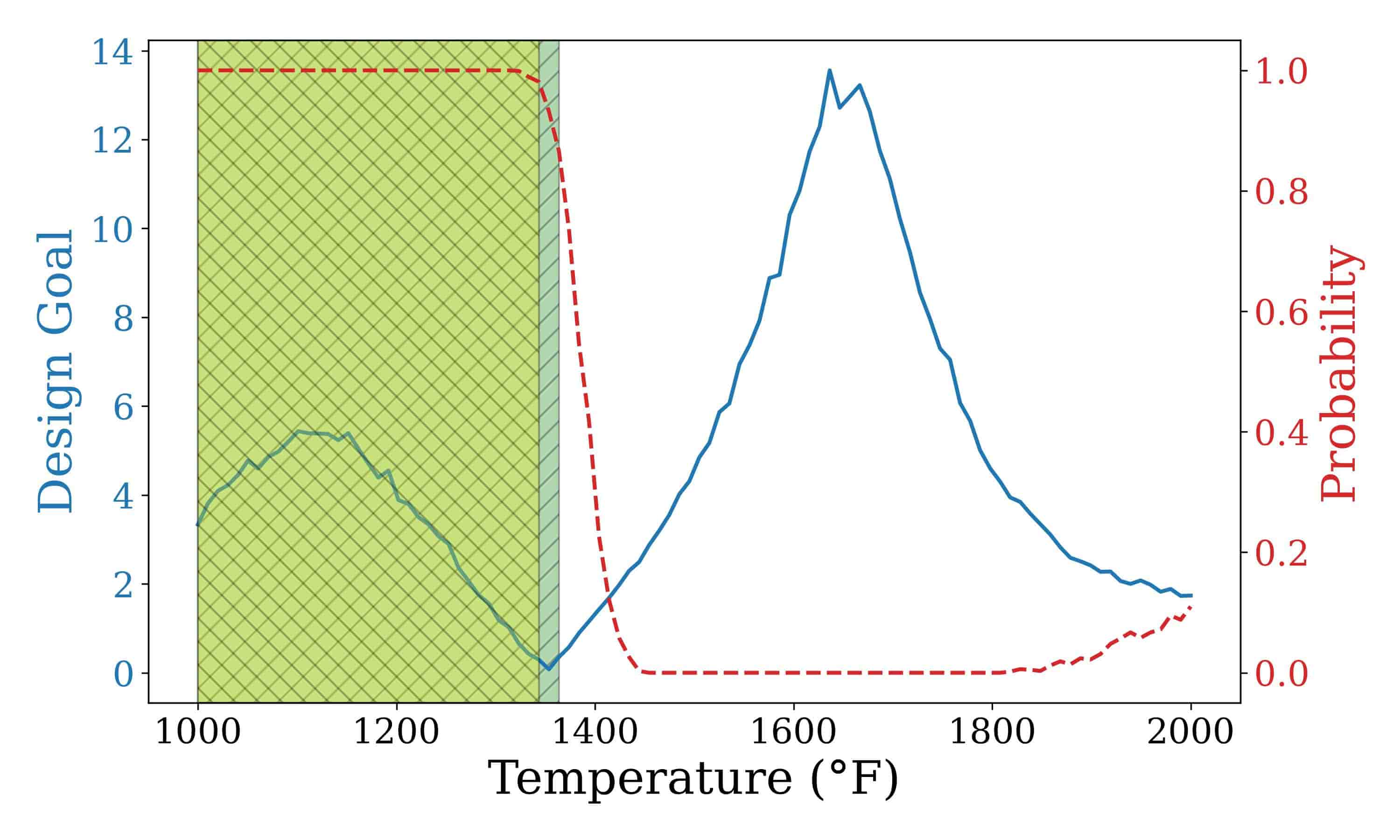}
    \end{subfigure}
    \hfill
    \begin{subfigure}[b]{0.3\textwidth}
      \includegraphics[width=\textwidth]{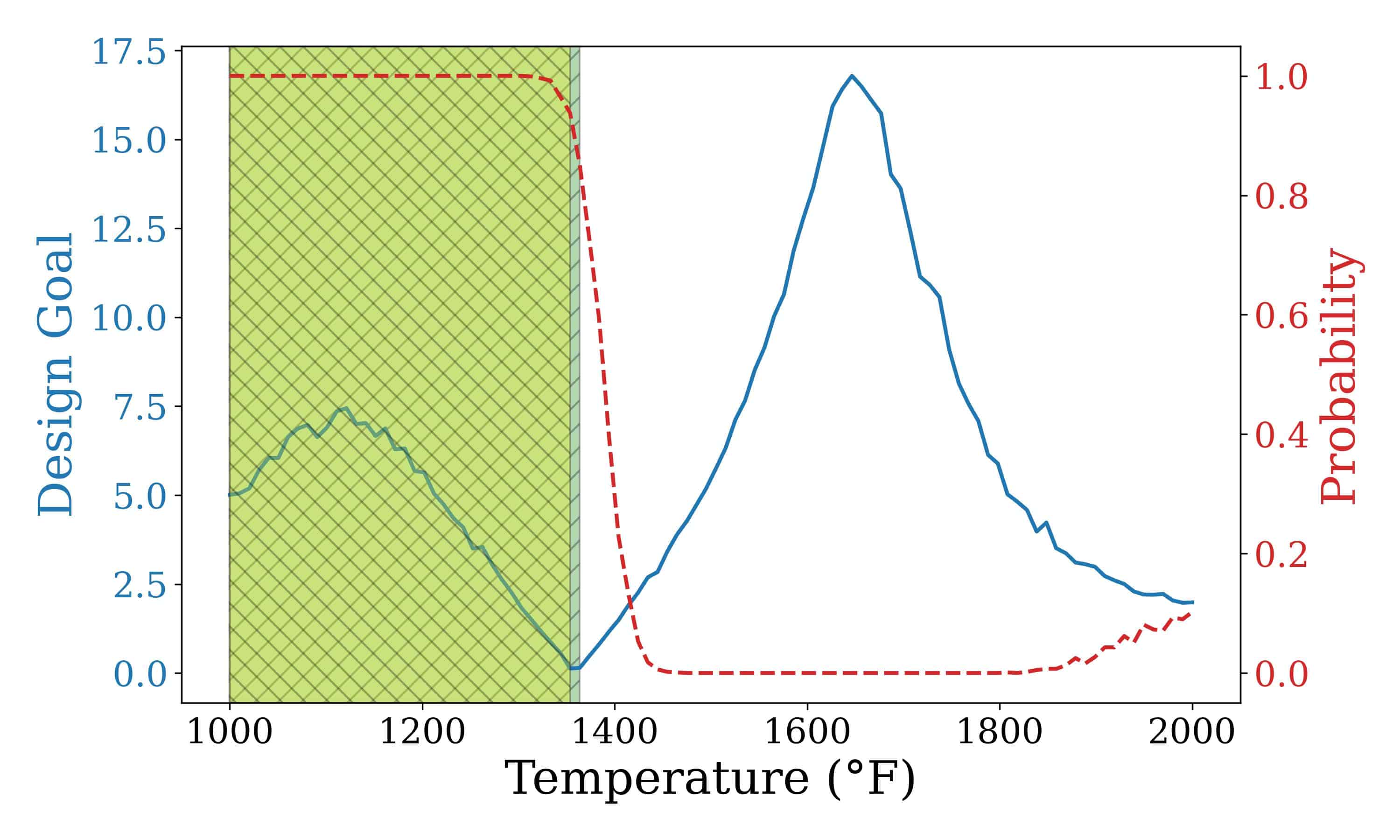}
    \end{subfigure}

    \vspace{0.5cm}
   \begin{subfigure}[b]{0.05\textwidth} 
    \centering
    \rotatebox{90}{\textcolor{black}{\parbox{4cm}{\centering\large \textbf{LRL = 280 MPa}}}}
\end{subfigure}
    \begin{subfigure}[b]{0.3\textwidth}
      \includegraphics[width=\textwidth]{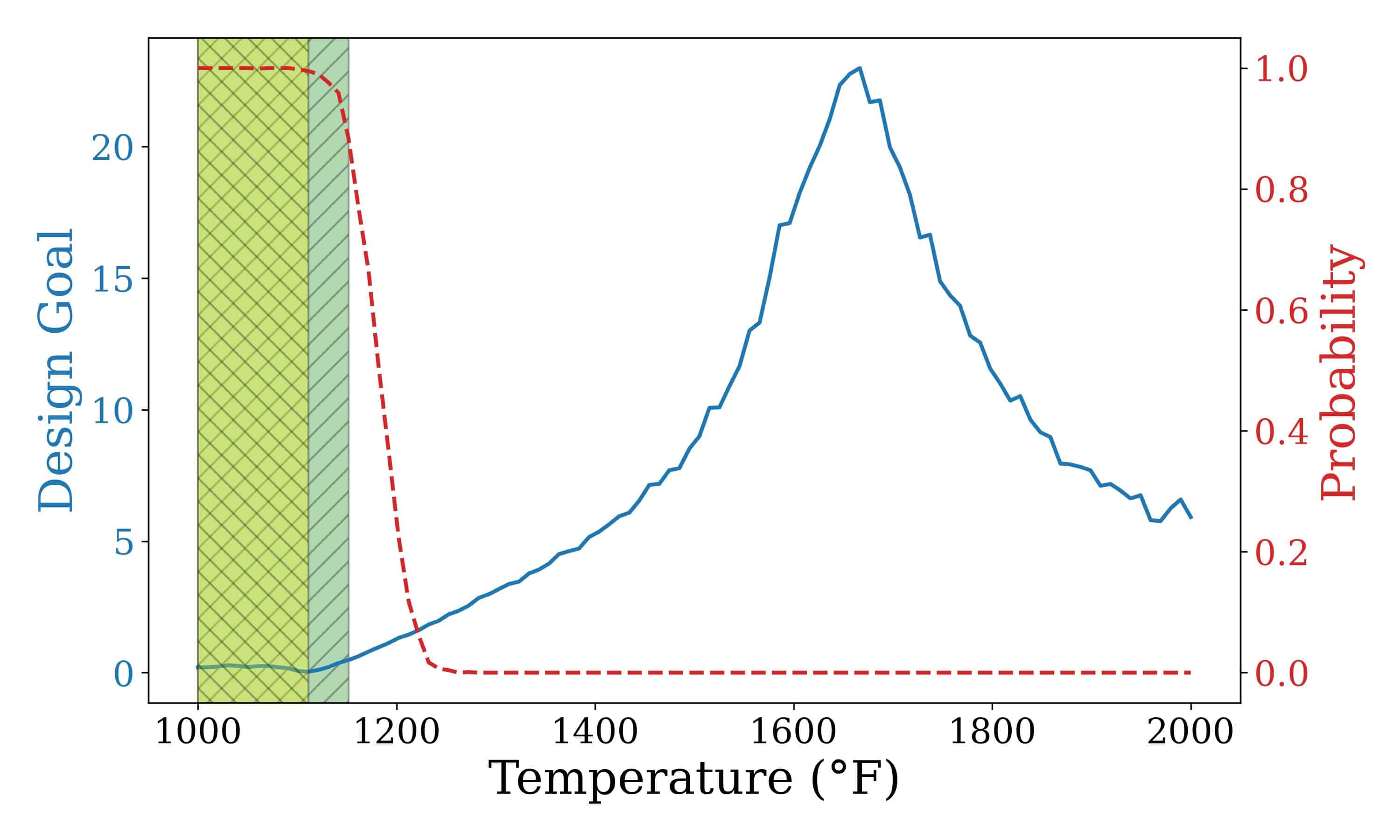}
    \end{subfigure}
    \hfill
    \begin{subfigure}[b]{0.3\textwidth}
      \includegraphics[width=\textwidth]{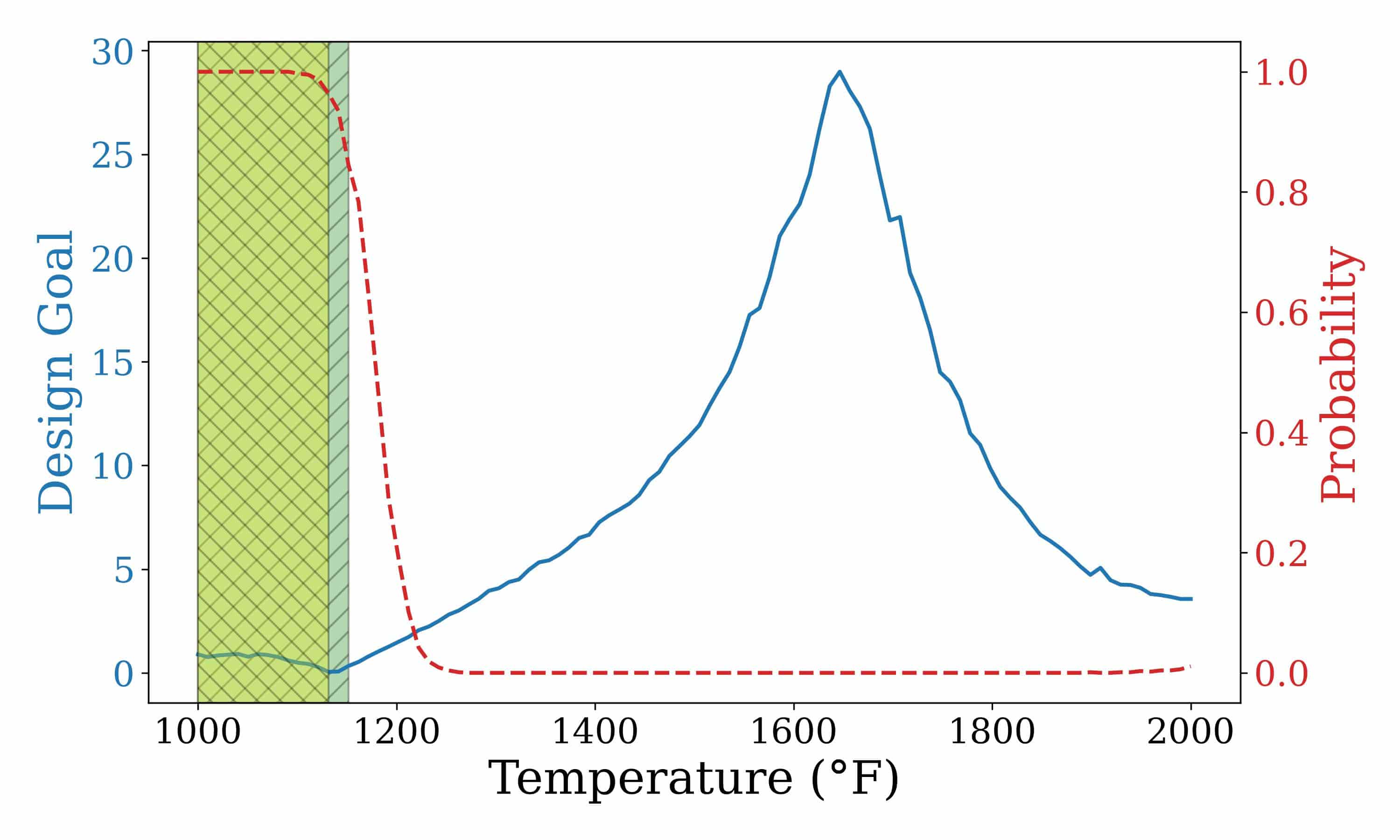}
    \end{subfigure}
    \hfill
    \begin{subfigure}[b]{0.3\textwidth}
      \includegraphics[width=\textwidth]{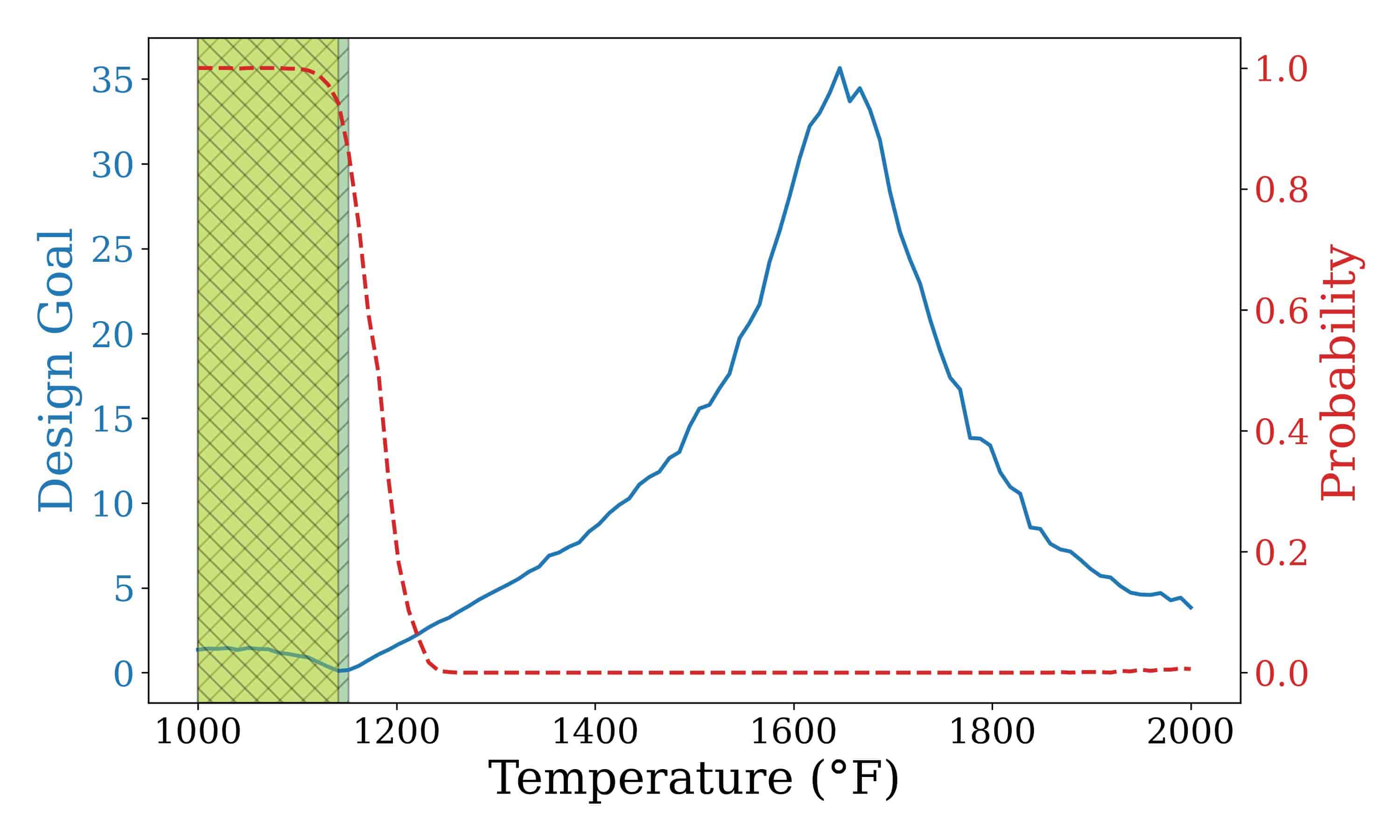}
    \end{subfigure}

    \vspace{0.3cm}
    \begin{subfigure}[b]{0.05\textwidth}
    \end{subfigure}
    \begin{subfigure}[b]{0.3\textwidth}
      \parbox{8cm}{\centering \large \textbf{(a) $\alpha = 0.99$, $EMI_{\text{target}} = 2.32$}}
    \end{subfigure}
    \hfill
    \begin{subfigure}[b]{0.3\textwidth}
      \parbox{7cm}{\centering \large \textbf{(b) $\alpha = 0.95$, $EMI_{\text{target}} = 1.644$}}
    \end{subfigure}
    \hfill
    \begin{subfigure}[b]{0.3\textwidth}
      \parbox{6cm}{\centering \large \textbf{(c) $\alpha = 0.9$, $EMI_{\text{target}} = 1.28$}}
    \end{subfigure}

  \end{subfigure}

  \caption{Comparison of $\text{RcDSP}$  and $\text{rcDSP}$ Methodologies Across Different LRL Values and Various Reliability Indexes for Case A}
  \label{CASEA_RB_vs_RBD}
\end{figure*}
\begin{figure*}[ht]
  \centering
  \includegraphics[width=0.7\textwidth]{title.jpg}
  \begin{subfigure}{\textwidth}
    \centering
   \begin{subfigure}[b]{0.05\textwidth} 
    \centering
    \rotatebox{90}{\textcolor{black}{\parbox{4cm}{\centering\large \textbf{LRL = 270 MPa}}}}
\end{subfigure}
    \begin{subfigure}[b]{0.3\textwidth}
      \includegraphics[width=\textwidth]{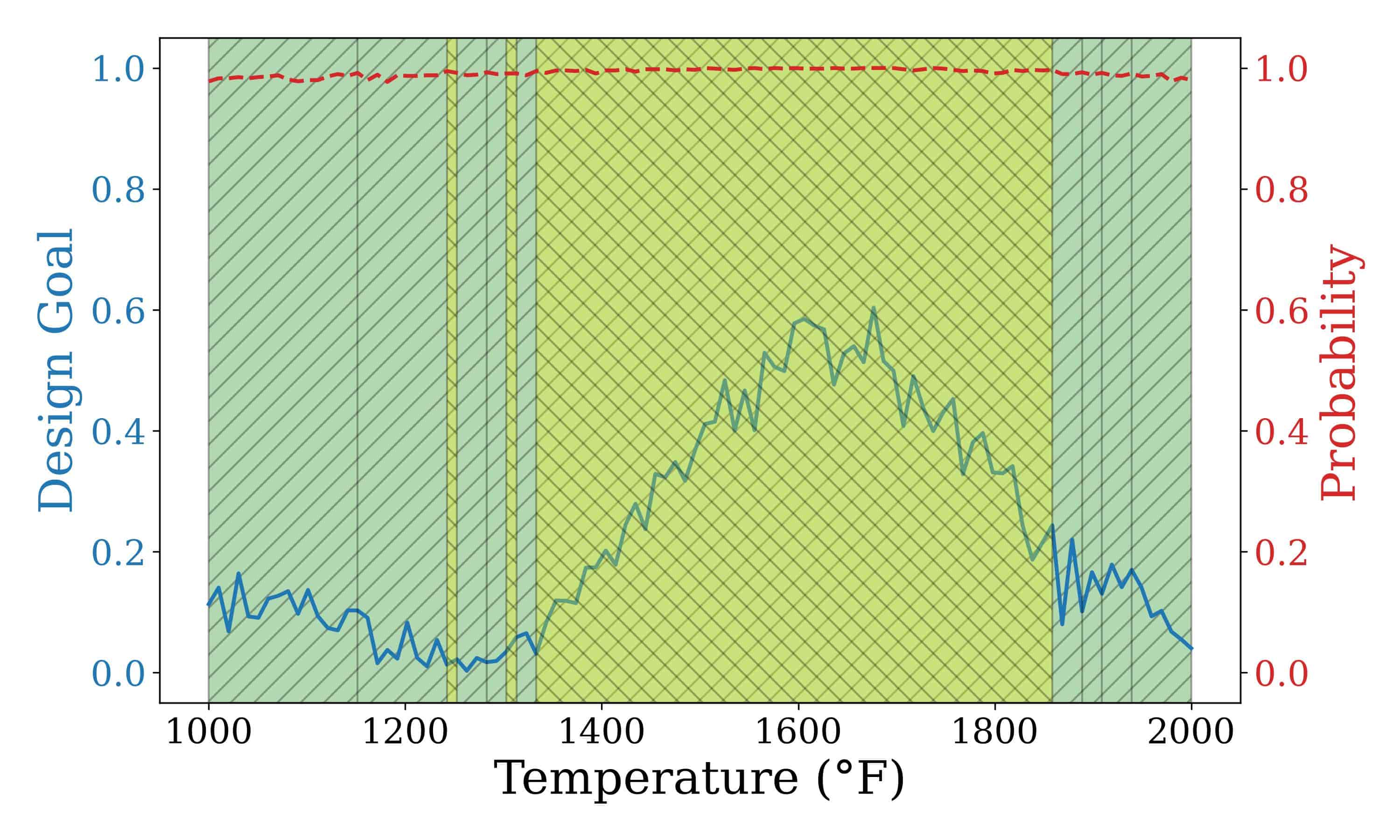}
    \end{subfigure}    
    \hfill
    \begin{subfigure}[b]{0.3\textwidth}
      \includegraphics[width=\textwidth]{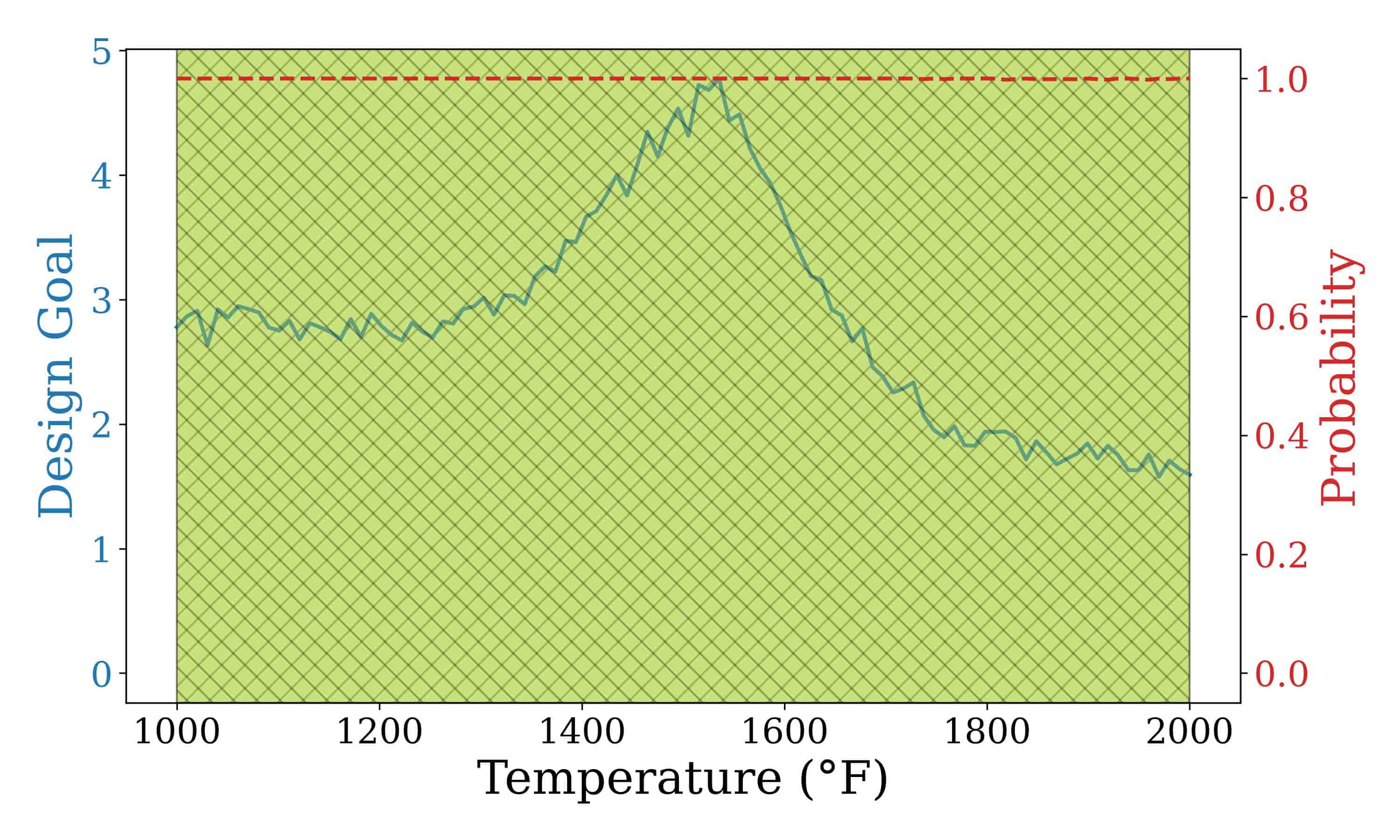}
    \end{subfigure}
    \hfill
    \begin{subfigure}[b]{0.3\textwidth}
       \includegraphics[width=\textwidth]{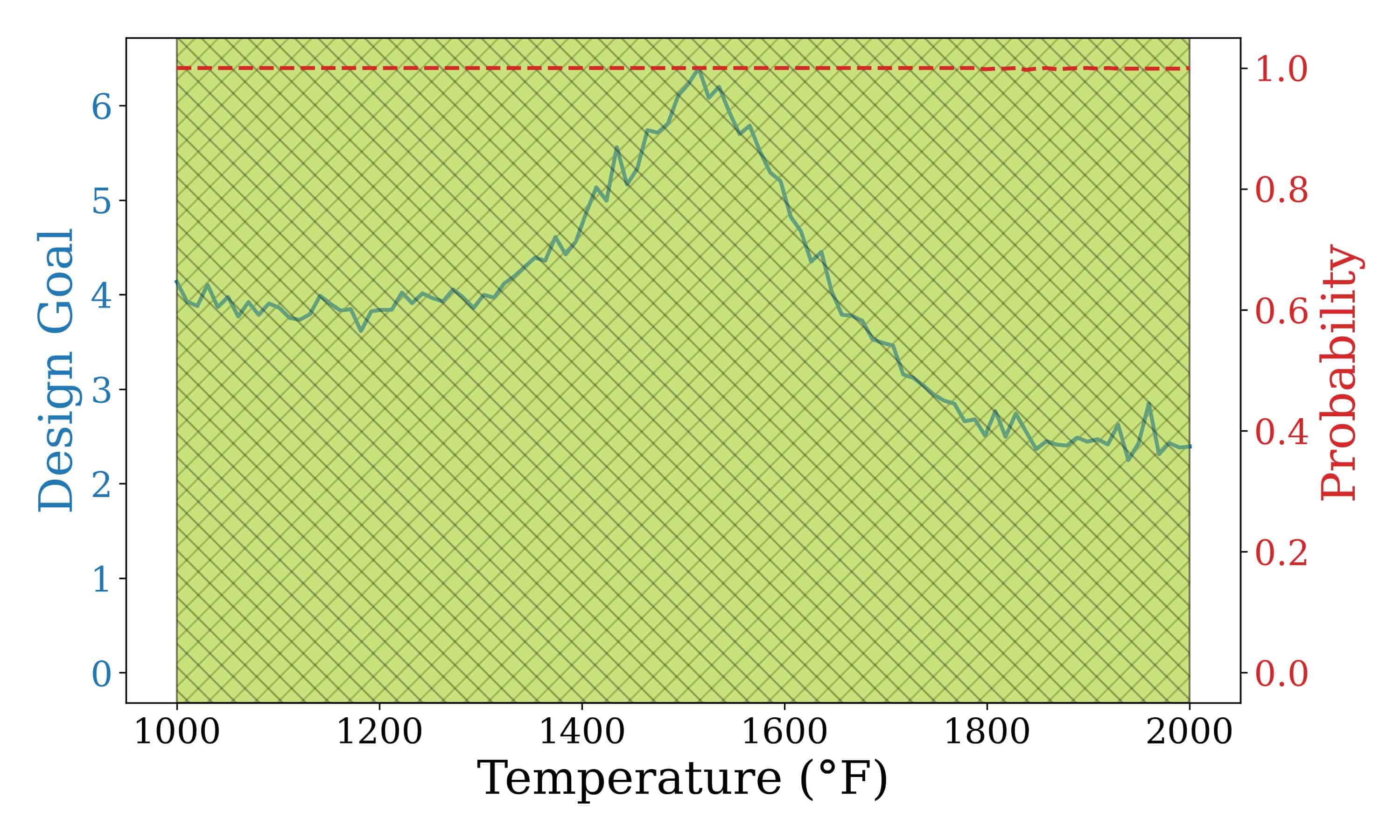}
    \end{subfigure}

    \vspace{0.5cm}
    \begin{subfigure}[b]{0.05\textwidth} 
    \centering
    \rotatebox{90}{\textcolor{black}{\parbox{4cm}{\centering\large \textbf{LRL = 280 MPa}}}}
\end{subfigure}
    \begin{subfigure}[b]{0.3\textwidth}
       \includegraphics[width=\textwidth]{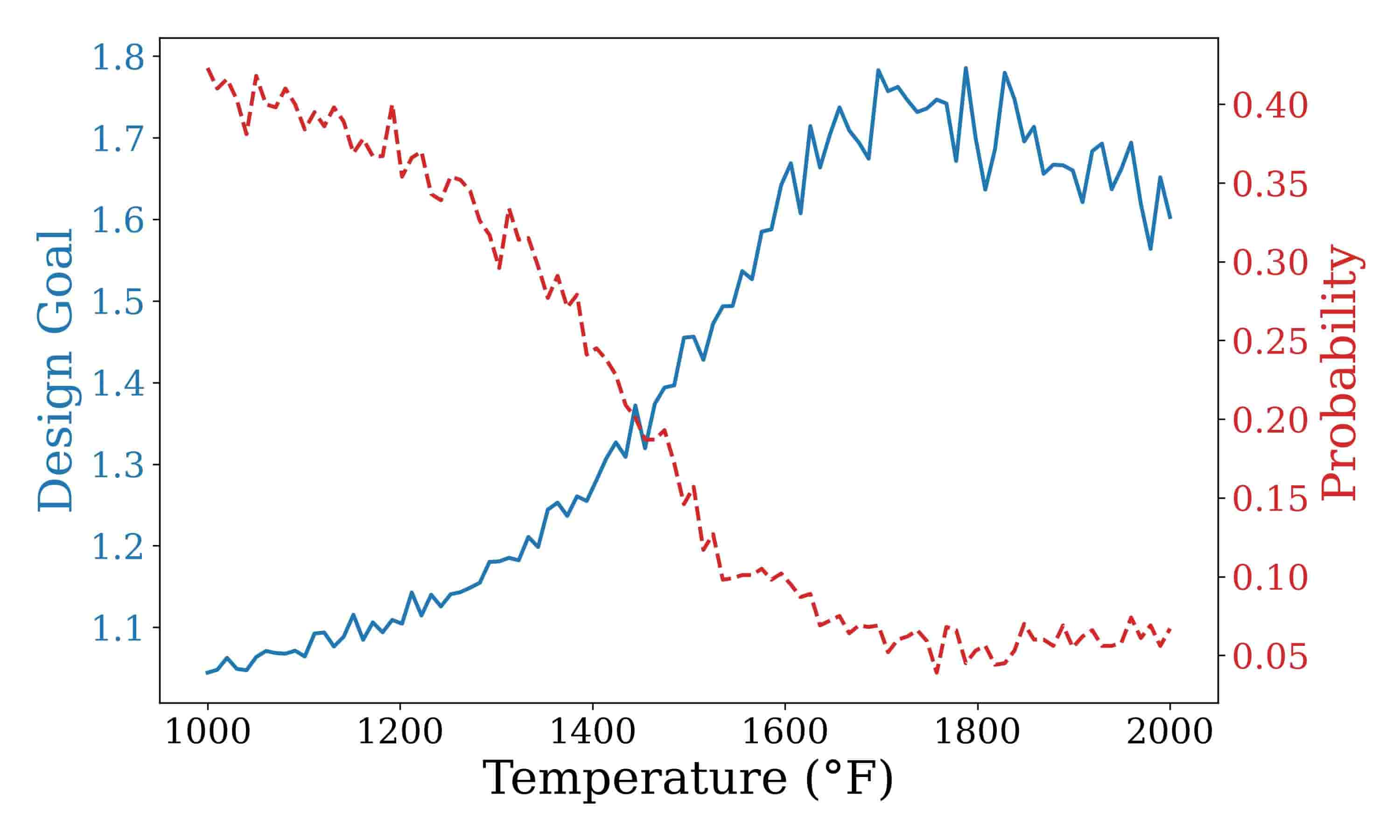}
    \end{subfigure}
    \hfill
    \begin{subfigure}[b]{0.3\textwidth}
      \includegraphics[width=\textwidth]{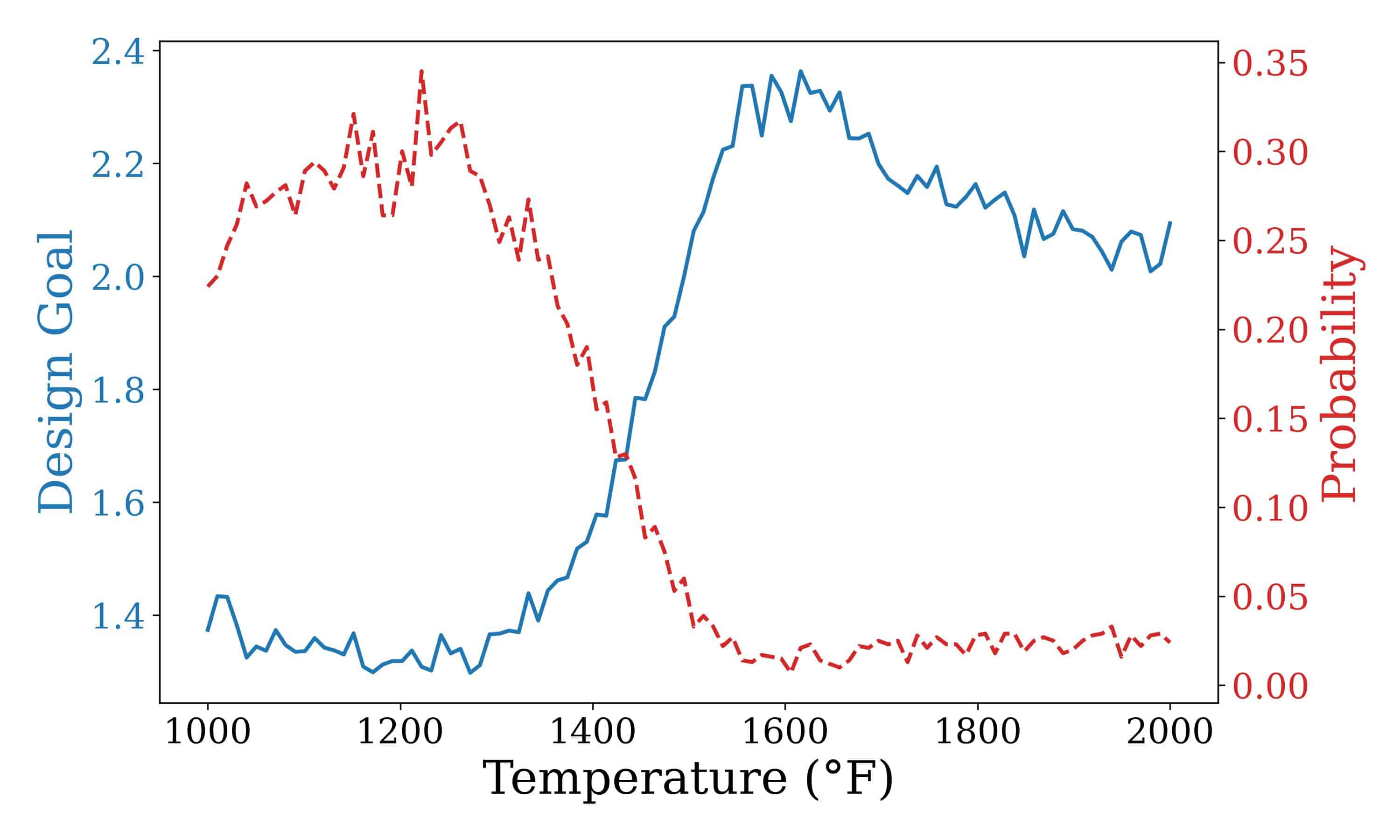}
    \end{subfigure}
    \hfill
    \begin{subfigure}[b]{0.3\textwidth}
      \includegraphics[width=\textwidth]{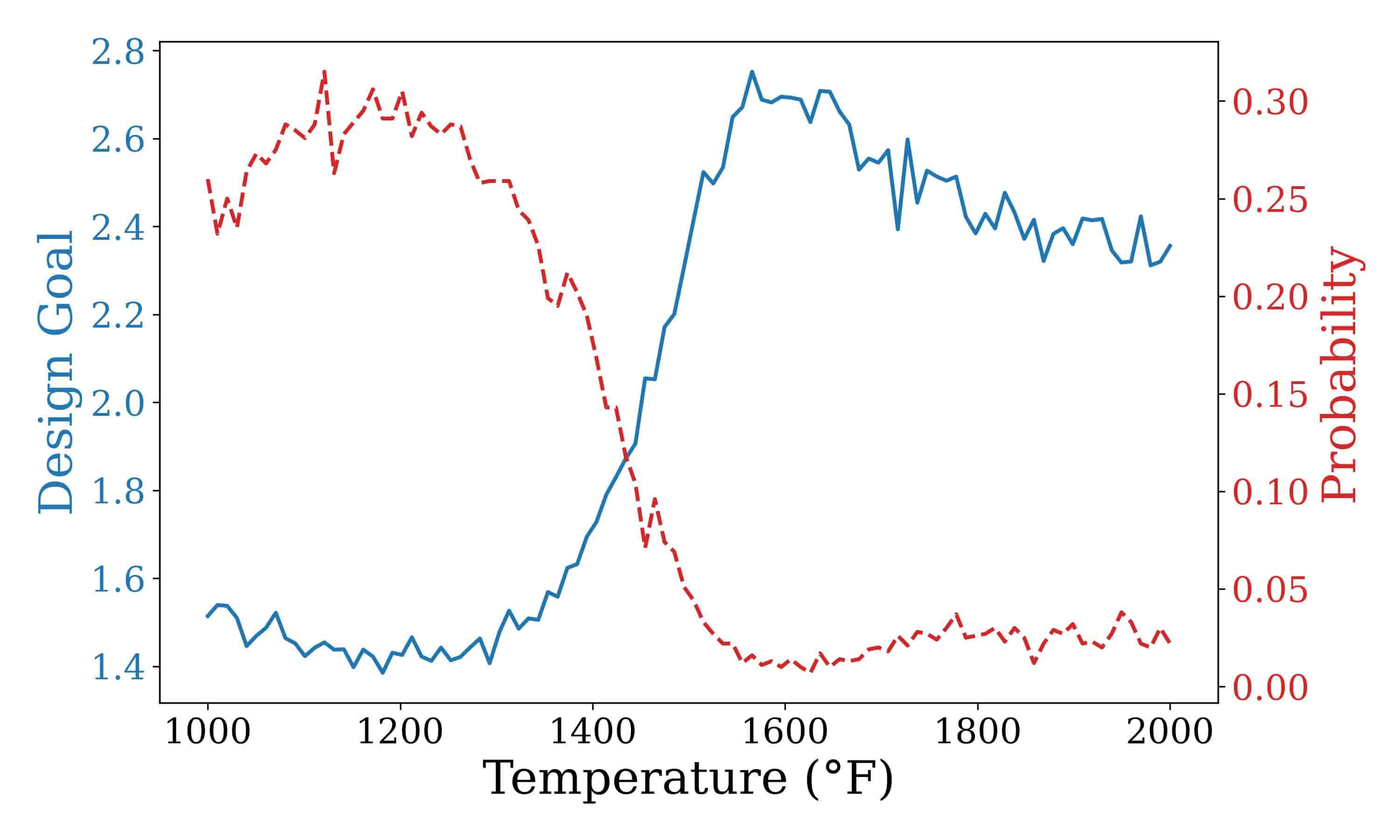}
    \end{subfigure}

    \vspace{0.5cm}
   \begin{subfigure}[b]{0.05\textwidth} 
    \centering
    \rotatebox{90}{\textcolor{black}{\parbox{4cm}{\centering\large \textbf{LRL = 200 MPa}}}}
\end{subfigure}
    \begin{subfigure}[b]{0.3\textwidth}
      \includegraphics[width=\textwidth]{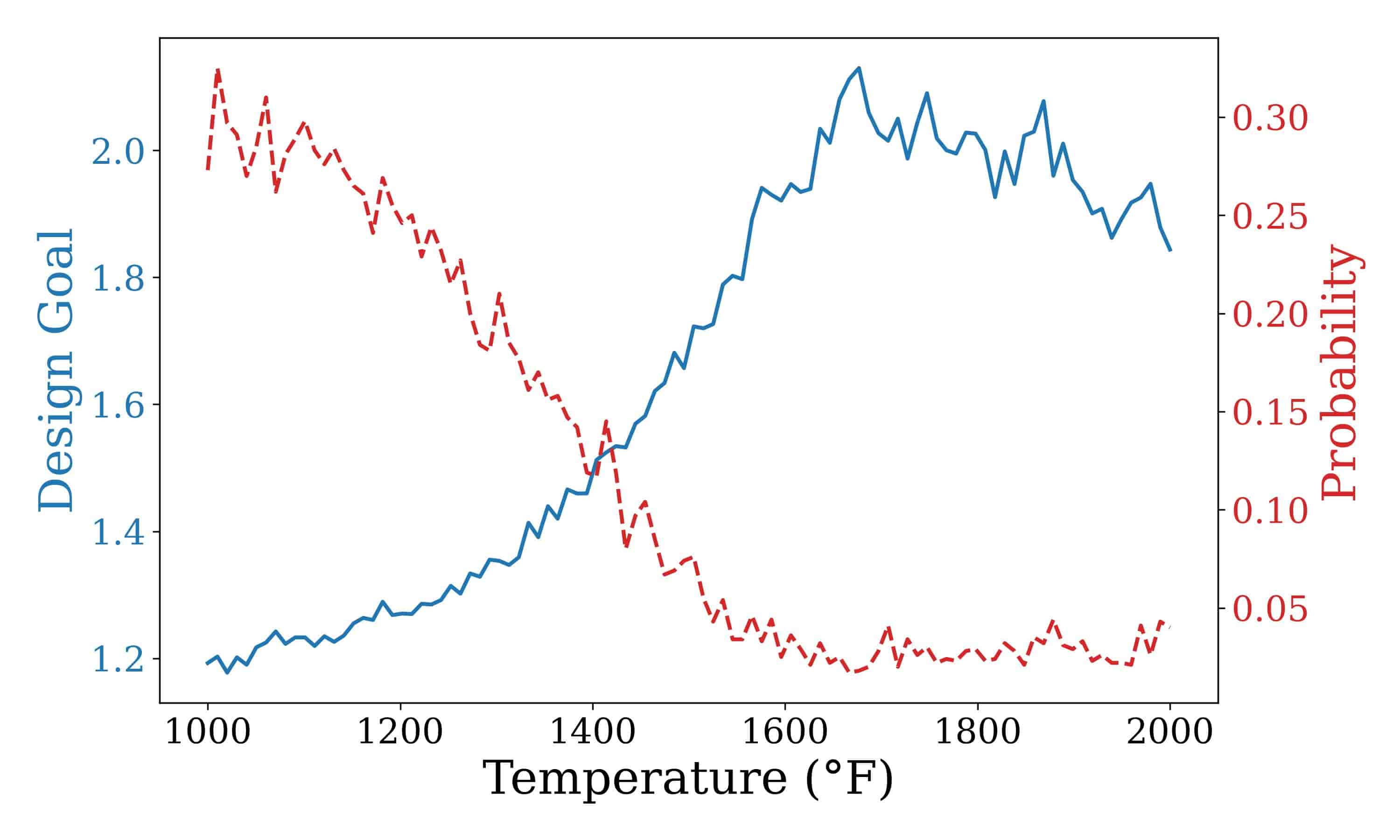}
    \end{subfigure}
    \hfill
    \begin{subfigure}[b]{0.3\textwidth}
      \includegraphics[width=\textwidth]{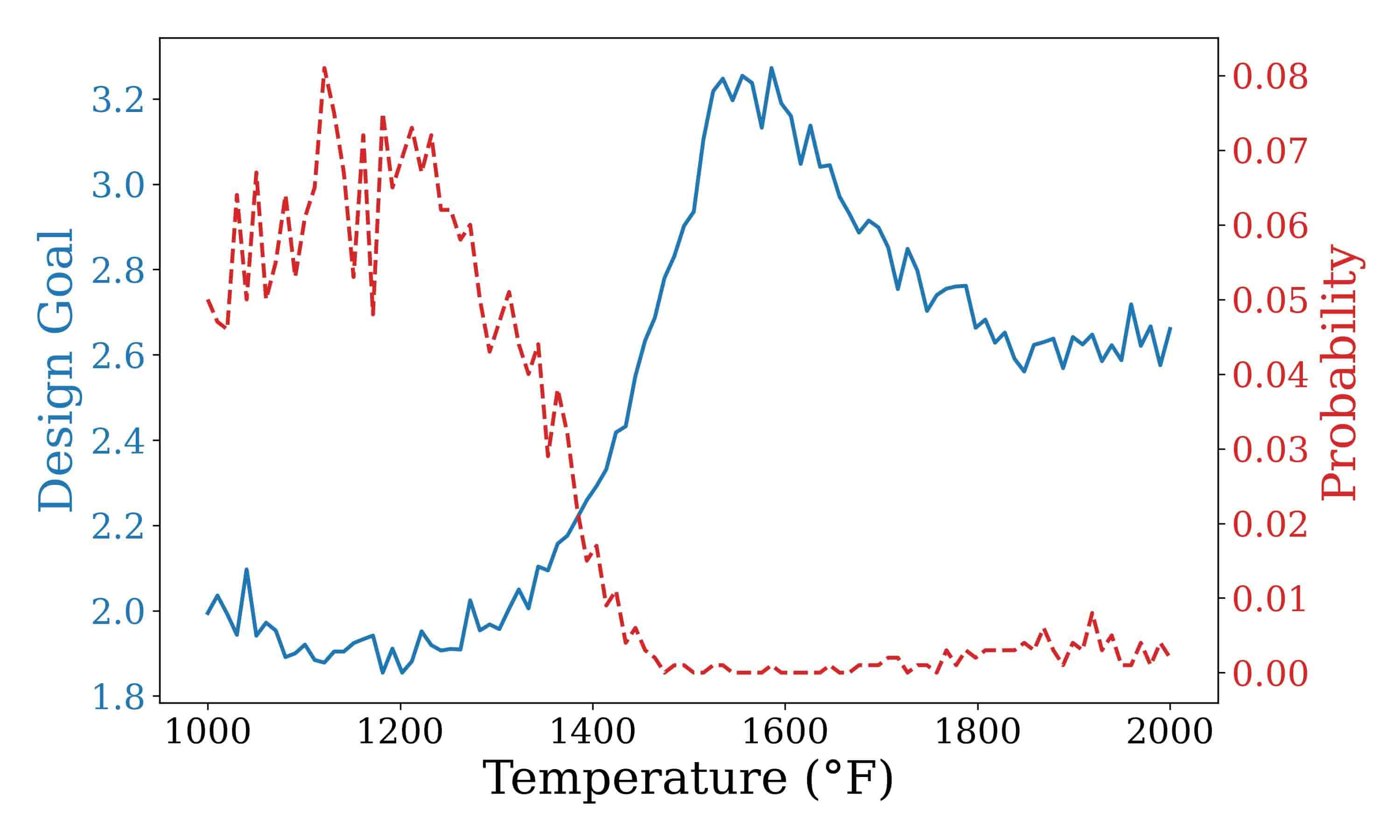}
    \end{subfigure}
    \hfill
    \begin{subfigure}[b]{0.3\textwidth}
      \includegraphics[width=\textwidth]{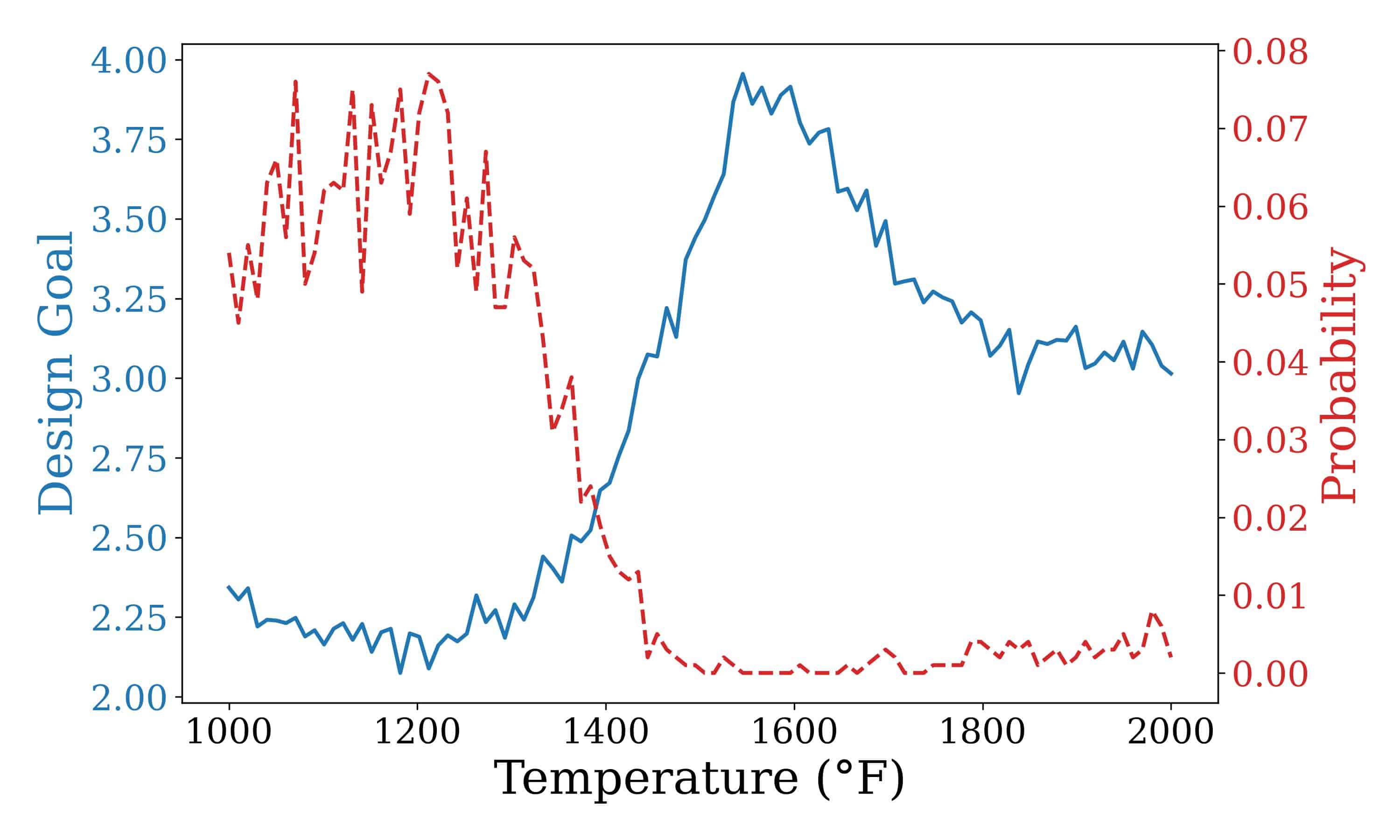}
    \end{subfigure}

    \vspace{0.3cm}
    \begin{subfigure}[b]{0.05\textwidth}
    \end{subfigure}
    \begin{subfigure}[b]{0.3\textwidth}
      \parbox{8cm}{\centering \large \textbf{(a) $\alpha = 0.99$, $EMI_{\text{target}} = 2.32$}}
    \end{subfigure}
    \hfill
    \begin{subfigure}[b]{0.3\textwidth}
      \parbox{7cm}{\centering \large \textbf{(b) $\alpha = 0.95$, $EMI_{\text{target}} = 1.644$}}
    \end{subfigure}
    \hfill
    \begin{subfigure}[b]{0.3\textwidth}
      \parbox{6cm}{\centering \large \textbf{(c) $\alpha = 0.9$, $EMI_{\text{target}} = 1.28$}}
    \end{subfigure}

  \end{subfigure}

  \caption{Comparison of $\text{RcDSP}$  and $\text{rcDSP}$ Methodologies Across Different LRL Values and Various Reliability Indexes for Case B}
  \label{CASEB_RB_vs_RBD}
\end{figure*}

Figure~\ref{yield_distributions} presents the normalized histogram of the resulting yield strength for each case at the specified temperature of \( T = \SI{1450}{\degree F} \).
Case~A (Figure~\ref{yield_distributions}a) exhibits a near-normal yield strength distribution. The normal characteristics of this distribution drive convergence between the admissible design spaces of rcDSP and RcDSP, yielding nearly identical optimal solution regions, as demonstrated in Figure~\ref{CASEA_RB_vs_RBD}. As the reliability index increases, the divergence between rcDSP and RcDSP solutions widens due to heightened non-normal characteristics at the tail of the distribution. This emerging discrepancy stems from the deliberate incorporation of uncertainty in tail behavior through RcDSP, which becomes increasingly pronounced under distributional asymmetry (\(\gamma > 0\)) and excess kurtosis (\(\kappa > 0\)) \cite{madsen2006methods,hasofer1974exact}.
Case~B (Figure~\ref{yield_distributions}b) highlights the compounded effects of performance uncertainty due to limited training data. The resulting yield strength distribution exhibits pronounced non-normal characteristics, particularly in the tail regions, fundamentally altering the design space topology. As shown in Figure~\ref{CASEB_RB_vs_RBD}, these distributional complexities lead to significant divergence between rcDSP and RcDSP outcomes at elevated values for $\alpha_T$. 

Under these conditions, the admissible design space for RcDSP becomes fragmented, reflecting sensitivity to higher-order statistical moments governing tail probabilities. This discontinuity arises at specific design temperatures where uncertainty is significantly amplified, driven by sudden spikes in variance interacting with the inherent asymmetry of the distribution. In contrast, rcDSP maintains continuous feasibility boundaries, as it relies solely on the first- and second-moment statistics \cite{hasofer1974exact} (i.e., mean and variance) through the EMI formulation. This fundamental distinction arises because the feasibility threshold of rcDSP remains invariant to distribution shape, whereas RcDSP explicitly accounts for tail probability deviations from normal assumptions.  As shown in Figure \ref{CASEB_RB_vs_RBD}  there is no admissible solution space for either RcDSP or rcDSP at elevated LRL values ($LRL>200$).

\begin{figure*}[ht]
  \centering
  \includegraphics[width=0.7\textwidth]{title.jpg}
  \label{top2}
  \begin{subfigure}{\textwidth}
    \centering

   \begin{subfigure}[b]{0.05\textwidth} 
    \centering
    \rotatebox{90}{\textcolor{black}{\parbox{4cm}{\centering\large \textbf{LRL = 150 MPa}}}}
\end{subfigure}
    \begin{subfigure}[b]{0.3\textwidth}
      \includegraphics[width=\textwidth]{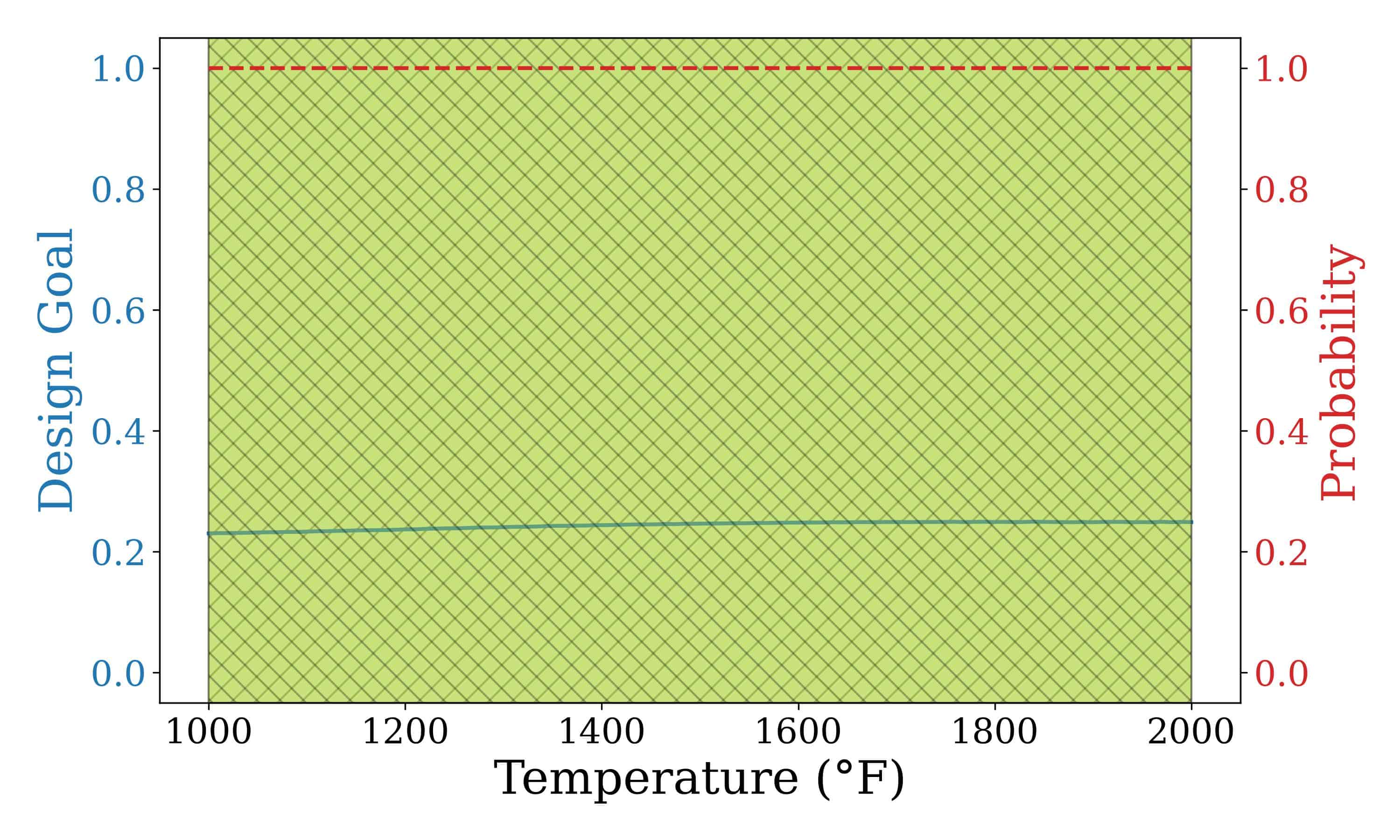}
    \end{subfigure}    
    \hfill
    \begin{subfigure}[b]{0.3\textwidth}
      \includegraphics[width=\textwidth]{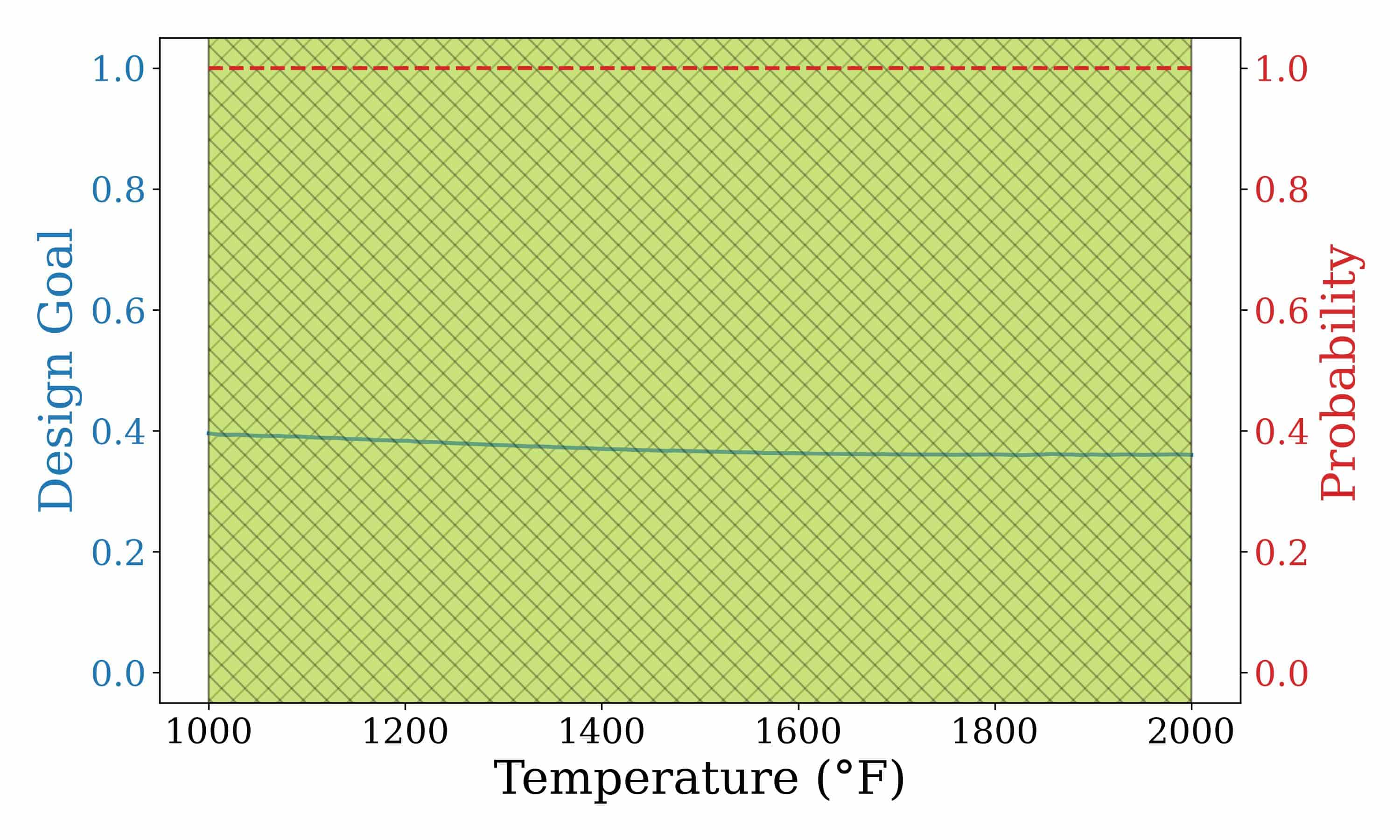}
    \end{subfigure}
    \hfill
    \begin{subfigure}[b]{0.3\textwidth}
       \includegraphics[width=\textwidth]{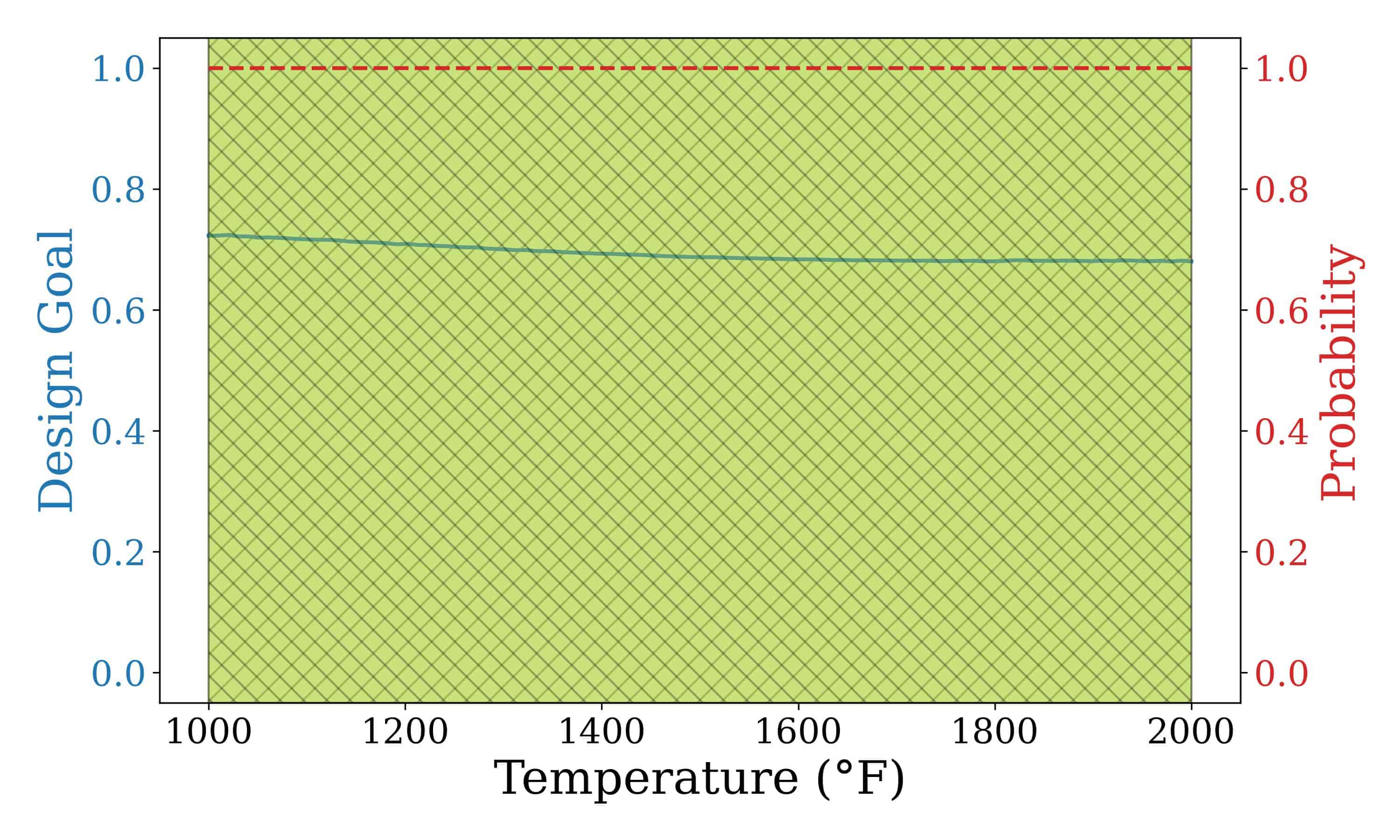}
    \end{subfigure}

    \vspace{0.5cm}
    \begin{subfigure}[b]{0.05\textwidth} 
    \centering
    \rotatebox{90}{\textcolor{black}{\parbox{4cm}{\centering\large \textbf{LRL = 180 MPa}}}}
\end{subfigure}
    \begin{subfigure}[b]{0.3\textwidth}
       \includegraphics[width=\textwidth]{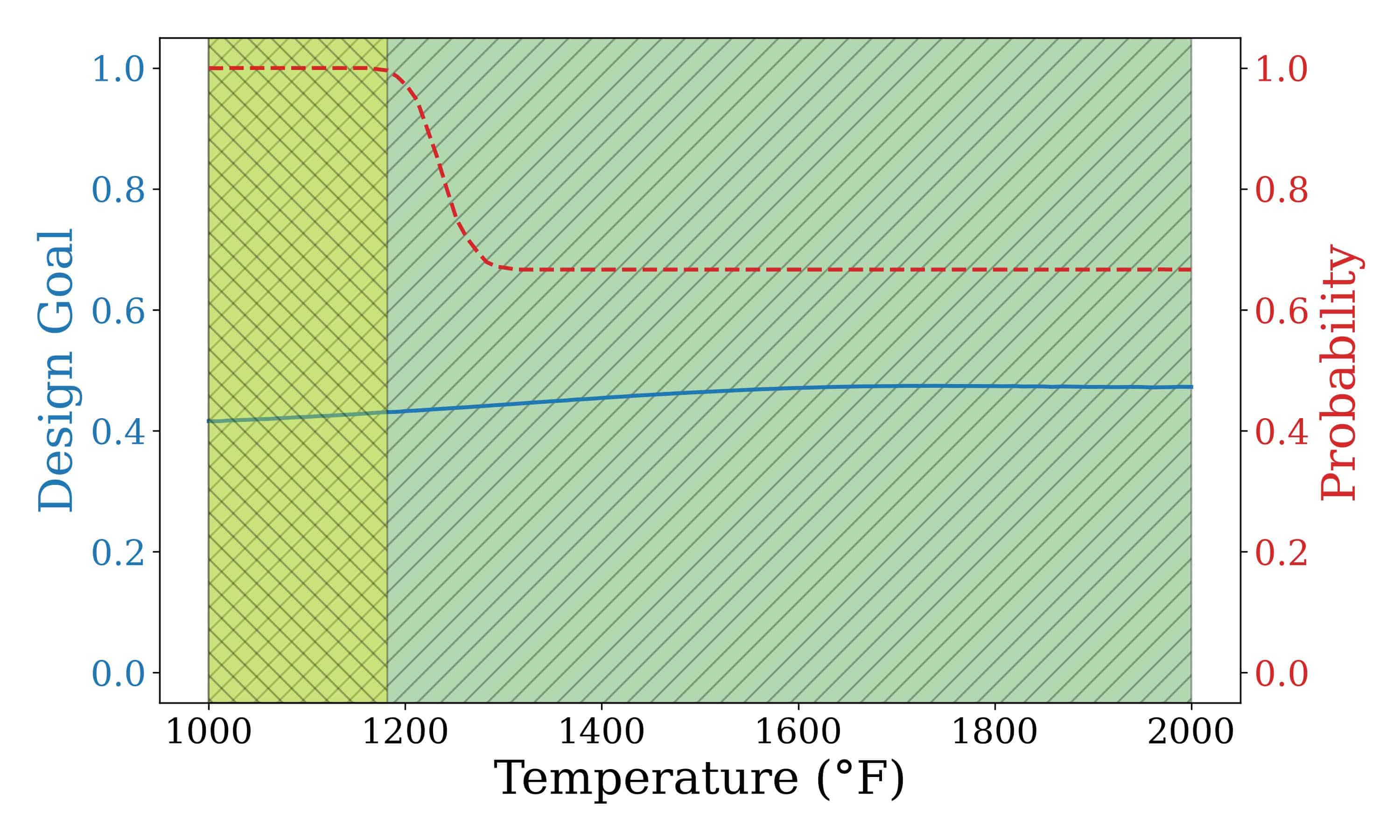}
    \end{subfigure}
    \hfill
    \begin{subfigure}[b]{0.3\textwidth}
      \includegraphics[width=\textwidth]{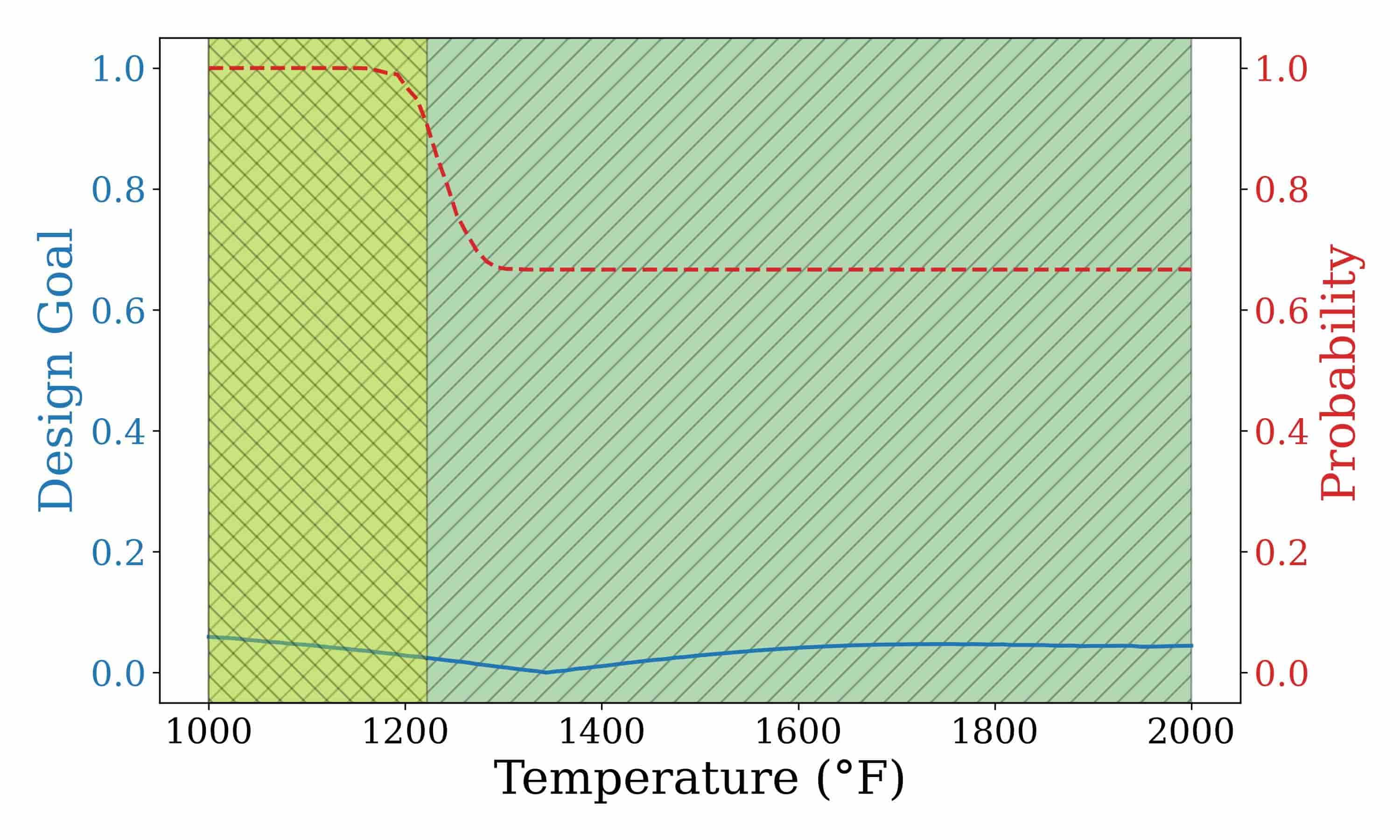}
    \end{subfigure}
    \hfill
    \begin{subfigure}[b]{0.3\textwidth}
      \includegraphics[width=\textwidth]{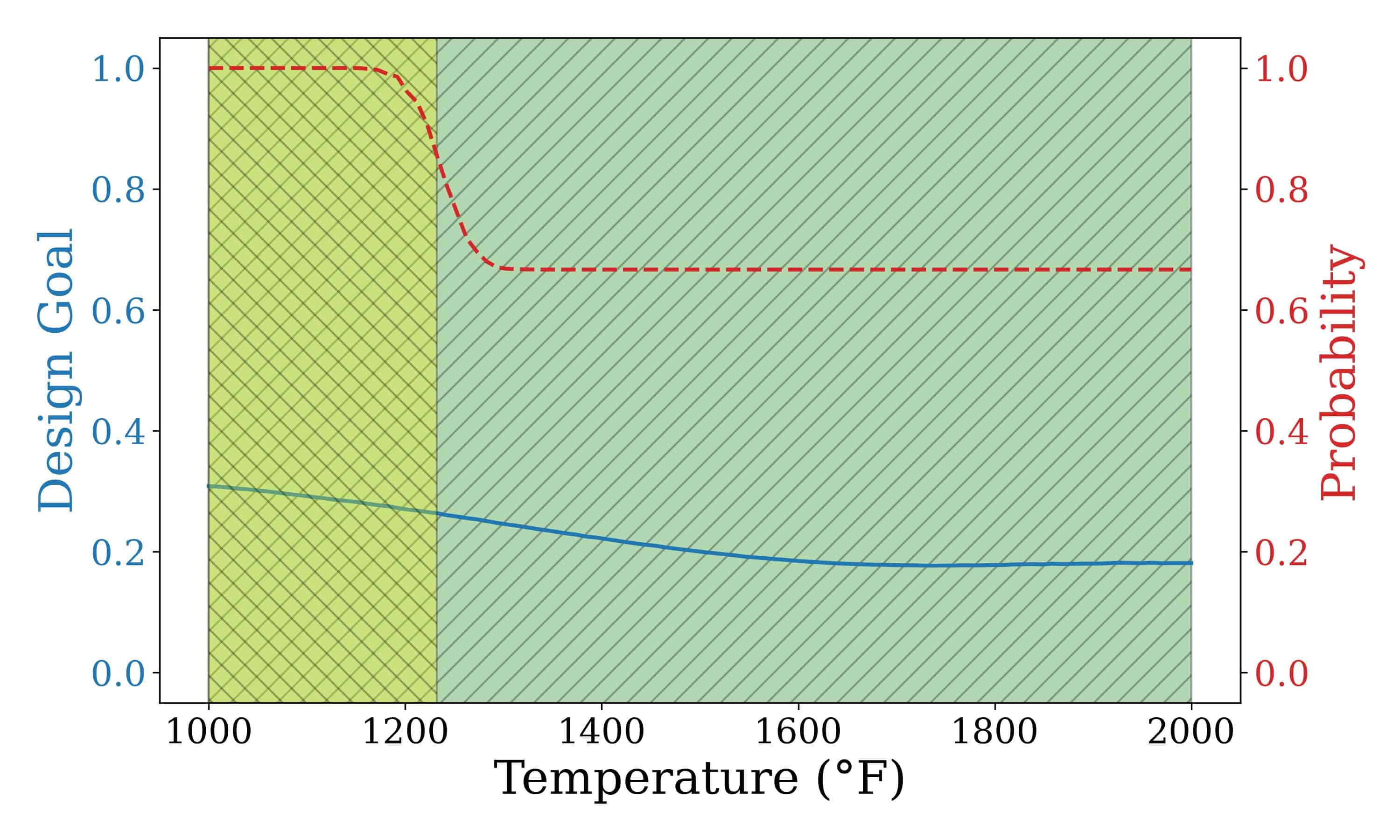}
    \end{subfigure}

    \vspace{0.5cm}
   \begin{subfigure}[b]{0.05\textwidth} 
    \centering
    \rotatebox{90}{\textcolor{black}{\parbox{4cm}{\centering\large \textbf{LRL = 200 MPa}}}}
\end{subfigure}
    \begin{subfigure}[b]{0.3\textwidth}
      \includegraphics[width=\textwidth]{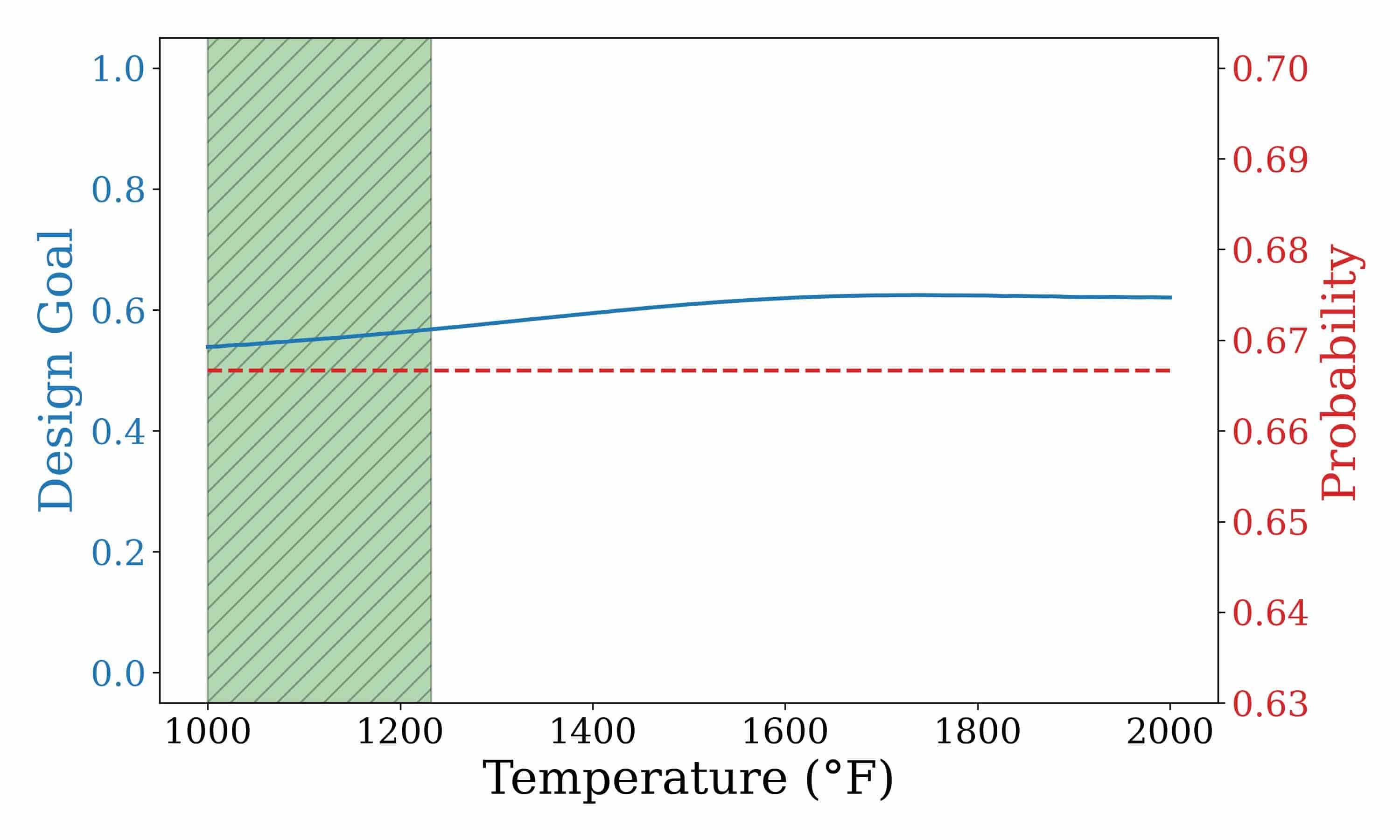}
    \end{subfigure}
    \hfill
    \begin{subfigure}[b]{0.3\textwidth}
      \includegraphics[width=\textwidth]{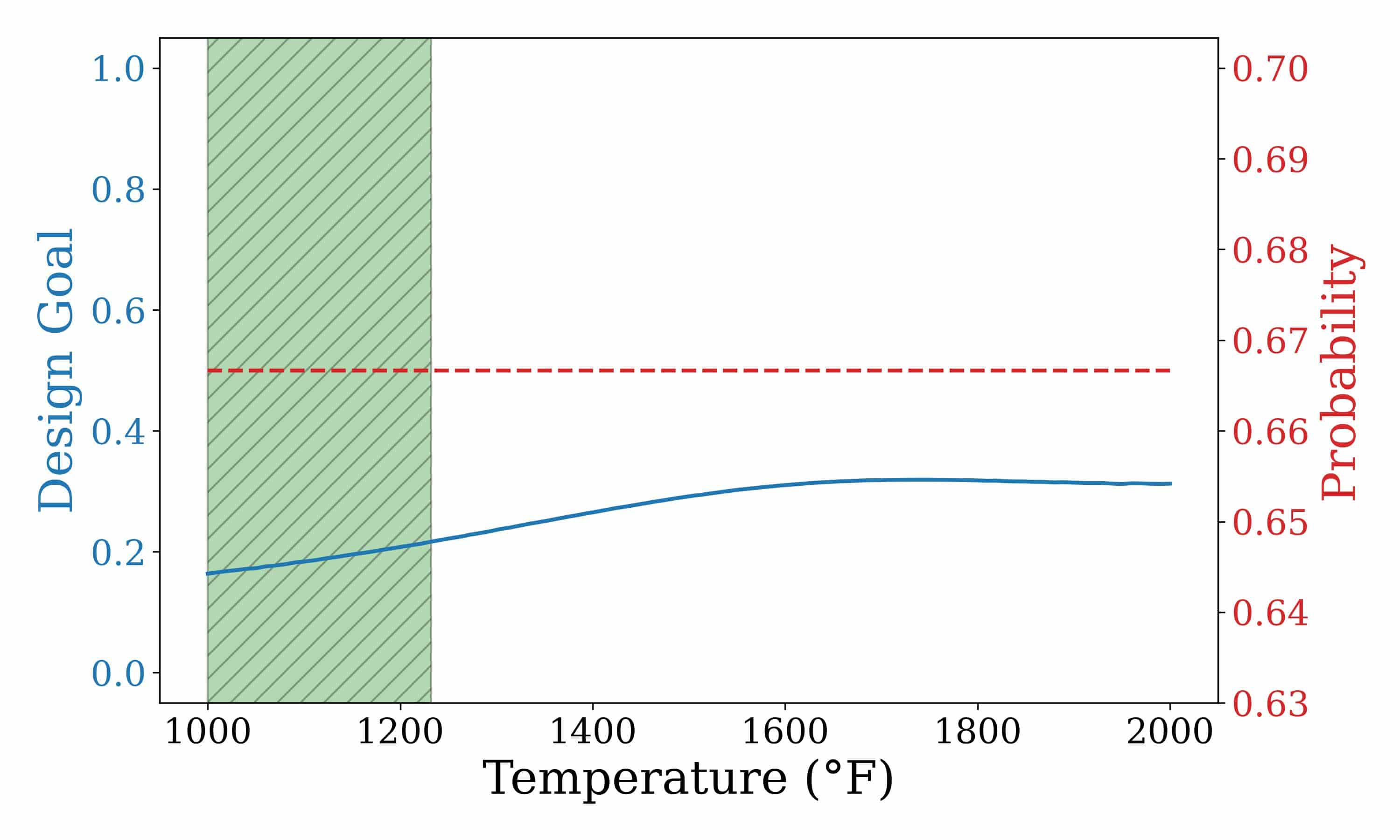}
    \end{subfigure}
    \hfill
    \begin{subfigure}[b]{0.3\textwidth}
      \includegraphics[width=\textwidth]{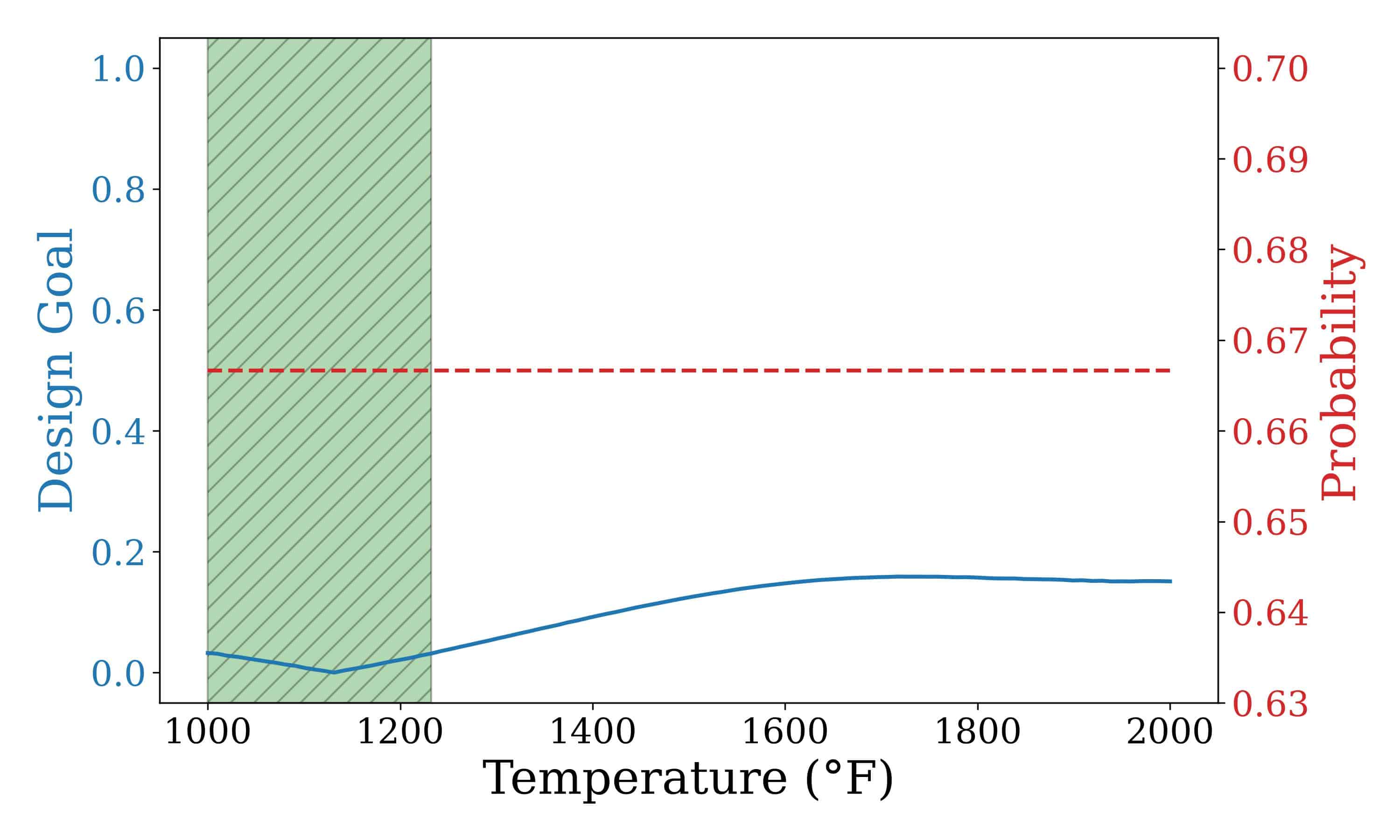}
    \end{subfigure}

    \vspace{0.3cm}
    \begin{subfigure}[b]{0.05\textwidth}
    \end{subfigure}
    \begin{subfigure}[b]{0.3\textwidth}
      \parbox{8cm}{\centering \large \textbf{(a) $\alpha = 0.99$, $EMI_{\text{target}} = 2.32$}}
    \end{subfigure}
    \hfill
    \begin{subfigure}[b]{0.3\textwidth}
      \parbox{7cm}{\centering \large \textbf{(b) $\alpha = 0.9$, $EMI_{\text{target}} = 1.28$}}
    \end{subfigure}
    \hfill
    \begin{subfigure}[b]{0.3\textwidth}
      \parbox{6cm}{\centering \large \textbf{(c) $\alpha = 0.85$, $EMI_{\text{target}} = 1.036$}}
    \end{subfigure}

  \end{subfigure}

  \caption{Comparison of $\text{RcDSP}$  and $\text{rcDSP}$ Across Different LRL Values and Various Reliability Indexes for Case C}
  \label{CASEC_RB_vs_RBD}
\end{figure*}

\begin{figure*}[ht]
  \centering
  \includegraphics[width=0.7\textwidth]{title.jpg}
  \label{top}
  \begin{subfigure}{\textwidth}
    \centering

   \begin{subfigure}[b]{0.05\textwidth} 
    \centering
    \rotatebox{90}{\textcolor{black}{\parbox{4cm}{\centering\large \textbf{LRL = 150 MPa}}}}
\end{subfigure}
    \begin{subfigure}[b]{0.3\textwidth}
      \includegraphics[width=\textwidth]{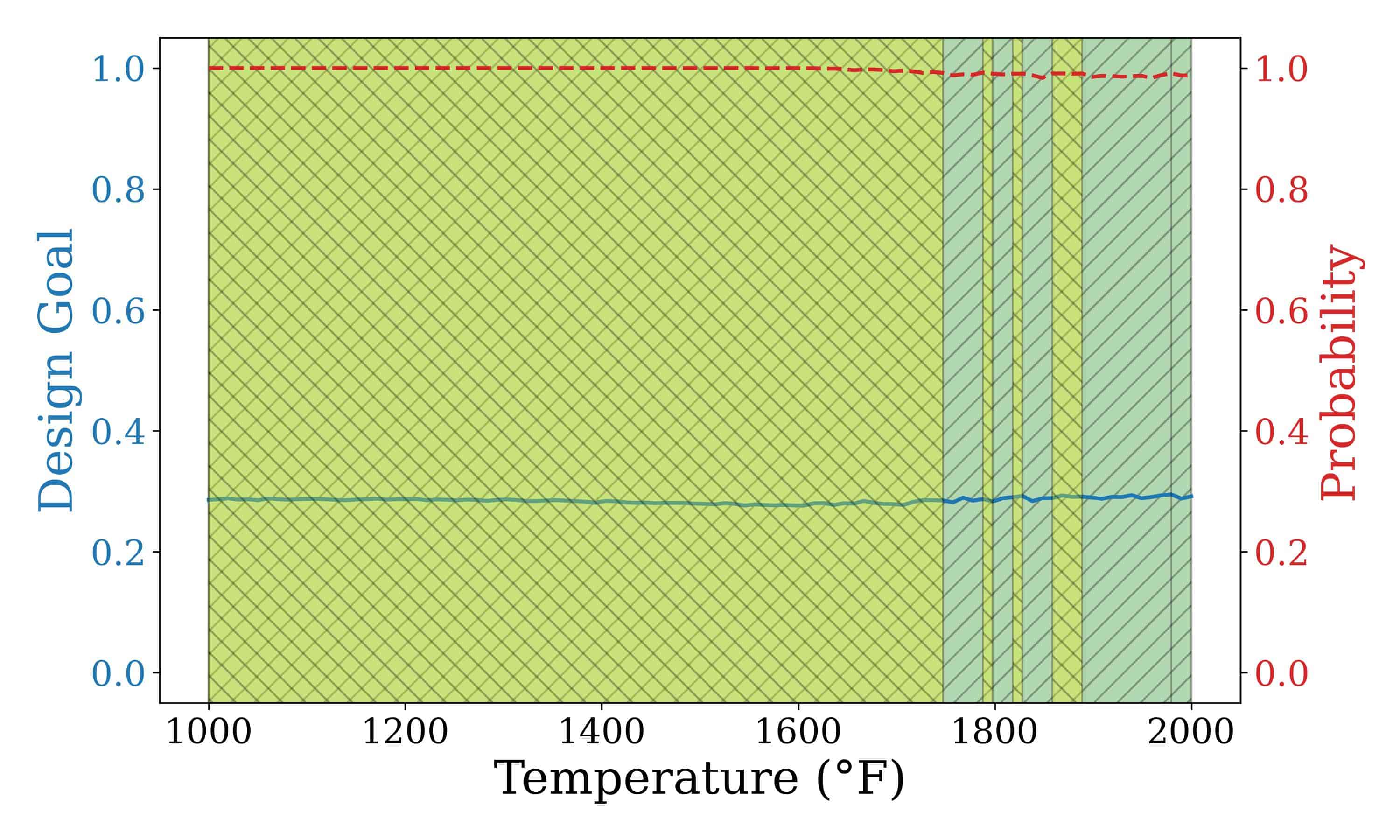}
    \end{subfigure}    
    \hfill
    \begin{subfigure}[b]{0.3\textwidth}
      \includegraphics[width=\textwidth]{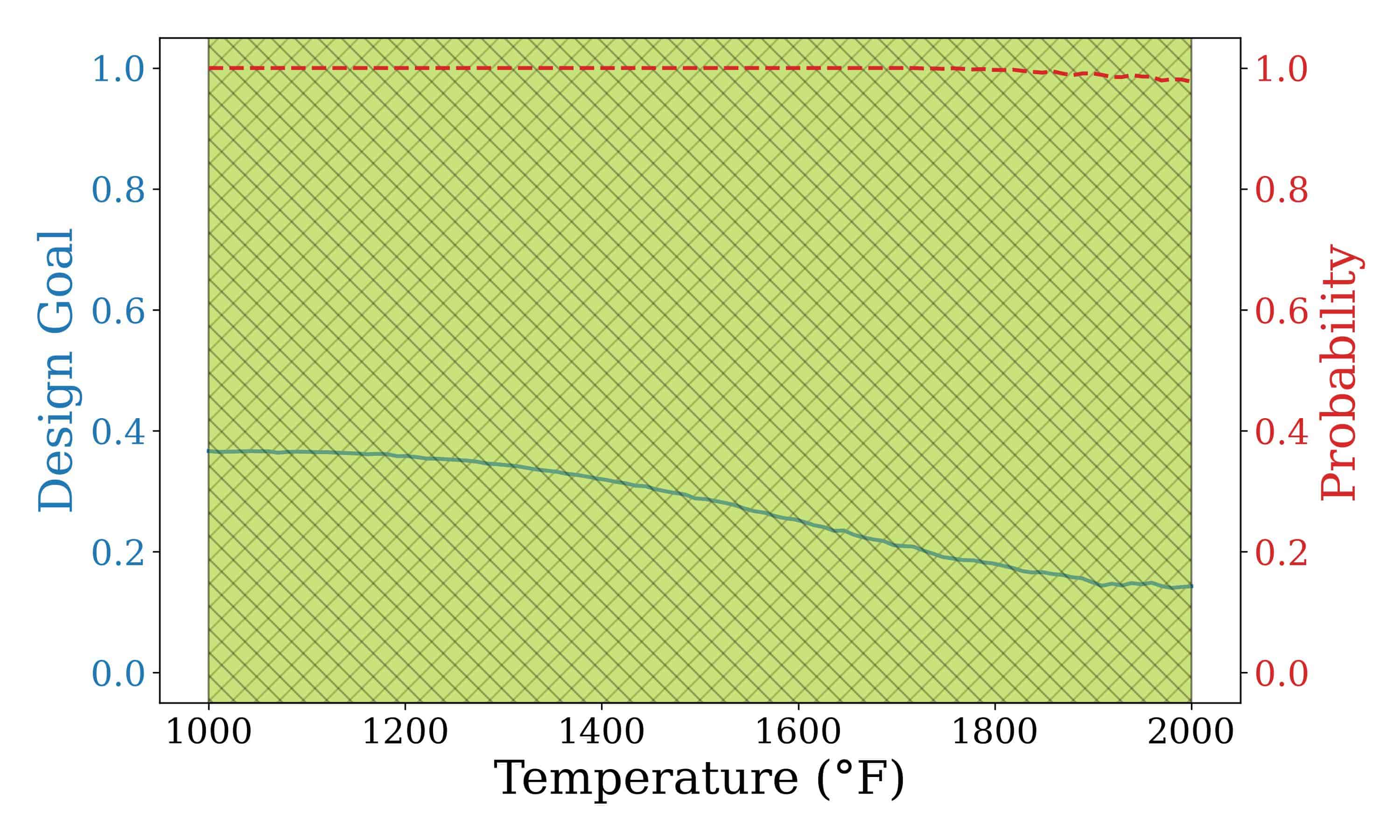}
    \end{subfigure}
    \hfill
    \begin{subfigure}[b]{0.3\textwidth}
       \includegraphics[width=\textwidth]{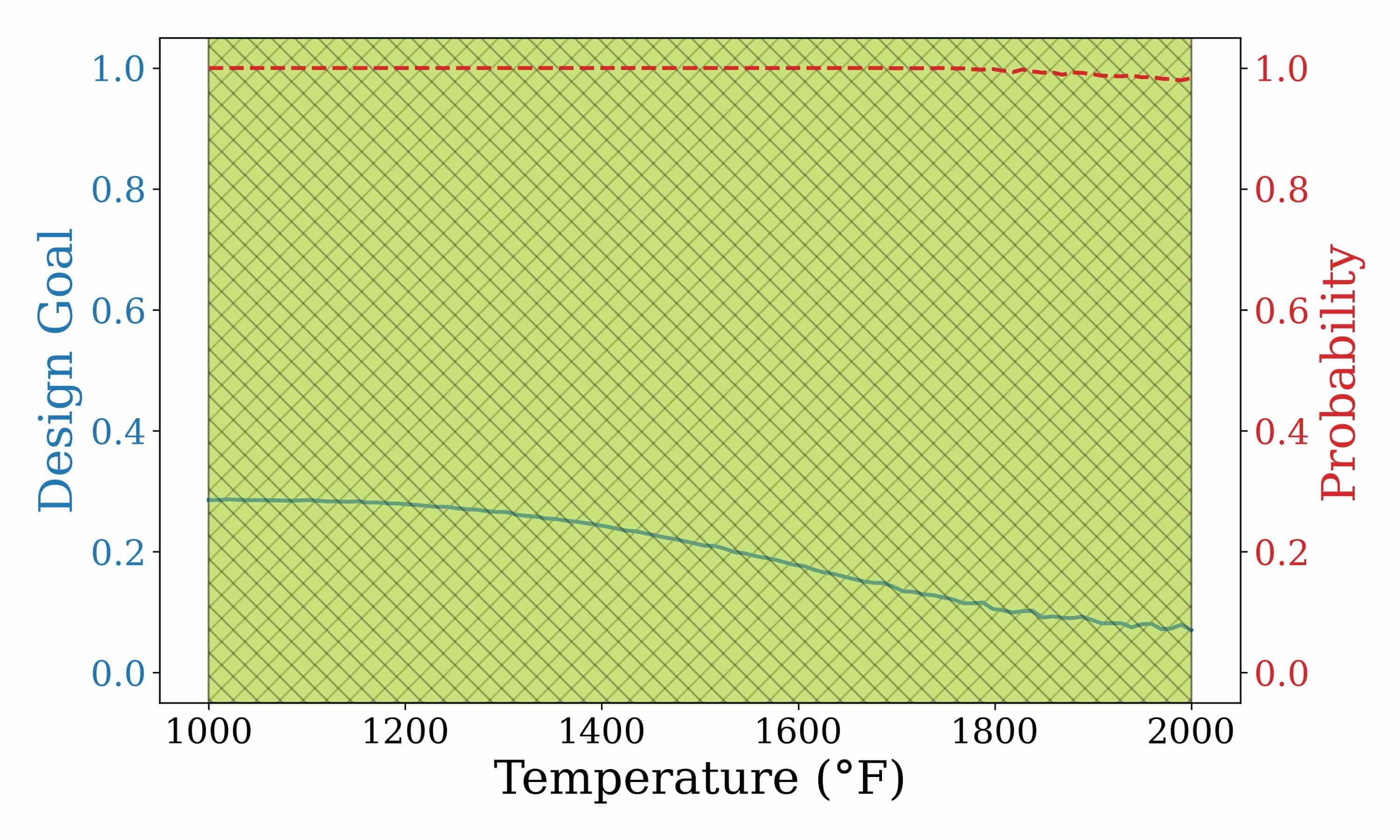}
    \end{subfigure}

    \vspace{0.5cm}
    \begin{subfigure}[b]{0.05\textwidth} 
    \centering
    \rotatebox{90}{\textcolor{black}{\parbox{4cm}{\centering\large \textbf{LRL = 180 MPa}}}}
\end{subfigure}
    \begin{subfigure}[b]{0.3\textwidth}
       \includegraphics[width=\textwidth]{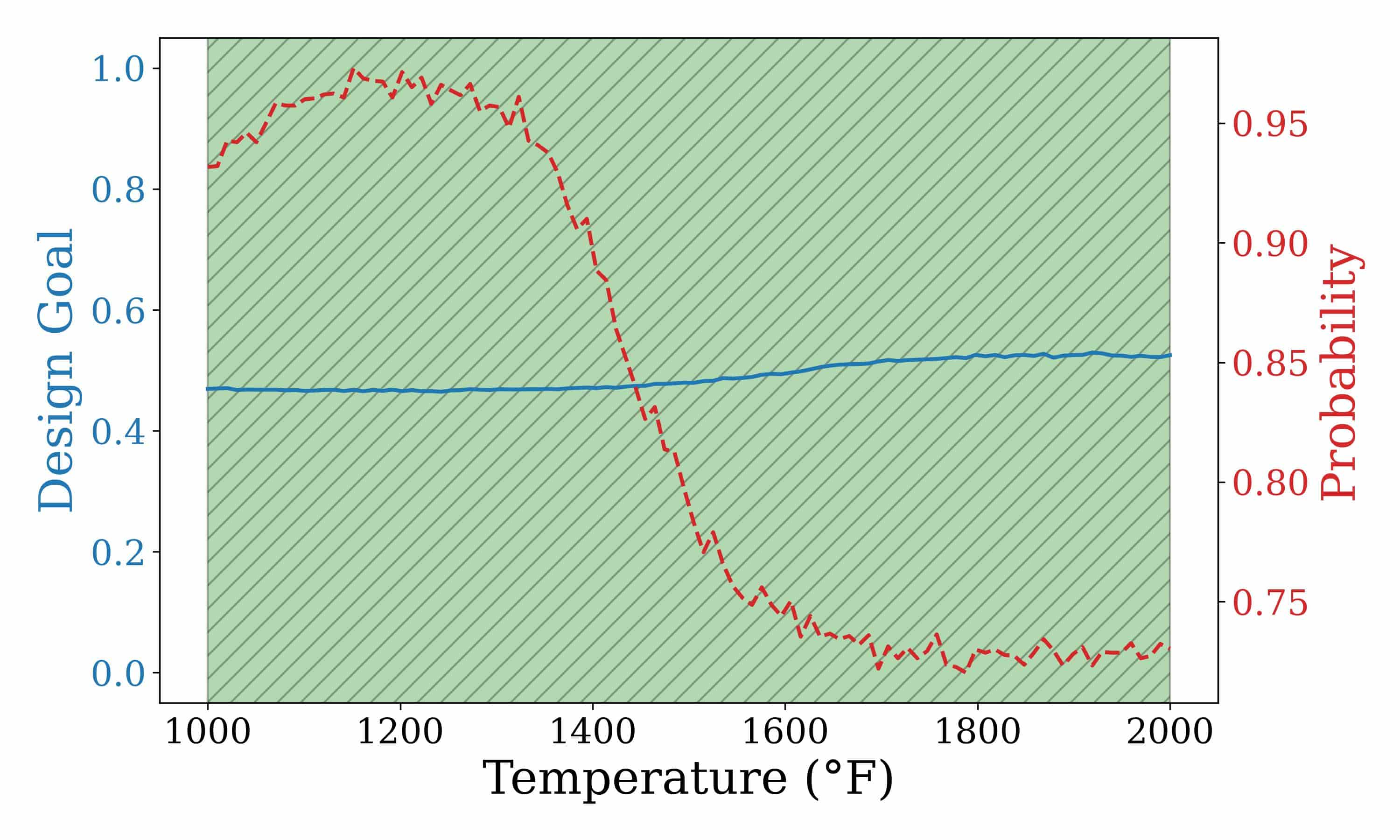}
    \end{subfigure}
    \hfill
    \begin{subfigure}[b]{0.3\textwidth}
      \includegraphics[width=\textwidth]{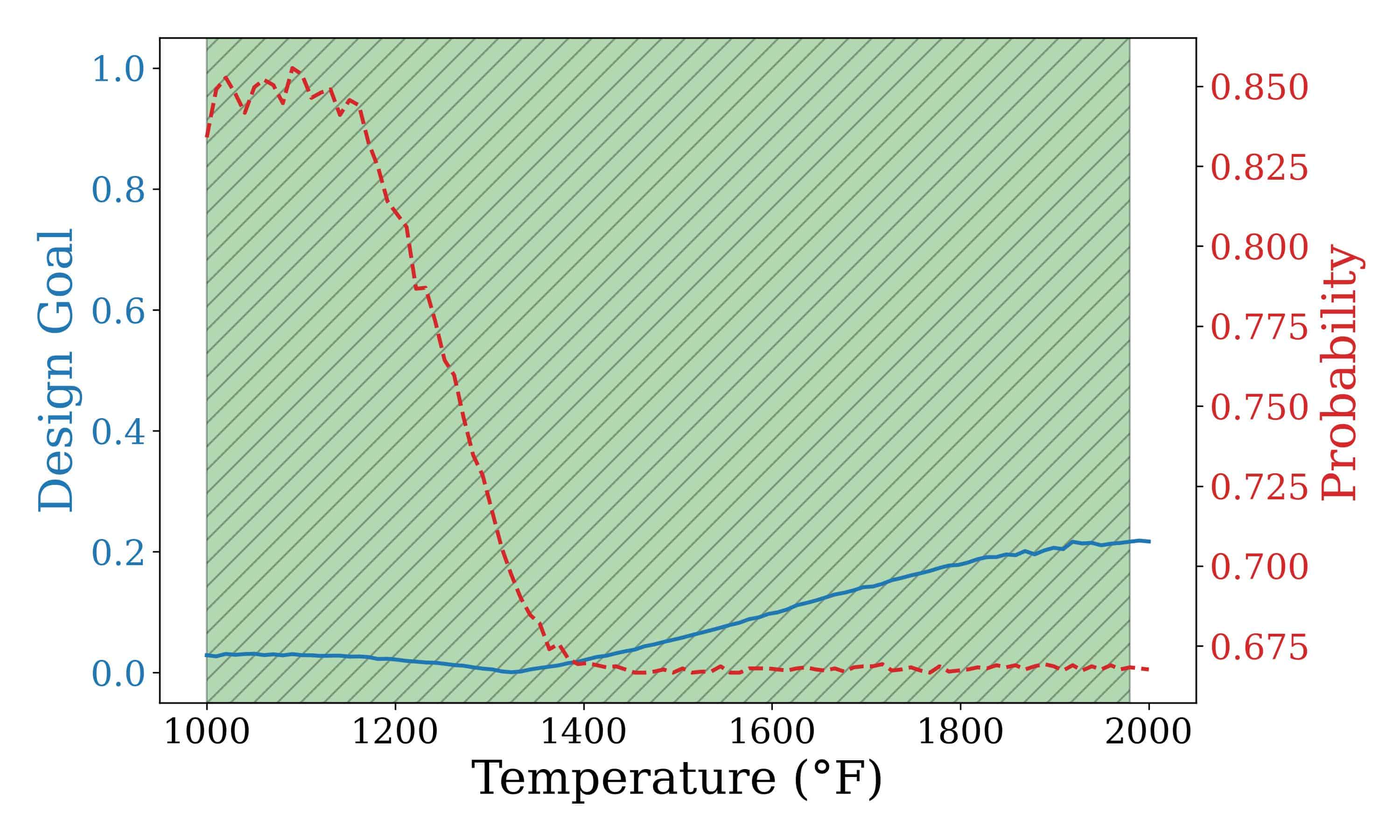}
    \end{subfigure}
    \hfill
    \begin{subfigure}[b]{0.3\textwidth}
      \includegraphics[width=\textwidth]{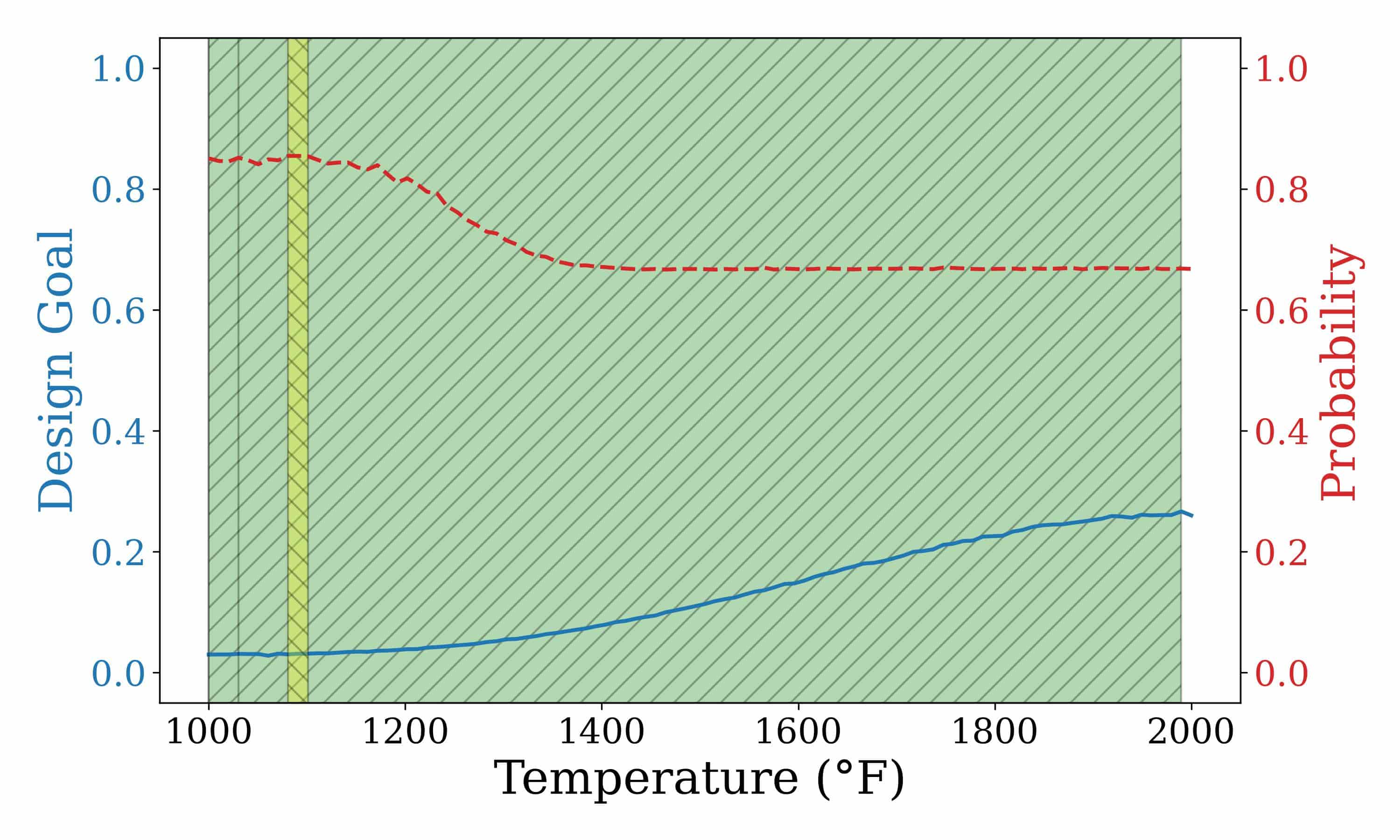}
    \end{subfigure}

    \vspace{0.5cm}
   \begin{subfigure}[b]{0.05\textwidth} 
    \centering
    \rotatebox{90}{\textcolor{black}{\parbox{4cm}{\centering\large \textbf{LRL = 200 MPa}}}}
\end{subfigure}
    \begin{subfigure}[b]{0.3\textwidth}
      \includegraphics[width=\textwidth]{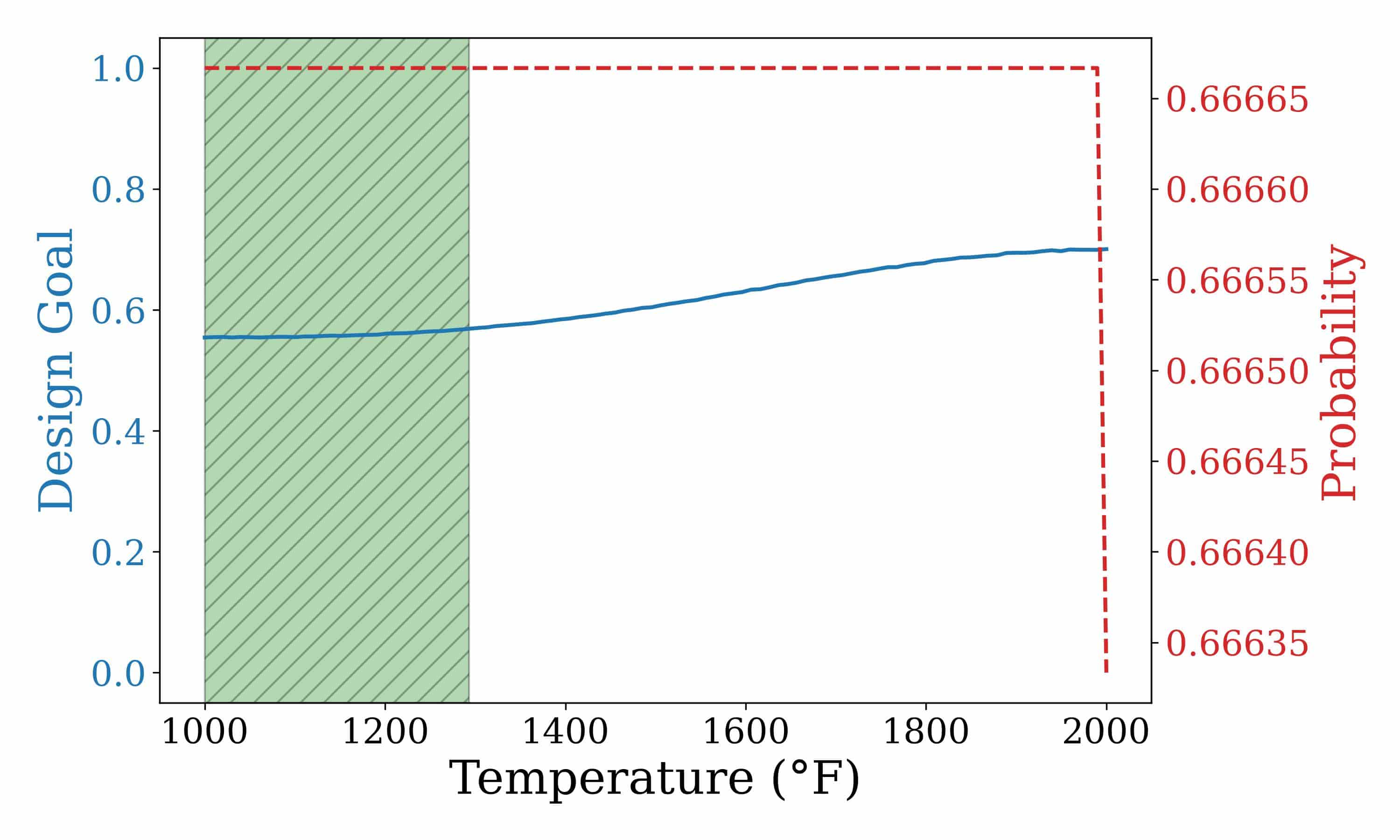}
    \end{subfigure}
    \hfill
    \begin{subfigure}[b]{0.3\textwidth}
      \includegraphics[width=\textwidth]{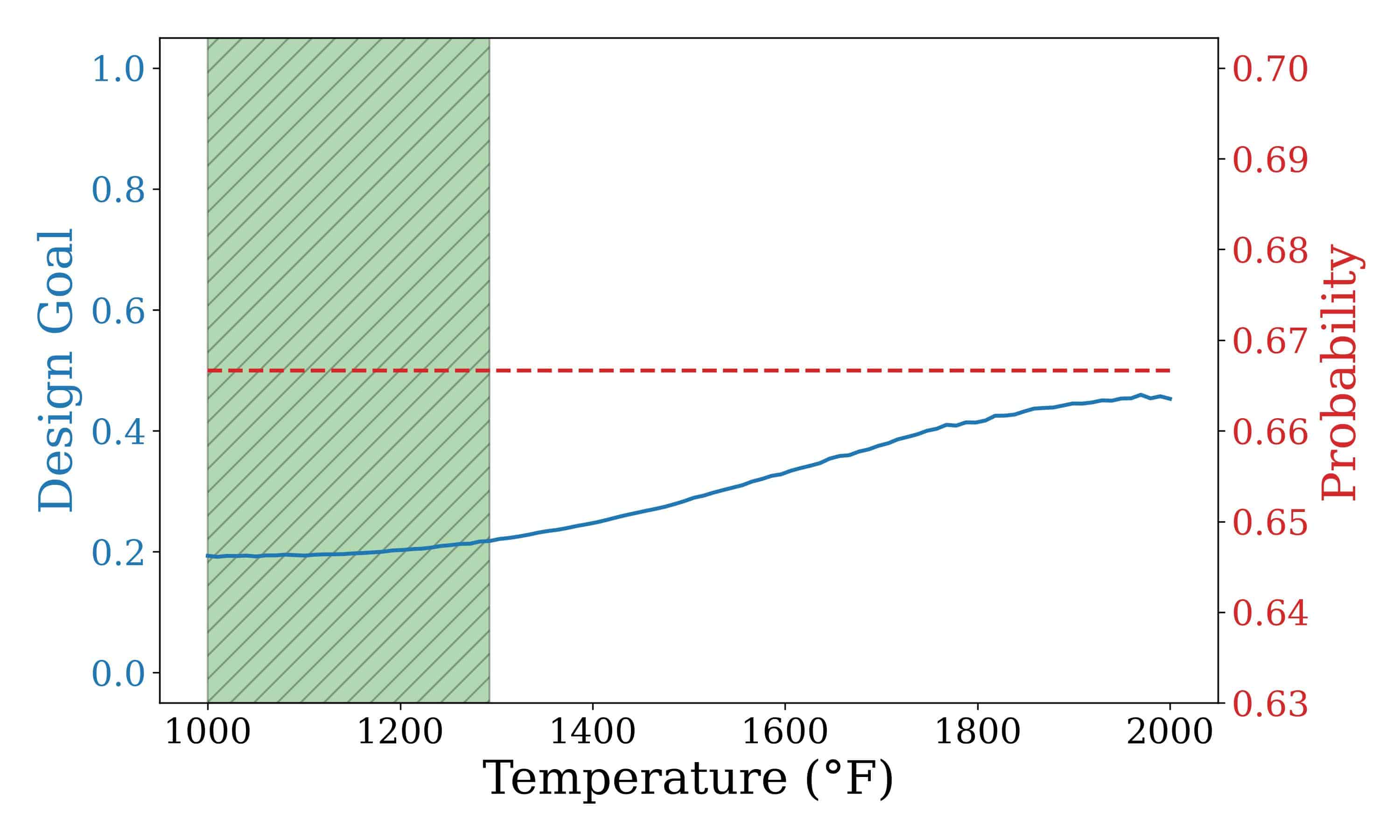}
    \end{subfigure}
    \hfill
    \begin{subfigure}[b]{0.3\textwidth}
      \includegraphics[width=\textwidth]{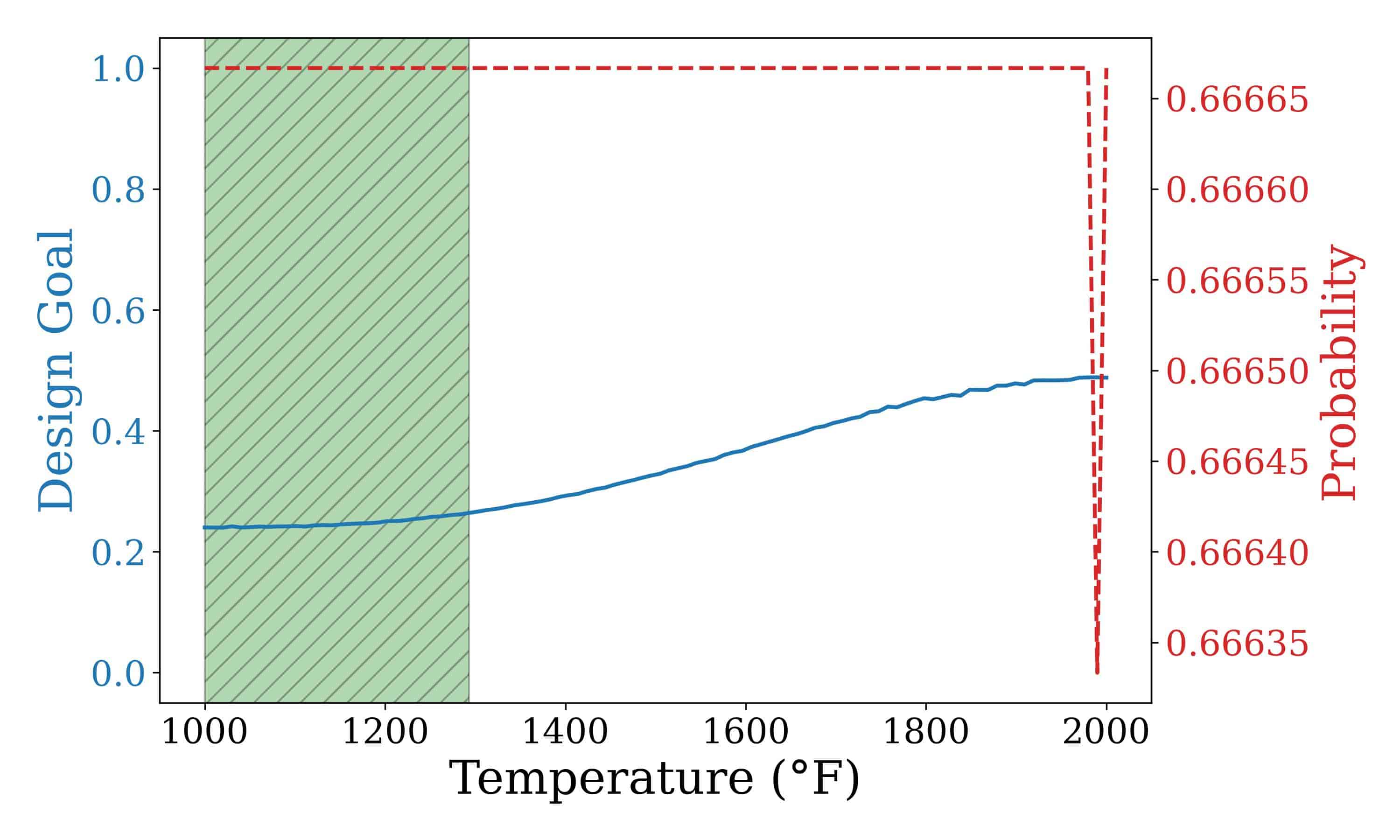}
    \end{subfigure}

    \vspace{0.3cm}
    \begin{subfigure}[b]{0.05\textwidth}
    \end{subfigure}
    \begin{subfigure}[b]{0.3\textwidth}
      \parbox{8cm}{\centering \large \textbf{(a) $\alpha = 0.99$, $EMI_{\text{target}} = 2.32$}}
    \end{subfigure}
    \hfill
    \begin{subfigure}[b]{0.3\textwidth}
      \parbox{7cm}{\centering \large \textbf{(b) $\alpha = 0.90$, $EMI_{\text{target}} = 1.28$}}
    \end{subfigure}
    \hfill
    \begin{subfigure}[b]{0.3\textwidth}
      \parbox{6cm}{\centering \large \textbf{(c) $\alpha = 0.85$, $EMI_{\text{target}} = 1.036$}}
    \end{subfigure}

  \end{subfigure}

  \caption{Comparison of $\text{RcDSP}$  and $\text{rcDSP}$ Methodologies Across Different LRL Values and Various Reliability Indexes for Case D}
  \label{CASED_RB_vs_RBD}
\end{figure*}

Figures \ref{CASEC_RB_vs_RBD} and \ref{CASED_RB_vs_RBD} demonstrate a significant divergence in the admissible design space between rcDSP and RcDSP as LRL increases. This discrepancy arises due to a pronounced right-skewed distribution of the outputs, driven by performance and model uncertainties tied to type 2 robustness. As illustrated in Cases C and D of Figure \ref{yield_distributions}, these uncertainties result in strongly non-normal behavior, where the distribution of the yield strength deviates strongly from symmetric, bell-shaped patterns (our case studies might appear unrealistic to some readers, but they are intentionally designed to emphasize the differences between the two design scenarios). For higher LRL values (e.g., LRL = 200), an admissible solution space exists for rcDSP, but the same solution is not acceptable for RcDSP due to the absence of a sufficiently reliable design space. This highlights the impact of skewness and higher-order distribution effects \cite{madsen2006methods}, which rcDSP does not account for, whereas RcDSP explicitly considers tail probabilities \cite{rohatgi2015introduction}.
Furthermore, the probability of achieving the target exhibits a sudden drop when transitioning from the admissible space of RcDSP to a non-admissible space, reinforcing the impact of distributional skewness, a factor that is overlooked in rcDSP. The results highlight a key limitation of rcDSP: it overlooks the asymmetric, non-normal uncertainties that dominate at higher LRLs. In contrast, RcDSP explicitly incorporates these effects, revealing stricter constraints on feasible designs. The summary of results will further detail the implications of this behavior for optimal design goals and temperature profiles.

\begin{table*}[t]
    \centering
    \caption{Comparison of the Achieved Design Goal of cDSP Under Reliability-Based and Robust Constraints}
    \label{tab:ComprehensiveResult}  
    \begin{small}
    \begin{tabular}{l *{8}{c}}  
    \toprule
    \multirow{2}{*}{Experiments} & 
    \multicolumn{3}{c}{Conditions} & 
    \multicolumn{3}{c}{$rcDSP$} & 
    \multicolumn{2}{c}{$RcDSP$} \\
    \cmidrule(lr){2-4} \cmidrule(lr){5-7} \cmidrule(lr){8-9}  
    & Case & $LRL(MPa)$ &$\alpha_{T}$ &$d$  & $\alpha_A$ & $T_{O}(\,\text{\textdegree}\mathrm{F})$& $d$ &$T_{O}(\,\text{\textdegree}\mathrm{F})$ \\
    \midrule
    1 & A & 200 &0.99 & 10.74 & 0.99 & 1000 & 10.74 & 1000 \\
    2 & A & 270 &0.99 & 0.083 & 0.982 & 1343& 0.134& 1333 \\
    3 & A & 280 &0.99 & 0.0021 & 0.99 & 1121& 0.0021& 1121 \\
     4 & A & 200 &0.95 & 16.34 & 1 & 1010& 16.34& 1010 \\
     5 & A & 270 &0.95 & 0.2068 & 0.95 & 1353& 0.2068& 1353 \\
     6 & A & 280 &0.95 & 0.05451 & 0.95 & 1151& 0.05451& 1151 \\
     7 & A & 200 &0.90 & 21.35 & 1 & 1000& 21.35& 1000 \\
     8 & A & 270 &0.90 & 0.0438 & 0.898 & 1363& 0.3931& 1353 \\
      9 & A & 280 &0.90 & 0.0268 & 0.90 & 1161& 0.0268& 1161 \\
      10 & B & 200 &0.99 & 0.0066 & 0.99 & 1292& 0.0066& 1292 \\
      11 & B & 270 &0.99 & NA & - & -& NA& - \\
      12 & B & 280 &0.99 & NA & - & -& NA& - \\
      13 & B & 200 &0.95 & 1.5764 & 1 & 1969& 1.5764& 1964 \\
      14 & B & 270 &0.95 & NA & - & -& NA& - \\
      15 & B & 280 &0.95 & NA & - & -& NA& - \\
      16 & B & 200 &0.90 & 2.2490 & 0.99 & 1939& 2.2490& 1939 \\
      17 & B & 270 &0.90 & NA & - & -& NA& - \\
      18 & B & 280 &0.90& NA & - & -& NA& - \\
      19 & C & 150 &0.99 & 0.2303 & 1 & 1000& 0.2303& 1000 \\
      20 & C & 180 &0.99 & 0.4157 & 0.99 & 1010& 0.4157& 1010 \\
      21 & C & 200 &0.99 & 0.5388 & 0.66 & 1000& NA& - \\
      22 & C & 150 &0.90 & 0.6595 & 1 & 1888& 0.6595& 1888 \\
      23 & C & 180 &0.90 & 2.7E-6 & 0.66 & 1343& 0.0243& 1222 \\
       24 & C & 200 &0.90 & 0.1637 & 0.66 & 1000& NA& - \\
       25 & C & 150 &0.85 & 0.68 & 1 & 1979& 0.68& 1979 \\
       26 & C & 180 &0.85 & 0.177 & 0.66 & 1737& 0.2637& 1232 \\
       27 & C & 200 &0.85& 0.00012 & 0.66 & 1131& NA& - \\
       28 & D & 150 &0.99 & 0.2763 & 0.99 & 1606& 0.2763& 1606 \\
       29 & D & 180 &0.99 & 0.4646 & 0.966 & 1242& NA& - \\
        30 & D & 200 &0.99 & 0.5542 & 0.66 & 1000& NA& - \\
    31 & D & 150 &0.9 & 0.1401 & 0.98 & 1979& 0.1401& 1979 \\
     32 & D & 180 &0.9 & 0.0007 & 0.6973 & 1323& NA& - \\
     33 & D & 200 &0.9 & 0.1915 & 0.66 & 1000& NA& - \\
     34 & D & 150 &0.85 & 0.0702 & 0.9833 & 2000& 0.0702& 2000 \\
     35 & D & 180 &0.85 & 0.0280 & 0.84 & 1060& 0.0298& 10540 \\
     36 & D & 200 &0.85 & 0.2398& 0.66 & 1020& NA& - \\
    \bottomrule
    
    \end{tabular}
    \end{small}
    \label{Comparative results}
\end{table*}

\section{Summary of Results and Considerations}\label{sect:summary}

Our analysis of the hot rod rolling process for the 36 cases is presented in Table~\ref{Comparative results} compares the design goal \( d \) by evaluating it against robust and reliable constraints. This comparison is conducted for different values of the LRL and target reliability \( \alpha_T \) across various cases. For each experiment, the $EMI_{target}$ and $\alpha_T$ have been selected so that the reliability and robust-based approach would result in the same admissible design space when the system level performance is normally distributed. The goal is to assess how well \( d \) satisfies the robust and reliable criteria under varying conditions. In this section, we analyze the results and highlight key considerations that emerge from the analysis.

We introduce \(\alpha_A\), the maximum reliability attained through the rcDSP process. If \(\alpha_A \geq \alpha_T\), the rcDSP process meets the desired reliability goal. Conversely, if \(\alpha_A < \alpha_T\), the design falls short of the target reliability and is deemed unreliable based on the specified criteria. Here, the role of the designer is emphasized in choosing whether to proceed with a robust design while compromising reliability or to refine the design to achieve the desired reliability.

For example, as illustrated in Figure~\ref{yield_distributions}, the presence of model uncertainty (Cases C and D) introduces significant skewness in the output distribution, causing it to deviate from normal behavior. However, as demonstrated in Figures~\ref{CASEC_RB_vs_RBD} and~\ref{CASED_RB_vs_RBD}, rcDSP maintains an acceptable admissible space, indicating that model uncertainty alone does not critically compromise robustness. This suggests that rcDSP is largely indifferent to the shape of the overall uncertainty distribution, provided the resulting model meets key performance criteria such as stability and accuracy. However, for RcDSP the shape of the overall output distribution influenced by different sources of uncertainty has a stronger impact. This often reduces or eliminates the admissible design space. This highlights that RcDSP is highly sensitive to the accuracy of the assumed model. Deviations from normal assumptions can cause significant shifts in the optimal solution or render reliable designs infeasible. On the other hand, this relationship between the shape of the final distribution and RcDSP can lead to designs that, while not robust, exhibit fully reliable behavior. This is largely influenced by the behavior of subsystems, which ultimately determine the overall system performance. However, in the hot rod rolling case studied here, these findings reveal a key distinction: rcDSP tends to be less conservative, which can result in solutions that may fail to meet stricter reliability requirements.
According to Table \ref{Comparative results}, RcDSP consistently outperforms rcDSP in achieving higher design goals across various cases, provided that an admissible design space exists. This highlights the conservative nature of RcDSP in this case study. For instance, in Experiment 23, the design goal obtained using RcDSP is an order of magnitude larger than that of rcDSP. While rcDSP aims for optimality under nominal assumptions, its solutions frequently fall short of reliability requirements. This is evident in the same experiment, where the design goal achieved by rcDSP corresponds to an unacceptably low-reliability index.

Our results for the hot rod rolling case study highlight a critical limitation of rcDSP solutions in this specific context. While rcDSP solutions in this case study may satisfy robust constraints, they often fail to meet reliability targets, revealing a significant gap between robustness and reliability. For example, in Experiment 24, the design achieved under rcDSP conditions satisfies all robust constraints ($EMI > 1$) but resulted in a reliability index of 0.66, significantly below the target reliability of 0.9. This gap underscores that, in this specific case study, the emphasis of rcDSP on the robustness of the design can lead to designs that are statistically unreliable, even if they appear robust under idealized assumptions. Such outcomes demonstrate that solutions of rcDSP, while mathematically valid within its framework, may mislead the designer into having undue trust that their design will meet requirements. This highlights a potential limitation of relying solely on rcDSP when reliability is critical, as it may prioritize efficiency over adequately addressing risk.

The findings presented in this study are specific to the case of hot rod rolling and should not be generalized without further investigation. While our results highlight the conservative nature of the RcDSP in this particular context, it is important to note that this behavior is not universal. In some scenarios, rcDSP may also exhibit conservative tendencies, depending on the system and constraints involved. For example, the output distribution for Case D shows long tail probabilities toward low yield strength values as shown. However, if these long tail probabilities were to be on the other side of the distribution the rcDSP would suggest designs with improved average predicted performance whereas the RcDSP approach would suggest overly conservative designs. Consequently, these results suggested that designers would benefit from using the rcDSP formulation when their system-level outputs are expected to be non-normally distributed, a scenario that is typically the case when dealing with multidisciplinary systems.

An additional consideration is that evaluation of the RcDSP is computationally more efficient than rcDSP as the mean is a direct output of the system-level simulation and the variance can be estimated through techniques such as first-order Taylor series approximation. This is especially true for Type 1 robust design problems. Conversely, the rcDSP formulation will typically require many hundreds of function evaluations to approximate the reliability index, this is especially true when high levels of reliability are required (e.g., 99\% or more). 

Finally, a comprehensive understanding of the system response, in terms of both reliability and robustness, provides designers with valuable feedback to refine and improve the design in alignment with desired goals. This approach not only enhances the design process but also reduces costs by preventing the allocation of resources to unnecessary or less critical improvements. By focusing on the most impaction aspects of the system, designers can optimize performance while efficiently managing resources.

\section{Concluding Remarks}
In this paper, we presented a comparative framework for robust and reliability-based design for compromise decision support problems by applying it to a multidisciplinary hot rod rolling process. By integrating Gaussian process surrogate modeling, we enabled systematic uncertainty propagation across subsystems, enabling a comprehensive evaluation of design trade-offs under varying conditions including the shape of uncertainty sources. Our results demonstrate that the reliability-based approach offers a broader admissible design space and is highly sensitive to the shape of the uncertainty distribution, making it a practical choice when model assumptions are uncertain. In contrast, the robust approach works well when the system-level output is normally distributed and has a reduced computational cost. The key insight from this study is that while robustness-focused approaches may optimize performance under nominal conditions, they do not always ensure reliability, especially in cases with non-normally distributed system-level outputs. Conversely, reliability-based design methods enforce stricter constraints, making them more suitable for safety-critical applications where the consequences of failure are severe. The practical implications of these findings suggest that designers should carefully assess the nature of uncertainty in their systems before selecting an optimization strategy.

Future work can include the extension of this framework to more complex engineering applications, including scenarios with dynamic uncertainties and multi-fidelity models. Additionally, refining the integration of data-driven methods with physics-based models will further enhance the predictive capabilities of surrogate modeling in multidisciplinary system design. By bridging the gap between robustness and reliability, this research provides a structured approach for managing uncertainty in engineering decision-making, facilitating the realization of designs with reduced sensitivity to uncertainty and improved performance.


\section{Acknowledgments}
Anton van Beek and Maryam Ghasemzadeh acknowledge the support provided by the Chang'an-Dublin International College of Transportation (CDIC). Additionally, Anand Balu Nellippallil and HM Dilshad Alam Digonta extend their gratitude for the support received through NSF Award 2301808.

\bibliographystyle{asmeconf}  
\bibliography{main}
\end{document}